\documentclass[useAMS,usenatbib]{mn2e}
\usepackage[colorlinks,urlcolor=blue,citecolor=blue,linkcolor=blue]{hyperref}
\usepackage{graphicx}
\usepackage{epstopdf,color,verbatim}
\usepackage{threeparttable}
\usepackage[]{txfonts}
\def\simless{\mathbin{\lower 3pt\hbox
{$\rlap{\raise 5pt\hbox{$\char'074$}}\mathchar"7218$}}}   
\def\simmore{\mathbin{\lower 3pt\hbox
{$\rlap{\raise 5pt\hbox{$\char'076$}}\mathchar"7218$}}}   
\newcommand{\eqb}{\begin{eqnarray}}
\newcommand{\eqe}{\end{eqnarray}}
\newcommand{\be}{\begin{eqnarray}}
\newcommand{\ee}{\end{eqnarray}}
\newcommand{\bi}{\begin{itemize}}
\newcommand{\ei}{\end{itemize}}

\newcommand{\lyub}{\citet{lyubarsky_05}}

\newcommand{\unit}[1]{\nobreak{\mathrm{\;#1}}} 

\def\comp{\,c/\omega_{\rm p}}

\newcommand{\sect}[1]{Sect.~\ref{sec:#1}}
\newcommand{\app}[1]{Appendix~\ref{app:#1}}
\newcommand{\fig}[1]{Fig.~\ref{fig:#1}}
\newcommand{\tab}[1]{Table~\ref{tab:#1}}
\newcommand{\eq}[1]{Eq.~(\ref{eq:#1})}

\newcommand{\lab}{_{\rm lab}}
\newcommand{\rhot}{\,r_{0,\rm hot}}
\newcommand{\alf}{Alfv\'en}

\newcommand{\lc}{{\,L/c}}

\usepackage{graphicx}
\usepackage{epsfig} 
\usepackage{epstopdf}

\title[]
{Plasmoids in relativistic reconnection, from birth to adulthood: \\first they grow, then they go}

\author[Lorenzo Sironi,  Dimitrios Giannios and Maria Petropoulou]
{Lorenzo Sironi,$^1$\thanks{E-mail: lsironi@cfa.harvard.edu.} 
 Dimitrios Giannios$^{2}$ and Maria Petropoulou$^{2}$\thanks{NASA Einstein Post-Doctoral Fellow.}\\
$^1$Harvard-Smithsonian Center for Astrophysics, 
60 Garden Street, MA 02138, USA\\
$^{2}$Department of Physics and Astronomy, Purdue University, 525 Northwestern
Avenue, West Lafayette, IN 47907, USA}

\begin{document}
\date{Received / Accepted}
\pagerange{\pageref{firstpage}--\pageref{lastpage}} \pubyear{2015}

\maketitle

\label{firstpage}

\begin{abstract}
Blobs, or quasi-spherical emission regions containing relativistic particles and magnetic fields, are often assumed {\it ad hoc} in emission models of relativistic astrophysical jets, yet their physical origin is still not well understood. Here, we employ a suite of large-scale two-dimensional particle-in-cell simulations in electron-positron plasmas to demonstrate that relativistic magnetic reconnection can naturally account for the formation of quasi-spherical plasmoids filled with high-energy particles and magnetic fields. Our simulations extend to unprecedentedly long temporal and spatial scales, so we can capture the asymptotic physics independently of the initial setup. We characterize the  properties of the plasmoids that are continuously generated as a self-consistent by-product of the reconnection process: they are in rough energy equipartition between particles and magnetic fields (with kinetic and magnetic energy densities proportional to the magnetization $\sigma$); the upper energy cutoff of the plasmoid particle spectrum is proportional to the plasmoid width $w$, corresponding to a Larmor radius $\sim 0.2 \,w$; the plasmoids grow in size at $\sim 0.1$ of the speed of light (roughly half of the reconnection inflow rate), with most of the growth happening while they are still non-relativistic (``first they grow''); their growth is suppressed once they get accelerated to relativistic speeds by the field line tension, up to a terminal four-velocity $\sim\!\sqrt{\sigma}\,c$ (``then they go''). The largest plasmoids, whose typical recurrence interval is $\sim 2.5\,L/c$, reach a characteristic size $w^{\rm max}\sim 0.2\, L$ independently of the system length $L$, they have nearly isotropic particle distributions and they contain the highest energy particles, whose  Larmor radius is $\sim 0.03 \, L$. The latter can be regarded as the \textit{Hillas criterion for relativistic reconnection}. We briefly discuss the implications of our results for the high-energy emission from relativistic jets and pulsar winds.
\end{abstract} 

\begin{keywords}
galaxies: jets --- magnetic reconnection --- gamma-ray burst: general --- MHD --- pulsars: general --- radiation mechanisms: non-thermal
\end{keywords}

\section{Introduction}\label{sec:intro}
It is generally thought that pulsar winds and the relativistic jets of blazars and gamma-ray bursts are launched hydromagnetically \citep[][]{spruit_10}. The strong magnetic fields threading a rotating compact object or the associated accretion disk serve to convert the rotational energy of the central engine into the power of the relativistic outflow. Since the energy is carried initially in the form of Poynting flux, it is a fundamental question how and where the energy in the  fields is transferred to the plasma, and then radiated away to power the observed emission. Because of their tightly wound-up magnetic fields, jets may be susceptible to magnetohydrodynamic (MHD) kink instabilities. At their non-linear stages, these MHD instabilities introduce small-scale magnetic field reversals that lead to dissipation of magnetic energy through magnetic reconnection \citep{begelman_98,spruit_01}. Alternatively, the outflow may contain current sheets from its base, as it is the case in pulsar winds, if the pulsar rotational and magnetic axes are not aligned \citep{lyubarsky_kirk_01}. 
In either case, field dissipation via magnetic reconnection has been often invoked to explain the non-thermal signatures of pulsar wind nebulae \citep[PWNe; e.g.,][]{lyubarsky_kirk_01,lyubarsky_03,kirk_sk_03,petri_lyubarsky_07,sironi_spitkovsky_11b}, jets from active galactic nuclei \citep[AGNs; e.g.,][]{romanova_92,giannios_09,giannios_10b,giannios_13} and gamma-ray bursts \citep[GRBs; e.g.,][]{thompson_94, thompson_06,usov_94,spruit_01,drenkhahn_02a,lyutikov_03,giannios_08}. 
Despite decades of research, we have no reliable theory built from first principles for the particle distribution, geometry and magnetic field to be expected in the radiating regions of a reconnection-dominated system. 

In relativistic astrophysical outflows, reconnection proceeds in the ``relativistic'' regime in which the magnetic energy per particle can exceed the rest mass energy (or equivalently, the magnetization $\sigma$ is larger than unity). The flow dynamics in relativistic reconnection has been satisfactorily described by analytical studies \citep[e.g.,][]{lyutikov_uzdensky_03,lyubarsky_05}, even though analytical models still have to make assumptions on the resistive processes at work, that critically affect the geometry of the layer. However, the acceleration process of the emitting particles can only be captured from first principles by means of fully-kinetic particle-in-cell (PIC) simulations. Energization of particles in relativistic reconnection of pair plasmas has been investigated in a number of PIC studies, both in two dimensions \citep[2D; e.g.,][]{zenitani_01,zenitani_07,jaroschek_04,bessho_05,bessho_07,bessho_10,bessho_12,hesse_zenitani_07,daughton_07,lyubarsky_liverts_08,cerutti_12b,ss_14,guo_14,guo_15a,liu_15,nalewajko_15,sironi_15,werner_16,kagan_16} and three dimensions \citep[3D; e.g.,][]{zenitani_05b,zenitani_08,yin_08,liu_11,sironi_spitkovsky_11b, sironi_spitkovsky_12,kagan_13,cerutti_13b,ss_14,guo_15a}. Recently, 2D PIC simulations have started to tackle the dynamics and acceleration capabilities of relativistic reconnection in electron-ion plasmas \citep[e.g.,][]{melzani_14,sironi_15,guo_16}.

PIC simulations can now reliably simulate the reconnection region, measure the reconnection speed and identify the mechanisms of particle acceleration (see \citealt{kagan_15} for a review). Yet, the separation between the microscopic plasma scales that PIC simulations need to resolve and the large astrophysical scales where the emission takes place is often precluding a direct application of PIC findings to astrophysical observations. It is only when PIC studies are performed with a domain much larger than the microscopic plasma scales, that the results can be properly employed to model astrophysical sources. With large-scale simulations, \citet{ss_14} (hereafter, SS14) have shown that non-thermal particle acceleration is a generic by-product of the long-term evolution of relativistic reconnection, in both 2D and 3D, and that the accelerated particles populate a a power law whose slope is harder than $-2$ for magnetizations $\sigma\gtrsim 10$. With large-scale PIC simulations, we have demonstrated that reconnection can satisfy all the basic conditions for the high-energy emission from blazar jets \citep[hereafter, SPG15]{sironi_15}: efficient dissipation, extended particle distributions, and rough equipartition between particles and magnetic field in the emitting region.

In this work, we employ a suite of large-scale 2D PIC simulations in electron-positron plasmas to follow the evolution of the reconnection layer to unprecedentedly long temporal and spatial scales, so we can capture the asymptotic physics independently of the initial setup of the current sheet. Earlier works were often limited to small domains, nearly one order of magnitude smaller than what we employ here. As a result, transient effects that depended on the initialization of the current sheet were artificially over-emphasized, while particle distributions did not have sufficient time to isotropize \citep[e.g.,][]{cerutti_13a}, or particle acceleration to the highest energies was artificially inhibited \citep[e.g.,][]{werner_16}. Also, the common choice of periodic boundary conditions in the outflow direction (as opposed to the absorbing/outflow boundary conditions that we employ here) limited the time that these simulations could run before the reconnection process was choked, which resulted in underestimating the terminal speed of the reconnection outflow \citep[e.g.,][]{guo_15a}.

With our large-scale simulations, we investigate the properties of the chain of plasmoids/magnetic islands that are constantly generated in the reconnection layer by the secondary tearing instability \citep{uzdensky_10}, as a self-consistent by-product of the system evolution. We argue that such plasmoids play the role of the ``blobs'' that are commonly invoked in phenomenological models of relativistic astrophysical jets, i.e., quasi-spherical emission regions containing relativistic particles and magnetic fields. We show that the plasmoids are indeed in rough energy equipartition between particles and magnetic fields (with kinetic and magnetic energy densities proportional to the magnetization $\sigma$), and that the upper energy cutoff of the plasmoid particle spectrum is proportional to the plasmoid width $w$, corresponding to a Larmor radius $\sim 0.2 \,w$. 

By following each individual plasmoid over time, we find that their life can be separated into two phases: \textit{first they grow, then they go}. The plasmoids grow in size at $\sim 0.1$ of the speed of light (roughly half of the reconnection inflow rate), with most of the growth happening while they are still non-relativistic; their growth is suppressed once they get accelerated to relativistic speeds by the field line tension, up to the terminal four-velocity $\sim\!\sqrt{\sigma}\,c$ expected from analytical models \citep{lyubarsky_05}. The largest plasmoids, occurring every $\sim 2.5\,L/c$, reach a characteristic size $w^{\rm max}\sim 0.2\, L$ independently of the system length $L$, they have nearly isotropic particle distributions and they contain the highest energy particles, whose  Larmor radius is $\sim 0.03 \, L$. The latter can be regarded as the \textit{Hillas criterion for relativistic reconnection}.

This work is organized as follows. In \sect{setup} we describe the simulation setup and our method for identifying and tracking the plasmoids. In \sect{struct} we discuss the overall structure of the reconnection layer, whereas \sect{chain} is devoted to the investigation of the plasmoid properties (i.e., their fluid properties, the particle population that they contain, the plasmoid growth and bulk acceleration). In \sect{size} we show how our conclusions depend on the system size $L$, and emphasize the artificial constraints imposed by small computational domains. This will allow us to extrapolate our results from our large-scale PIC simulations to the macroscopic scales relevant for the blazar emission (Petropoulou, Giannios and Sironi, in prep.). In \sect{summary} we summarize our findings and describe the astrophysical implications of our work.

\section{Simulation setup} \label{sec:setup}
We use the 3D electromagnetic PIC code TRISTAN-MP \citep{buneman_93, spitkovsky_05} to study relativistic reconnection in pair plasmas. Our simulations employ a 2D spatial domain, but we track all three components of the velocity and of the electromagnetic fields. We investigate the case of anti-parallel reconnection, i.e., in the absence of a guide field perpendicular to the alternating fields.
The reconnection layer is set up in Harris equilibrium, with the initial magnetic field $\bmath{B}_{\rm in}=-B_0\, \bmath{\hat{x}}\tanh\,(2\pi y/\Delta)$  reversing at $y=0$ over a thickness $\Delta$ that will be specified below. The field strength is parameterized by the magnetization $\sigma=B_0^2/4\pi m n_0 c^2=(\omega_{\rm c}/\omega_{\rm p})^2$, where $\omega_{\rm c}=eB_0/mc$ is the Larmor frequency and $\omega_{\rm p}=\sqrt{4\pi n_0 e^2/m}$ is the plasma frequency for the cold electron-positron plasma outside the layer (which is initialized with a small thermal spread of $k_B T/m c^2=10^{-4}$). The \alf\ speed is related to the magnetization as $v_A/c=\sqrt{\sigma/(\sigma+1)}$.
We focus on the regime $\sigma\gg1$ (i.e., $v_A/c\sim 1$) of relativistic reconnection, investigating three values of the magnetization: $\sigma=3$, 10 and 50 (see \tab{param}). 
The magnetic pressure outside the current sheet is balanced by the particle pressure in the sheet, by adding a component of hot plasma with overdensity $\eta=3$ relative to the number density $n_0$ of cold particles outside the layer. From pressure equilibrium, the temperature of the hot plasma in the sheet is $k_B T_{h}/m c^2=\sigma/2\eta$. The hot particles in the sheet are also initialized with a small drift speed along $z$ (electrons and positrons drifting in opposite directions), so that their electric current compensates the curl of the magnetic field.

\begin{table}
\centering
\caption{Physical Parameters of the Simulations}\label{tab:param}
\begin{tabular}{cccc}\hline\hline
$\;\;\sigma\;\;\;\;$ & $\;\;\;\;L/\comp\;\;\;\;\;\;$ & $L/\rhot$& ${\rm Duration} \;[L/c]$\\[4pt]
\hline\hline
3 & 1229 & 720 & 3.6 \\\hline
10 & 413 & 127 &  14.0 \\\hline
10 & 826 & 257 & 3.6 \\\hline
10 & 1651 & 518 & 3.6 \\\hline
10 & 3584 & 1130 & 3.6 \\\hline
50 & 3584 & 505 & 3.6 \\\hline\hline
\multicolumn{4}{l}{%
  \begin{minipage}{7cm}%
    We provide the system half-length $L$ in units of both the skin depth $\comp$ (second column), which is resolved with 10 cells, and of the Larmor radius of particles heated/accelerated by reconnection $\rhot=\sqrt{\sigma}\comp$ (third column). %
  \end{minipage}%
}\\
\end{tabular}
\end{table}
\vspace{0.35in}

We trigger reconnection near the center of the 2D computational domain, by removing the pressure of the hot particles initialized in the current sheet.\footnote{In essence, the current sheet particles around $x, \,y\sim 0$ are initialized with a small temperature, rather than $T_h$.} This triggers a local collapse of the current sheet, which generates an X-point at the center of the domain. After this initial perturbation, the system evolves spontaneously, i.e., we study spontaneous reconnection, as opposed to forced (or driven) reconnection. The initial perturbation results in the  formation of two ``reconnection fronts'' that propagate away from the center along $\pm\bmath{\hat{x}}$ (i.e., along  the current layer), at roughly the \alf\  speed $v_{A}=\sqrt{\sigma/({\sigma}+1)}\,c$ (see \sect{struct}). We choose the thickness of the current sheet $\Delta$ large enough such that reconnection does not get spontaneously triggered anywhere else in the current layer, outside of the region in between the two reconnection fronts. Taking $\rhot=\sqrt{\sigma}\comp$ as our unit of length, which corresponds to the Larmor radius of particles with energy $\sigma m c^2$ in the field $B_0$,\footnote{If reconnection were to transfer all of the field energy to the particles, the mean particle energy would be $\sim \sigma m c^2/2$. So, our definition of $\rhot$ corresponds, apart from a factor of two, to the Larmor radius of the particles heated/accelerated by reconnection.} the thickness $\Delta$ is chosen to be $\Delta\simeq29\rhot$ for $\sigma=3$, $\Delta\simeq22\rhot$ for $\sigma=10$, and $\Delta\simeq11\rhot$ for $\sigma=50$. We have tested that our results are insensitive to the value of $\Delta$ as long as it is appreciably larger than $\!\rhot$, so that active reconnection stays confined in between the two reconnection fronts.

After one \alf ic crossing time, the two reconnection fronts reach the $x$ boundaries of the computational box (see \sect{struct}). Here, we have explored two choices of boundary conditions: (\textit{i}) periodic boundary conditions in the $x$ direction, so that the particles outflowing from the center accumulate close to the boundaries, where a large magnetic island is formed; or (\textit{ii}) absorbing boundary conditions in the $x$ direction of the reconnection outflow, to mimic an open boundary in which no information is able to propagate back inward \citep{daughton_06,cerutti_15,belyaev_15}. The choice (\textit{i}) of periodic boundary conditions has two disadvantages: the large island contains the particles that were initialized in the current sheet, so the system still bears memory of the initial conditions; in addition, as the large island grows, the central region where reconnection stays active progressively shrinks, which prevents to study the long-term steady-state evolution of the system. For this reason, we adopt the choice (\textit{ii}) of absorbing boundary conditions, as soon as the two reconnection fronts reach the boundaries of the box (beforehand, periodic boundaries are used along the $x$ direction).
Even though we only present the results from simulations with absorbing boundaries, we have tested that our main conclusions are the same for both choices of boundary conditions.

In the case of absorbing boundary conditions in $x$, particles are removed from the simulation when they reach the two $x$ boundaries. In a region of width of 60 cells just inside of the two $x$ boundaries, the electromagnetic fields are set by hand to their initial values (i.e, $\bmath{B}_{\rm in}$ as specified above, and zero electric field). Further in, a finite width absorbing layer (with thickness $\Delta_{\rm abs}=50$ cells) is implemented, where Maxwell's equations contain an electric and a magnetic conductivity term, so that the fields are damped back to the initial conditions. In the absorbing layer, we solve
\be
\frac{\partial \bmath{B}}{\partial t}&=&-c\nabla\times\bmath{E}-\lambda(x)(\bmath{B}-\bmath{B}_{\rm in})\\
\frac{\partial \bmath{E}}{\partial t}&=&c\nabla\times\bmath{B}-4\pi \bmath{J}-\lambda(x)\bmath{E}
\ee
where the conductivity $\lambda(x)$ is a function of space within the absorbing layer. To minimize wave reflections off the inner edge of the absorbing layer, we gradually increase the conductivity toward the boundaries: if $x_{\rm 1}$ is the inner edge of the absorbing layer, the conductivity profile is $\lambda=(4/\Delta_{\rm abs}\delta t)[|x-x_1|/\Delta_{\rm abs}]^3$, where $\delta t$ is the simulation timestep. Outside of the absorbing layer, $\lambda=0$.

Along the $y$ direction of the reconnection inflow, we employ two Òmoving injectorsÓ (receding from $y=0$ at the speed of light along $\pm {\bmath{\hat{y}}}$) and an expanding simulation box, a technique that we have extensively employed in our studies of relativistic shocks \citep{sironi_spitkovsky_09,sironi_spitkovsky_11a,sironi_13} and relativistic reconnection (SS14). The two injectors constantly introduce fresh magnetized  plasma into the simulation domain. This permits us to evolve the system as far as the computational resources allow, retaining all the regions that are in causal contact with the initial setup. Such choice has clear advantages over the fully-periodic setup that is commonly employed, where the limited amount of particles and magnetic energy will necessarily inhibit the evolution of the system to long times, and the establishment of a steady state.

We resolve the plasma skin depth with $\comp=10$ cells, so that the Larmor gyration period $2\pi/\omega_{\rm c}=2\pi/\sqrt{\sigma}\,\omega_{\rm p}$ is resolved with at least a few timesteps, even for the largest magnetization $\sigma=50$ that we explore (the numerical speed of light is 0.45 cells/timestep).
 We investigate the long-term evolution of reconnection in large-scale computational domains. In the following, we shall call $L$ the half-length of the computational domain along the $x$ direction, i.e., along the reconnection layer. In units of the Larmor radius of hot particles $\rhot=\sqrt{\sigma}\comp$, the half-length $L$ for our fiducial runs is $L\simeq 720\rhot$ for $\sigma=3$, $L=1130\rhot$ for $\sigma=10$ and  $L=505\rhot$ for $\sigma=50$ (see \tab{param}). In units of the plasma skin depth, this amounts to $L/\comp\simeq1229$ for $\sigma=3$,  and $L/\comp\simeq3584$ for both $\sigma=10$ and $\sigma=50$. This corresponds to an overall box size along the $x$ direction (so, for the full length $2\,L$) of 24,576 cells for $\sigma=3$ and 71,680 cells for both $\sigma=10$ and $\sigma=50$.\footnote{The box size along $y$ increases over time, and at the end it is comparable or larger than the $x$ extent.} As we show below, such large domains are of paramount importance to reconcile the  results of PIC simulations with the analytical theory of relativistic magnetic reconnection by \citet{lyubarsky_05}, and to attain a sufficient dynamic range to separate effects happening on plasma scales  from the physics at macroscopic scales $\lesssim L$. 
 
For our reference case of $\sigma=10$, we investigate the dependence of our results on $L$, exploring also the cases $L\simeq127\rhot$, $L\simeq257\rhot$ and $L\simeq518\rhot$. We evolve the system up to a few $L/c$ (typically, up to $\simeq3.6 \,L/c$), corresponding to $\sim 95,000$ timesteps for $\sigma=3$ and $\sim 270,000$ timesteps for both $\sigma=10$ (as regard to our reference case with $L=1130\rhot$) and $\sigma=50$. This is sufficient to study with enough statistics the steady state of the system, which is established after $\sim L/c$, as described below.

We typically employ four particles per cell (including both species), but we have extensively tested that the physics is the same when using up to 64 particles per cell (the tests have been performed for the case with magnetization $\sigma=10$ and box sizes $L\simeq127\rhot$ and $L\simeq257\rhot$). In order to reduce noise in the simulation, we filter
the electric current deposited to the grid by the particles, effectively mimicking the role of a larger number of particles per cell \citep{spitkovsky_05,belyaev_15}.


\subsection{Plasmoid identification and tracking}
In this subsection, we describe our technique to identify the plasmoids that are self-consistently generated by reconnection, and to follow individual plasmoids over time. As it is customary in magnetohydrodinamic (MHD) simulations \citep{fermo_10,loureiro_12,huang_12,murphy_13}, the O-points at the center of plasmoids and the X-points in between neighboring plasmoids are identified in a 2D domain as local maxima and minima of the magnetic vector potential $A_z$. Apart from a minus sign, $A_z$ also corresponds to the magnetic flux function. Each maximum of $A_z$ identifies the O-point at the center of a plasmoid. The plasmoid area is defined as the region where the vector potential stays larger that the maximal value of $A_z$ of the two neighboring X-points (each one corresponding to a local minimum of $A_z$). In other words, the plasmoid contour corresponds to the equipotential line at the maximal value of $A_z$ of the two neighboring X-points. It is in this region that we measure the area-averaged properties of individual plasmoids, e.g., their density and magnetic and kinetic energy content. The logitudinal size of the plasmoid (or ``length'') is measured along the current sheet (i.e., along $x$ at $y=0$), whereas its transverse size (or ``width'') is measured along a cut in the $y$ direction taken at the location of the corresponding O-point. 

The plasmoid speed is defined as the local bulk velocity at the corresponding O-point. We have verified that an isolated plasmoid (i.e., not undergoing a merger with another plasmoid) moves nearly as a solid body, or equivalently that the electric field in the plasmoid comoving frame nearly vanishes. 
By knowing the plasmoid speed, we can readily measure the plasmoid properties in the comoving frame. Below, we shall indicate with a subscript ``lab'' all the quantities measured in the laboratory frame. Otherwise, we will refer to comoving quantities, with the exception of the spatial locations $x$ and $y$, the inflow speed $v_{\rm in}$ and the outflow four-velocity $\Gamma v_{\rm out}$, which are always measured in the laboratory frame. Here, the bulk Lorentz factor $\Gamma=(1-v_{\rm out}^2/c^2)^{-1/2}$.

By accounting for all the particles belonging to the plasmoid, we can compute the island energy and momentum spectrum, in the laboratory or comoving frame. 
Below, it will be convenient to have an estimate of the upper cutoff of the comoving momentum spectrum of individual plasmoids, which we measure as follows \citep[see also][]{bai_15}. If $p_i$ is the particle momentum in a given direction (below, we will primarily consider the positron momentum) measured in the plasmoid comoving frame, we define
\be\label{eq:pcut}
p_{i,\rm cut}=\frac{\sum_{\alpha \in P} p_{\alpha,i}^{n_{\rm cut} } }{\sum_{\alpha \in P} p_{\alpha,i}^{n_{\rm cut} -1}}
\ee
where the sum is extended over all the particles $\alpha$ belonging to the plasmoid $P$, and the power index $n_{\rm cut}$ is empirically chosen to be $n_{\rm cut}=6$. If the momentum distribution takes the form $dN/dp_i\propto p_i^{-s} \exp(-p_i/p_0)$ with power-law slope $s$ and exponential cutoff at $p_0$, then our definition yields $p_{i,\rm cut}\sim (n_{\rm cut}-s)\,p_0$. In the plasmoid comoving frame, a residual positron anisotropy might persist in the direction $+z$ of the reconnection electric field (with electrons having the opposite anisotropy). It will then be illuminating to distinguish the momentum spectrum of positrons having $p_{z}>0$ from positrons with $p_{z}<0$. The cutoffs of the corresponding momentum spectra will be indicated as $p_{+z,\rm cut}$ and $p_{-z,\rm cut}$, respectively. The quantity $p_{\rm cut}$ will refer to the cutoff momentum for the total comoving momentum $p={(p_x^2+p_y^2+p_z^2)}^{1/2}$. It will be convenient to cast the value of the momentum cutoffs in terms of the corresponding Larmor radius in the background field $B_0$, which will be indicated as $r_{0,\rm cut}=p_{\rm cut} c/eB_0$ for the total momentum $p_{\rm cut}$ and as $r_{0i,\rm cut}=p_{i,\rm cut} c/eB_0$ for the component along the direction $i$.

After having identified all the plasmoids in a given snapshot of the simulation, we describe how we follow the temporal trajectory of a given plasmoid, which allows to assess, e.g., how its growth proceeds over time. In our PIC code, each computational particle has a unique identifier. At any given time, we tag each new plasmoid that has been detected in the reconnection layer with a representative particle, whose Lorentz factor is typically chosen to be $1.2\leq\gamma_{\rm lab}\leq2.5$. This range ensures that we have enough particles to be able to tag each of the plasmoids (as we show below, the plasmoid spectrum starts at mildly relativistic energies), and at the same time we can be confident that the particle of our choice stays effectively trapped in the plasmoid (as opposed to a high-energy particle that was flying outward from the X-point and happened to lie within the plasmoid area at that particular time). At all subsequent times, the location of the chosen particle will allow to confidently track the temporal history of the given plasmoid.

At certain times, a single plasmoid might contain two or more of such ``tracer particles.'' This is a signature that a plasmoid merger has just occurred, and we can uniquely identify which plasmoids have been taking part in the merger event. The post-merger plasmoid will inherit the identifier from the largest of the pre-merger plasmoids (and inherit its corresponding tracer particle), while all the other pre-merger plasmoids will terminate their life history.

\begin{figure*}
\centering
 \resizebox{\hsize}{!}{\includegraphics{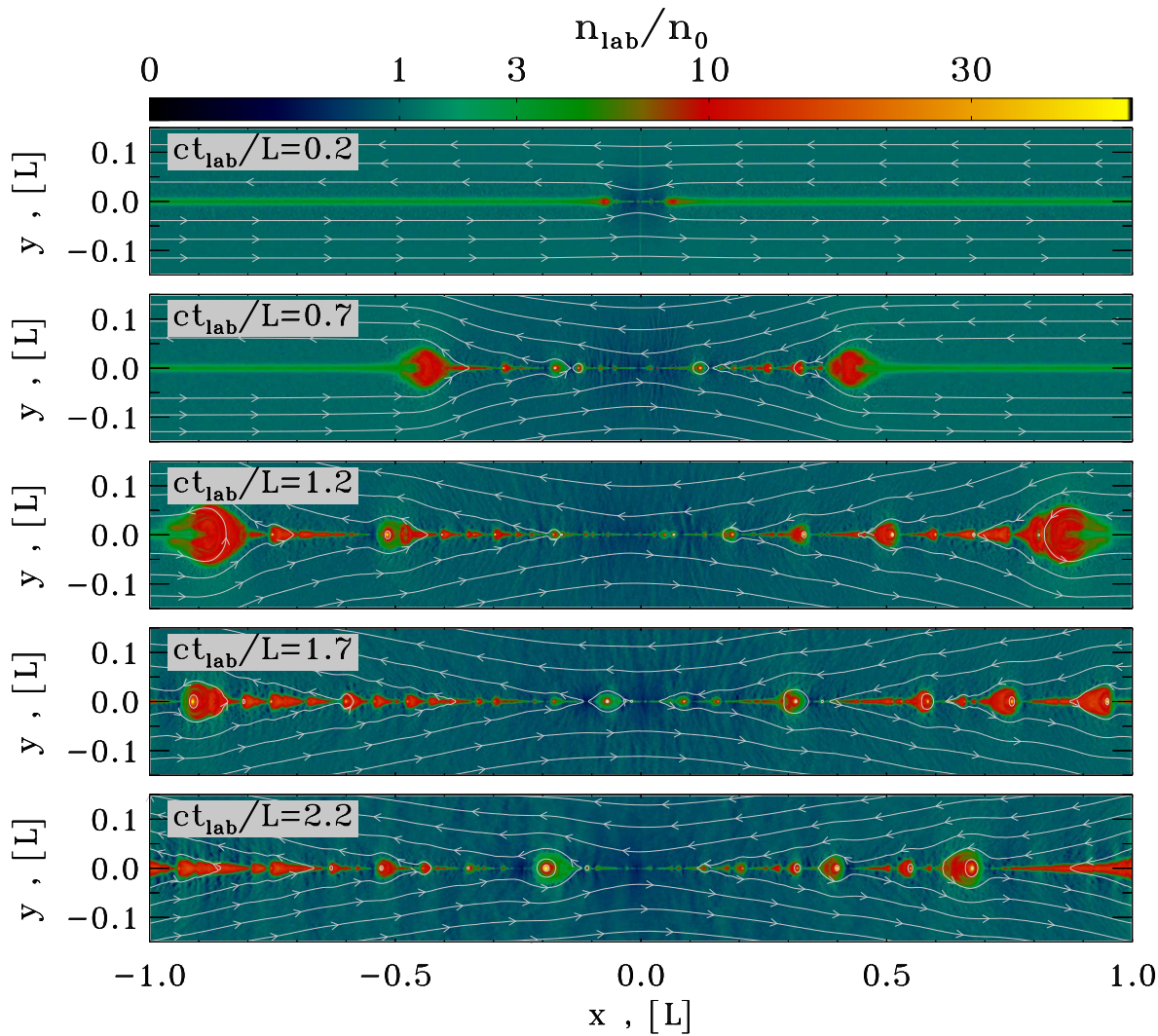}}
 \caption{2D structure of the particle number density in the lab frame $n_{\rm lab}$, in units of the lab-frame number density $n_0$ far from the reconnection layer, from a simulation with $\sigma=10$ and $L/r_{0,\rm hot}\simeq 518$. We only show the region $|y|/L<0.15$ to emphasize the small-scale structures in the reconnection layer (the extent of the computational box along $y$ increases at the speed of light, as described in \sect{setup}). The 2D density structure at different times (as marked on the plots) is shown in the panels from top to  bottom, with overplotted magnetic field lines. After triggering reconnection in the center of the current sheet ($x\simeq 0$ in the top panel), two ``reconnection fronts'' propagate to the right and to the left, reaching the boundaries of the box right after $ct_{\rm lab}/L= 1.2$ (see the two over-dense regions at $x\simeq -0.4\,L$ and $x\simeq 0.4\,L$ in the second panel, or at $x\simeq -0.85\,L$ and $x\simeq 0.85\,L$ in the third panel). The evolution of the reconnection layer after this time is completely independent of the initialization of the current sheet.}
 \label{fig:fluidtime}
\end{figure*}

\section{Structure and Evolution of the Reconnection Layer}\label{sec:struct}
We describe the evolution of the reconnection layer in our large-scale computational domain, from the initial setup until it reaches a statistical steady state. We focus on the overall structure of the current sheet, and we emphasize the dependence of the inflow and outflow speeds on the flow magnetization, demonstrating excellent agreement of the results of our PIC simulations with the analytical model by \citet{lyubarsky_05}.

\subsection{Towards a Steady-State}
\fig{fluidtime} illustrates the early phases of evolution of the system, presenting the 2D structure of the particle number density in the lab frame $n_{\rm lab}$ for a representative case with magnetization $\sigma=10$ and size $L/\rhot\simeq 518$. We only show the region $|y|/L<0.15$ closest to the current sheet, to emphasize the small-scale structures in the reconnection layer; the full extent of our computational box along $y$ is much larger, since it increases with time at the speed of light, as described in \sect{setup}. As anticipated in \sect{setup}, the reconnection process is initiated by hand at the center of the domain, by removing the thermal pressure of the hot particles in the current sheet near $x, y\sim0$. 
Our choice of driving the reconnection onset near the center--- rather than periodically modulating  the magnetic flux function, or letting the system go unstable via numerical noise --- would mimic the effect of a large-scale curvature of the field lines (over a scale $\sim L$), such that the current sheet is narrower near the center. The central region is then most likely to go unstable via the tearing mode,\footnote{For an unperturbed pair plasma that goes unstable via the tearing instability (seeded by numerical noise), the growth time of the  fastest growing mode increases with the current sheet thickness as $\Delta^{5/2}$ \citep[e.g.,][]{zenitani_07}, at fixed magnetization $\sigma$ and overdensity $\eta$.} and  the signal of ongoing reconnection will propagate toward the outer regions  (where the current sheet is broader) before they have time to become unstable.

The lack of pressure support  in the vicinity of $x,y\sim 0$ resulting from our initial perturbation triggers the collapse of the current sheet (top panel in \fig{fluidtime}), with the magnetic field lines that begin to get advected toward the center (the magnetic field lines are overplotted in grey in \fig{fluidtime}), where an X-point is formed. In the following, we indicate this X-point as the ``primary X-point.'' The primary X-point remains in the vicinity of $x\sim 0$ throughout the timespan of our simulations (deviating at most by $\sim 0.2\, L$), and it will be easy to identify by tracking the large-scale convergence of the  field lines, or the divergence of the outflowing plasma velocity.

On the two sides of the primary X-point, two ``reconnection fronts'' are formed (see the over-dense regions at $|x|/L\sim 0.1$ in the top panel of \fig{fluidtime}). Pulled by the tension force of the magnetic field lines, the two reconnection fronts move toward the ends of the current sheet at the \alf\ speed, which for relativistic reconnection approaches the speed of light (for the case of $\sigma=10$ in \fig{fluidtime}, the \alf\ speed is $v_A/c\simeq 0.95$). As they propagate, they sweep up the hot plasma that was initialized in the current sheet, so the ``heads'' of the two reconnection fronts become wider and wider with time (compare the over-dense structures at $|x|/L\sim 0.4$ for $ct_{\rm lab}/L\sim 0.7$ with those at $|x|/L\sim 0.85$ for $ct_{\rm lab}/L\sim 1.2$). After the two reconnection fronts exit through the two absorbing $x$ boundaries of the box (for the case in \fig{fluidtime}, this happens around $ct_{\rm lab}/L\sim 1.4$), the system retains no memory of the initialization of the current sheet (in contrast, in the case of periodic boundary conditions in the $x$ direction, the initial current sheet plasma would accumulate within a large island at the boundary of the box, see \sect{setup}). After the two reconnection fronts have left the domain, the system approaches a statistical steady state (compare the two bottom-most panels in \fig{fluidtime}). Even though our main results do not depend on the choice of boundary conditions along the $x$ direction (absorbing or periodic), it is only when we employ absorbing/outflow boundaries that we can study the steady state of the system for several \alf\ crossing times.

For our choice of a thick current sheet (i.e., $\Delta \gg \rhot$), the region in between the two reconnection fronts is the only portion of the domain where reconnection is active. In fact, we specifically choose the value of $\Delta$ such that reconnection does not spontaneously start anywhere in the region ahead of the two reconnection fronts, during the time it takes for the two fronts to reach the boundaries of the box. In contrast, in the case of a thinner current sheet, the tearing instability would periodically break the initial current layer into a series of magnetic islands, separated by X-points (e.g., see Fig.~1(a) in SS14). Such magnetic islands --- which we would call ``primary islands'' --- would still bear memory of the initialization of the current sheet, since their core has a negligible magnetic content (e.g., see Fig.~2 (c) in SPG15) and it is entirely supported by the pressure of the hot particles initialized in the current sheet\footnote{Similar conclusions would hold in the case that the current sheet is set up as a force-free layer \citep[e.g.,][]{guo_14}, rather than a Harris sheet. In such a case, the island cores would be dominated by the pressure of the out-of-plane field initially present in the current sheet.} \citep[see][for an investigation of the structure of primary islands]{nalewajko_15}. For our choice of a thick current sheet, no primary islands are formed in the reconnection layer.

The focus of our work is not on  primary islands, since their properties would still be sensitive to the details of the current sheet initialization, which would be impossible to constrain from astrophysical observations. Rather, we investigate the properties of the ``secondary islands'' resulting from the secondary tearing instability discussed by \citet{uzdensky_10}. Secondary plasmoids are continuously generated in between the two reconnection fronts (or in the whole domain, after the two reconnection fronts have exited the computational box) and then advected outwards by the tension force of the magnetic field lines (e.g., the plasmoid that was located at $x/L=-0.5$ for $ct_{\rm lab}/L=1.2$ has moved to $x/L=-0.9$ for $ct_{\rm lab}/L=1.7$). Secondary plasmoids are a self-consistent by-product of the long-term evolution of the system (see \citealt{daughton_07}, for similar conclusions in non-relativistic reconnection), and their properties only depend on the flow conditions far from the current sheet, i.e., only on the magnetization $\sigma$, for our case of anti-parallel reconnection in pair plasmas. This allows to make robust predictions on the observational implications of relativistic reconnection in pulsar winds and jets of blazars and GRBs.

When two plasmoids merge, a current sheet forms in between, along the $y$ direction. Similarly to the picture described above, this current sheet  results in a second generation of secondary plasmoids moving in the $y$ direction. Further mergers between these plasmoids gives a third generation of secondary plasmoids in the $x$ direction, with a fractal framework that is expected to continue down to microscopic plasma scales. In this work we only focus on the first generation of secondary plasmoids, but we expect that our results are equally applicable to all subsequent generations. The only caveat is that, while the plasma flowing into the first-generation current sheet is cold, in all subsequent phases the inflowing plasma (which already belongs to a ``parent'' plasmoid) is relativistically hot.

\begin{figure*}
\centering
 \resizebox{\hsize}{!}{\includegraphics{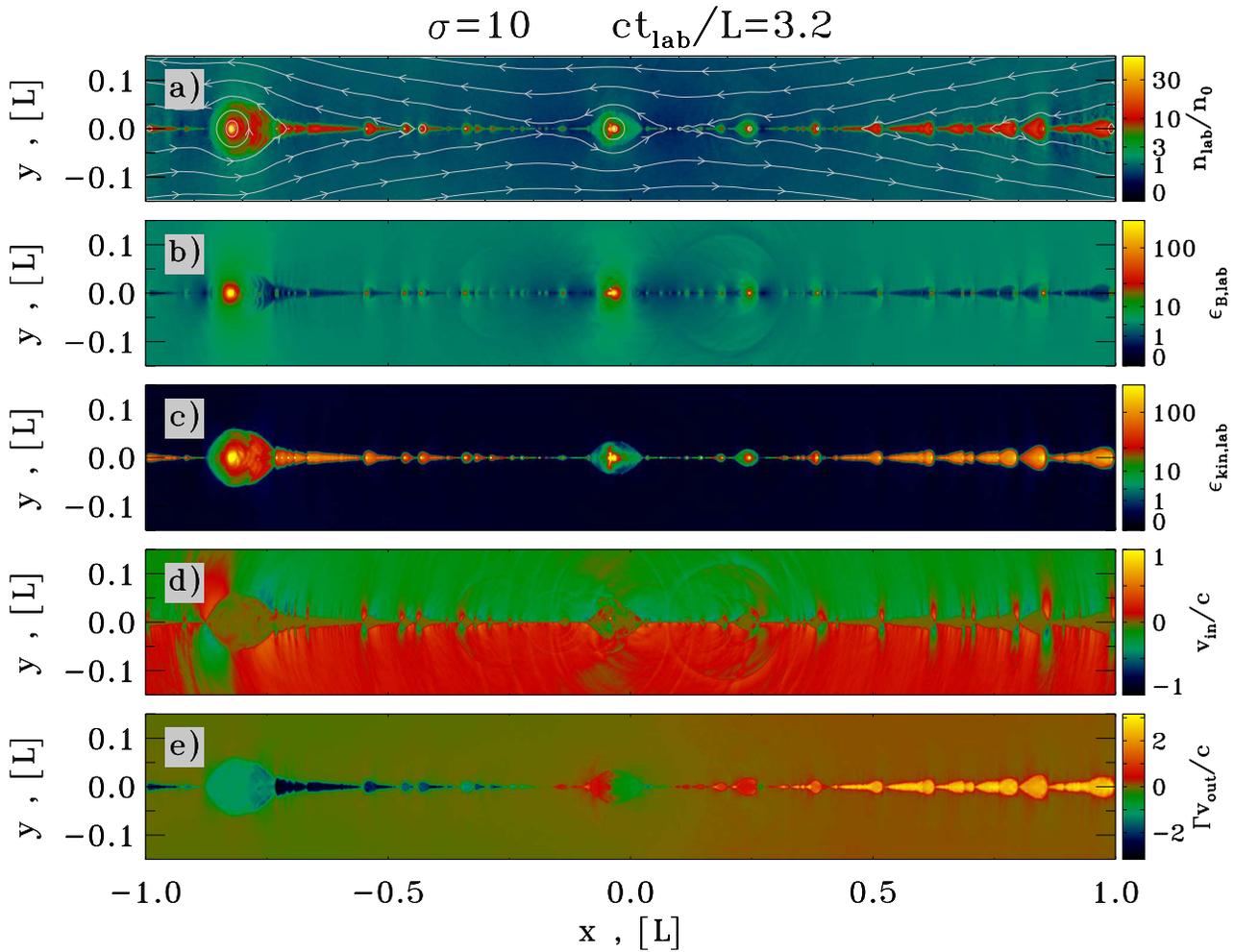}}
 \caption{2D structure of the reconnection layer at $ct\lab/L=3.2$, from a simulation with $\sigma=10$ and $L/\rhot\simeq518$. Along the $y$ direction, we only show the region $|y|/L<0.15$ to emphasize the small-scale structures in the reconnection layer (the extent of the computational box along $y$ increases at the speed of light, as described in \sect{setup}). From the top to the bottom panel, we show: (a) the particle number density in the lab frame $n_{\rm lab}$, in units of the number density $n_0$ far from the reconnection layer, with overplotted magnetic field lines; (b) the magnetic energy fraction $\epsilon_{\rm B,lab}=B\lab^2/8\pi n_0 m c^2$, where the magnetic field $B\lab$ is measured in the lab frame; (c) the kinetic energy fraction $\epsilon_{\rm kin,lab}=(\langle\gamma\rangle\lab-1)\, n_{\rm lab}/n_0$, where $\langle\gamma\rangle\lab$ is the mean particle Lorentz factor in the lab frame; (d) the inflow bulk speed $v_{\rm in}=\bmath{v}\cdot \hat{\bmath{y}}$ in units of the speed of light; (e) the outflow bulk four-velocity $\Gamma v_{\rm out}=\Gamma \,\bmath{v}\cdot \hat{\bmath{x}}$ in units of the speed of light. The reconnection layer is fragmented into a series of over-dense magnetized hot plasmoids, which are propagating away from the center at ultra-relativistic speeds. The inflow speed (i.e., the reconnection rate) is mildly relativistic ($v_{\rm in}/c\sim 0.1$), and remarkably uniform along the current sheet.}
 \label{fig:fluid}
\end{figure*}
\subsection{The Steady-State Reconnection Layer}
The continuous formation and ejection of secondary plasmoids characterizes the steady-state appearance of the reconnection layer, which is presented in \fig{fluid} at $ct\lab/L=3.2$ for the same simulation as in \fig{fluidtime}. We plot the 2D structure of various quantities, as measured in the lab frame of our simulations. In \sect{chain}, we will quantify how the comoving density and magnetic and kinetic energy content of the secondary plasmoids depend on the plasmoid size. The secondary plasmoids appear as over-dense structures (\fig{fluid}(a)) containing strong fields (\fig{fluid}(b))  and hot particles (\fig{fluid}(c)).  \fig{fluid}(b) shows that the cores of secondary plasmoids are significantly magnetized, in contrast with the structure of primary plasmoids (see Fig.~1(c) in SS14), whose core would be populated by the unmagnetized hot particles that were initialized in the current layer. In the center of secondary plasmoids, the magnetic and kinetic densities are roughly comparable, i.e., secondary plasmoids are nearly in equipartition of magnetic and kinetic energy (see SPG15, and also \sect{chain} below). At the center of the plasmoids, the magnetic energy fraction reaches $\epsilon_{\rm B,lab}=B\lab^2/8\pi n_0 m c^2\sim 100$. This should be compared with the value expected in the inflow region, where $\epsilon_{\rm B,lab}\sim\sigma/2=5$. So, the magnetic field in the plasmoid core is compressed by a factor of $\sim 4$, with respect to the initial $B_0$. 

The magnetic field strength is nearly uniform in the inflow region, with the exception of a few areas with weaker fields (in blue in \fig{fluid}(b); see, e.g., at $x/L\sim 0.1$ and $|y|/L\lesssim 0.05$). Such regions lie close to the current sheet, in the vicinity of a large secondary island (for the case indicated above, see the island at $x\sim 0$). As the magnetic field lines advect into the current sheet during the reconnection process, they wrap around large magnetic islands. The same bundle of field lines that are now accumulating on the outskirts of a large island have to make their way to the current sheet, on the two sides of the large island. It follows that the density of field lines near a large magnetic island will be reduced, resulting in weaker fields. In view of flux freezing, this will also correspond to a locally lower value of the number density of inflowing particles (e.g., see the same region at $x/L\sim 0.1$ and $|y|/L\lesssim 0.05$ in \fig{fluid}(a)). As we demonstrate in \sect{chain}, this will have implications for the magnetic energy content of small plasmoids, that reside in regions where the inflowing magnetic field is weaker. 

The inflow rate of particles is nearly uniform along the current sheet, as shown by the plot of the inflow velocity  $v_{\rm in}/c=\bmath{v}/c\cdot \hat{\bmath{y}}$ in \fig{fluid}(d). The only exceptions are the regions just ahead of the secondary plasmoids, where the plasmoids plunge into the inflowing plasma and they push it aside (see, e.g., to the left of the plasmoid at $x/L\sim -0.8$ in \fig{fluid}(d), where $v_{\rm in}$ has opposite sign than in the bulk of the inflow).
The inflow speed is non-relativistic, $v_{\rm in}/c\sim 0.15$. In \citet{lyubarsky_05}'s analytical model of relativistic reconnection, the inflow speed is closely related to the opening angle $\theta$ of the magnetic field in the inflow region (with respect to the $x$ axis), with $v_{\rm in}/c=\tan\theta$. From \fig{fluid}(a), the inclination of the magnetic field lines is such that $\tan\theta\sim 0.15$, which is in excellent agreement with the inflow speed that we measure in \fig{fluid}(d). In steady state, both the obliquity of the field lines and the inflow velocity (or ``reconnection rate'') stay remarkably constant in time.

The 2D plot in \fig{fluid}(d) also reveals the presence of spherical waves propagating back into the inflow region (e.g., see the spherical front centered at $(x,y)\sim(0.2,0)\,L$). They appear most clearly in the plots of  inflow velocity (\fig{fluid}(d)) and magnetic energy fraction (\fig{fluid}(b)), and they are generated by the merger event of two plasmoids.
Plasmoid mergers are rather frequent, as a  result of the fact that different plasmoids propagate at different speeds along the layer, with larger plasmoids typically moving slower. \fig{fluid}(e) presents the structure of the outflow bulk four-velocity $\Gamma v_{\rm out}=\Gamma \,\bmath{v}\cdot \hat{\bmath{x}}$, in units of the speed of light. The large plasmoid at $x/L\sim-0.8$ moves slower than the smaller plamoids in its wake, which will eventually accrete onto the large plasmoid, further increasing its mass. In addition to such ``minor mergers'' between plasmoids of unequal sizes, occasional ``major mergers'' of equal-mass plasmoids might occur, as it is happening in the central region of \fig{fluid}(e). The convergence of the velocity flow in between merging plasmoids  is believed to play an important role for particle acceleration \citep[SS14,][]{drake_06,guo_14,guo_15a,nalewajko_15}. 
The dependence of the plasmoid speed on its size will be extensively quantified in \sect{chain}. Yet, \fig{fluid}(e) already suggests that, while large islands exit the layer  with only trans-relativistic velocities, small plasmoids can reach ultra-relativistic speeds. Small plasmoids approach the terminal four-velocity $\Gamma v_{\rm out}/c\sim \sqrt{\sigma}\sim 3.3$ expected for the bulk outflow from relativistic  $\sigma=10$  reconnection \citep{lyubarsky_05}.\footnote{We remind that \citet{lyubarsky_05}'s predictions on the bulk outflow speed from relativistic reconnection are based on a steady-state model of the reconnection layer, i.e., neglecting its fragmentation into plasmoids.}

\begin{figure}
\centering
 \resizebox{\hsize}{!}{\includegraphics{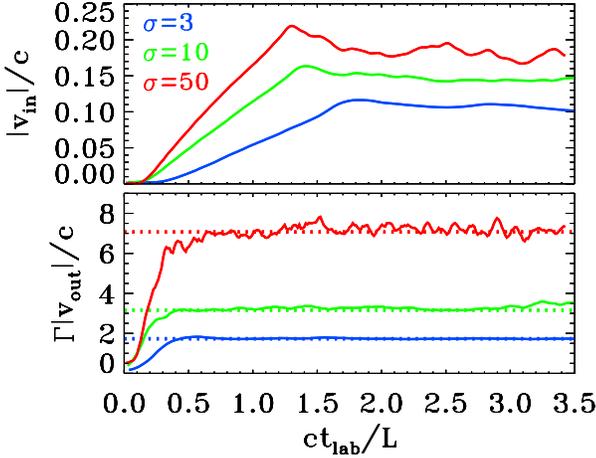}}
 \caption{Inflow and outflow speeds as a function of time in the lab frame, for three values of the magnetization $\sigma$, as indicated in the legend of the top panel ($\sigma=3$ in blue, $\sigma=10$ and $L\simeq 518\rhot$ in green, and $\sigma=50$ in red). Top panel: inflow bulk speed in units of the speed of light, averaged over a region of width $|y|\lesssim 0.5\,L$ across the reconnection layer (but our results are nearly independent of this choice). The steady-state inflow speed, after the two reconnection fronts have exited the computational domain (i.e., at $ct\lab/L\gtrsim 1.5$), has a weak dependence on the magnetization. Bottom panel: maximum outflow four-velocity, in units of the speed of light. For each value of the magnetization, we take the slice at $y=0$ and plot the fifth largest value of the ouflow four-velocity (but we have tested that the variation from the first largest to the tenth largest is minimal). The resulting peak outflow four-velocity in in excellent agreement with the prediction $\Gamma\, |v_{\rm out}|/c\sim\sqrt{\sigma}$ by \citet{lyubarsky_05}, indicated by the dotted lines.}
 \label{fig:outflows}
\end{figure}

In \fig{outflows}, we demonstrate that the predictions of \citet{lyubarsky_05}'s theory, as regard to the inflow and outflow speeds in relativistic reconnection, hold for the whole range of magnetizations $\sigma=3-50$ explored in our work. In the top panel, we present the inflow bulk speed in units of the speed of light, averaged over a region of width $|y|\lesssim 0.5\,L$ along the $y$ direction and extending along the whole current sheet in the $x$ direction. The top panel in \fig{outflows} shows that, after the system reaches a steady state (i.e., at $ct\lab/L\gtrsim 1.5$), the inflow speed (or equivalently, the reconnection rate) has only a weak dependence on the magnetization, varying from $|v_{\rm in}|/c\sim 0.10$ for $\sigma=3$ up to $|v_{\rm in}|/c\sim 0.18$ for $\sigma=50$. The weak dependence of the inflow speed on the magnetization is consistent with earlier works \citep[SS14,][]{guo_15a} and with the analytical model of \citet{lyubarsky_05}, that predicted that the reconnection rate should saturate in the limit $\sigma\gg1$ at a value around $\sim 0.1 \,c$. The inflow speeds presented in \fig{outflows} are obtained by averaging over a macroscopic region of size $2\,L\times L$, and are therefore representative of the mean dissipation rate of the magnetic field. The value of the inflow speed can be much larger in the vicinity of X-points, approaching the speed of light \citep[e.g.,][]{liu_15}, but such large values only extend over microscopic skin-depth scales. 

The inflow speeds presented in the top panel of \fig{outflows} are about a factor of two larger than we reported in SS14, just due to a different choice of the area where we average the inflow rate. In SS14, reconnection proceeded from numerical noise, resulting in the formation of a number of primary plasmoids. The inflow speed nearly vanished at the location of the primary plasmoids, whereas in between primary plasmoids it resembled the values presented in \fig{outflows}. At any given time, about half of the current sheet length was occupied by primary plasmoids. By averaging over the entire domain (and not just over the active regions in between primary plasmoids), we then obtained a value that is indeed expected to be half of what we quote here. 

The linear increase in the inflow speed at early times ($ct\lab/L< 1.5$) is simply driven by the fact that the distance between the two reconnection fronts is progressively increasing, as they move from the center toward the boundaries of the box. As we have explained above, it is only in the region in between the two  fronts that the reconnection process is active. At all times, the inflow velocity in between the two fronts is the same as the steady-state value that we read from the top panel of \fig{outflows} at late times ($ct\lab/L\gtrsim 1.5$). Ahead of the two reconnection fronts, the plasma is still at rest. One would then expect that the reconnection rate presented in the top panel of \fig{outflows}, which is averaged over the whole extent of the current sheet, would increase linearly from zero up to the steady-state value, as the two fronts move outward at nearly the  \alf\ speed. The time for the two fronts to propagate to the boundaries of the box is $\sim L/v_{A}$, which explains why it takes slightly longer for $\sigma=3$, where $v_{A}\simeq0.85$, than for $\sigma=10$ and $50$, where the \alf\ speed is nearly the speed of light.

While the inflow speed in relativistic reconnection is non-relativistic, the outflowing plasma can reach ultra-relativistic velocities. This is illustrated in the bottom panel of \fig{outflows}, where we present the maximal value of the outflow four-velocity $\Gamma\, |v_{\rm out}|/c$ as a function of time.\footnote{The maximal value plotted in \fig{outflows} is defined as the fifth largest value of $\Gamma\, |v_{\rm out}|/c$ computed along the 1D slice at $y=0$, but we have tested that the variation from the first largest to the tenth largest  is minimal.} For all the magnetizations we explore, the peak outflow four-velocity is in excellent agreement with the prediction $\Gamma\, |v_{\rm out}|/c\sim\sqrt{\sigma}$ by \citet{lyubarsky_05}, which is indicated by the dotted lines (the color coding is described in the legend in the top panel). Or equivalently, the outflow velocity approaches the \alf\ speed $v_A/c=\sqrt{\sigma/(\sigma+1)}$. This conclusion also holds in the case of periodic boundary conditions, as long as the computational box is sufficiently large (as argued in SS14). 

The box length needed to capture the asymptotic value of $\Gamma\, |v_{\rm out}|/c$ can be estimated from the early increase in the curves in the bottom panel of \fig{outflows}. This shows that the outflow accelerates to the terminal speed on a timescale that is shorter than the \alf\ crossing time of the box, yet not much shorter. For the most extreme magnetization of $\sigma=50$ (red curve), the acceleration time is  $\sim 0.4 \,L/c$. Since the overall box length in this case is $2L\sim 1000\rhot$, we argue that a simulation whose $x$ extent is smaller than $\sim 400\rhot$ would not be able to capture the terminal outflow speed of $\sigma=50$ reconnection. While this estimate is appropriate for the case of absorbing (or outflow) boundary conditions, in the case of periodic boundaries the same requirement should be imposed over the distance in between two neighboring primary islands, resulting in a much more constraining condition on the overall box length (which typically includes many primary plasmoids). This might explain why earlier works  \citep{cerutti_13a,guo_15a,kagan_16} claimed that the outflow speed from relativistic reconnection seems systematically lower than \citet{lyubarsky_05}'s prediction. With a sufficiently large domain, we are able to demonstrate that the outflow four-velocity from relativistic reconnection can be as fast as $\sim\sqrt{\sigma}\,c$, in full agreement with \citet{lyubarsky_05}'s model.

\begin{figure*}
\centering
 \resizebox{\hsize}{!}{\includegraphics{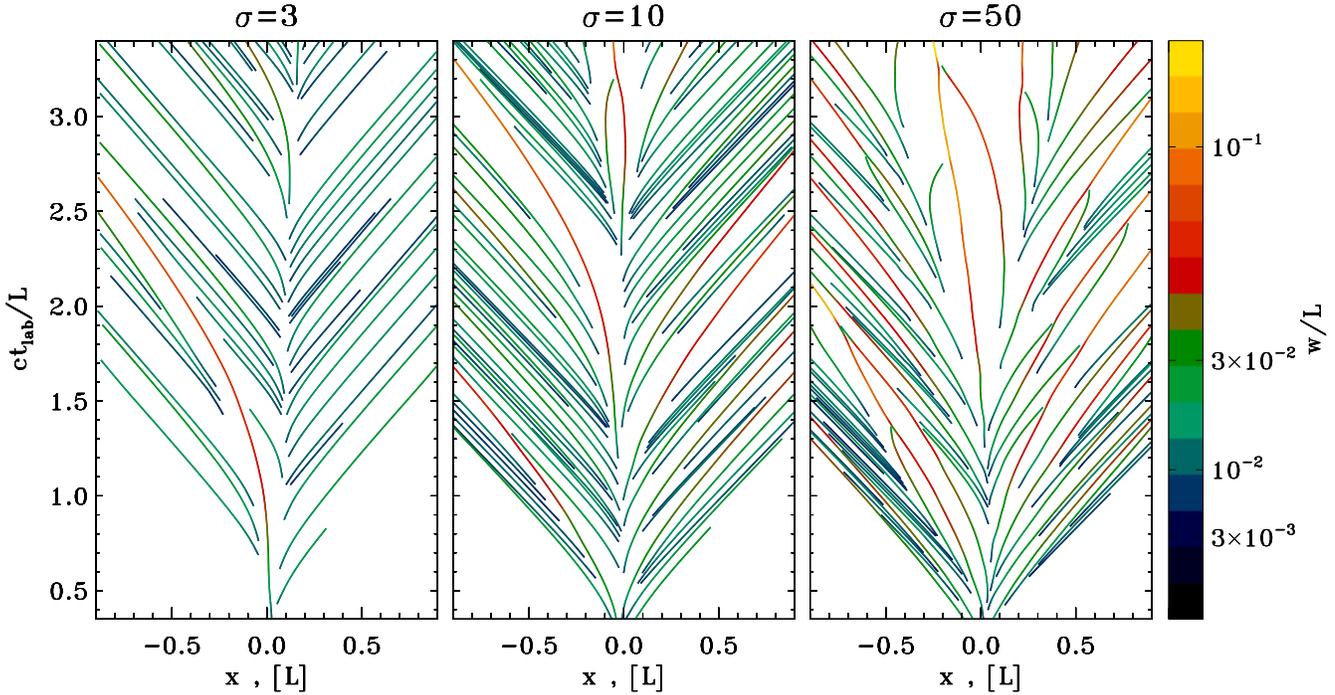}}
 \caption{Position-time diagram of magnetic islands, for three values of the magnetization, as indicated at the top ($\sigma=3$ in the left panel, $\sigma=10$ and $L/\rhot\simeq 518$ in the middle panel, and $\sigma=50$ in the right panel). The islands start from the center of the layer ($x\simeq 0$), where reconnection is initially triggered, and while propagating outwards they grow in size (as indicated by the colors, the plasmoid width $w$ is in units of the system length $L$) and they accelerate, approaching the speed of light (which would correspond to lines oriented at $45^\circ$ in the plots). Their tracks terminate when they either leave the reconnection layer at $|x|= L$ or they merge with a bigger plasmoid, as it frequently happens at large magnetizations. For the sake of clarity, we only plot the evolutionary tracks of plasmoids whose lifetime (in the lab frame) is longer than $\simeq0.35 L/c$.}
 \label{fig:isltracks}
\end{figure*}

\section{The Plasmoid Chain}\label{sec:chain}
In this section, we focus on the chain of secondary plasmoids that are continuously generated in the reconnection layer as a result of the secondary tearing instability \citep{uzdensky_10}. As a function of their size, in \sect{fluid} we quantify their fluid properties (averaged over the plasmoid volume), and  in \sect{spec} the spectrum and anisotropy of the particle population that they contain. We also follow individual plasmoids over time, quantifying their growth (in \sect{growth}) and acceleration (in  \sect{accel}) as they propagate from the center toward the boundaries.

\fig{isltracks} presents the position-time diagram of secondary plasmoids, for the three values of  magnetization that we explore in this work ($\sigma=3$ on the left, $\sigma=10$ in the middle and $\sigma=50$ on the right). The various tracks follow the trajectories of individual plasmoids (more precisely, of their center), as they move along the reconnection layer. The plasmoid width $w$ in the direction transverse to the current sheet is indicated by the color scale on the right (red and yellow for the largest plasmoids). For the sake of clarity, we only plot the evolutionary tracks of plasmoids whose lifetime (in the lab frame) is longer than $\simeq0.35 L/c$.

For all the values of magnetization we explore, we see that the vicinity of the primary X-point, where reconnection is initially triggered, is always a preferred region for the formation of long-lived secondary plasmoids. They form around the center with an initial width of a few plasma skin depths, and as they propagate outwards they grow in size, as indicated by the colors (e.g., see the plasmoid that for $\sigma=3$ starts in the center at $ct\lab/L\sim 0.4$ and exits to the left at $ct\lab/L\sim 2.7$). In general, the longer they spend around the central region, the bigger they grow. Eventually, the tension force of the magnetic field accelerates them outwards, with the fastest plasmoids approaching the \alf\ speed. Since for all the cases explored in this work the \alf\ speed is close to the speed of light (with the only marginal exception of $\sigma=3$, that gives $v_{A}\simeq 0.85\,c$), plasmoids moving at the \alf\ speed would have position-time tracks oriented at $\pm 45^\circ$ in \fig{isltracks}, as indeed is observed for most of the small plasmoids (as we show in \sect{accel}, at a given distance from the center larger plasmoids always move slower than smaller ones).

Even though most of the plasmoids shown in \fig{isltracks} start from the central region, copious production of secondary islands occurs everywhere in the reconnection layer (see also \fig{fluid}). For example, the relatively large plasmoid that exits to the left at $ct\lab/L\sim 2.7$ for $\sigma=3$ is preceded by a series of smaller plasmoids, all generated at a distance of $\gtrsim 0.2 \, L$ from the center, and in some cases even further out. Plasmoids generated close to the boundaries exit the domain before reaching the threshold lifespan of $\simeq0.35 L/c$ adopted in \fig{isltracks}, so they do not appear in the figure.

The plasmoid trajectories in \fig{isltracks} terminate when either the plasmoid exits one of the two boundaries or it merges with a bigger plasmoid. For example, the small plasmoids that in the left panel are trailing the large plasmoid mentioned above will terminate their life by merging with it. It follows that smaller plasmoids typically have  shorter lives, since they will eventually encounter a bigger plasmoid moving ahead of them (which propagates slower) and merge with it. In addition to such minor mergers (indicated as ``minor'' because of the size difference of the two plasmoids), major mergers between plasmoids of comparable widths can also occur, typically close to the center. This is seen in the case of $\sigma=50$ (right panel) at $ct\lab/L\sim 3.3$, where two large islands merge  at $x\simeq -0.2\,L$. \fig{isltracks} shows that, in the course of the merger, the width of the two islands shrinks (see the point where the two tracks meet). This is just a consequence of our criterion for the identification of the plasmoid contour, that relies on the maximal value of the vector potential $A_z$ among the two neighboring X-points (each one being a local minimum of $A_z$). In a merger, the X-point in between the two islands will have the largest value of $A_z$, and it will set the plasmoid contours. As the two plasmoids approach in the course of the merger, the value of $A_z$ in the X-point will increase, so the two contours will shrink more and more, resulting in an apparent decrease of the width of the two islands. Soon after the merger, the surviving plasmoid (the larger of the two) recovers its proper width.

Mergers are most frequent at higher magnetizations. The total electric current in a plasmoid of a given size scales as $\propto \sqrt{\sigma} w$,\footnote{Here, we have assumed that the magnetic field at the plasmoid boundary is always equal to the initial $B_0$.} which results in a stronger interaction among neighboring plasmoids for higher magnetizations, and a consequent increase in the merger rate. Indeed, while in the case of $\sigma=3$ (left panel), mergers predominantly involve smaller (so, faster) plasmoids catching up with larger (so, slower) islands, the situation is much more diverse in the high-magnetization case $\sigma=50$ (right panel). There, it is quite frequent that large plasmoids merge both with trailing plasmoids (for the large plasmoid exiting on the left at $ct\lab/L\sim 2.1$, see the merger at $x/L\sim -0.7$ and $ct\lab/L\sim 1.9$) and with leading plasmoids that are pulled back by the attractive force of the large plasmoid (for the same plasmoid, see the merger at $x/L\sim -0.5$ and $ct\lab/L\sim 1.4$). In the latter case, the leading plasmoid that is pulled back might even reverse its velocity along the layer.

The high merger rate in high-$\sigma$ flows has two main consequences. First, the spherical waves emanating from each merger event (e.g., see \fig{fluid}(d)) will seed fluctuations in the current sheet, triggering the formation of additional secondary plasmoids. In fact, we find that, for the same timespan and domain size,  high-$\sigma$ flows result in a much larger number of secondary islands. Second, in the high-magnetization case, small plasmoids leading a large island will be decelerated by its attraction (in the extreme limit, to the point of being pulled back and merge with the large plasmoid). This will tend to inhibit their acceleration up to the \alf\ speed. As we show in \sect{accel}, for $\sigma=50$ a smaller number of plasmoids will be able to approach the expected terminal four-velocity $\sqrt{\sigma}\, c$, as compared to lower magnetizations.

Finally, \fig{isltracks} suggests that large plasmoids are rarer than smaller plasmoids. At any given time, several small plasmoids can co-exist in the reconnection layer, but only a few large plasmoids. In \sect{fluid}, we will quantify the plasmoid size distribution. Interestingly, in the right panel of \fig{isltracks} we notice a clear signature of quasi-periodicity in the ejection time of relatively large plasmoids (with final width between $0.05\,L$ and $0.1 \,L$). Looking at the plasmoids escaping on the right, we detect an island leaving at $ct\lab/L\sim 1.9$, followed by others of similar size at $ct\lab/L\sim 2.3$, 2.6, 3.1 and 3.3. In \sect{growth}, we demonstrate that such a quasi-periodicity also holds for the largest plasmoids generated in the layer, with width $\sim 0.2\,L$. By following a system with $\sigma=10$ and $L/\rhot\simeq127$ up to $ct\lab/L\sim14$, we will show that their typical recurrence time is $\sim 2.5\,L/c$.


\begin{figure*}
\centering
 \resizebox{0.87\hsize}{!}{\includegraphics{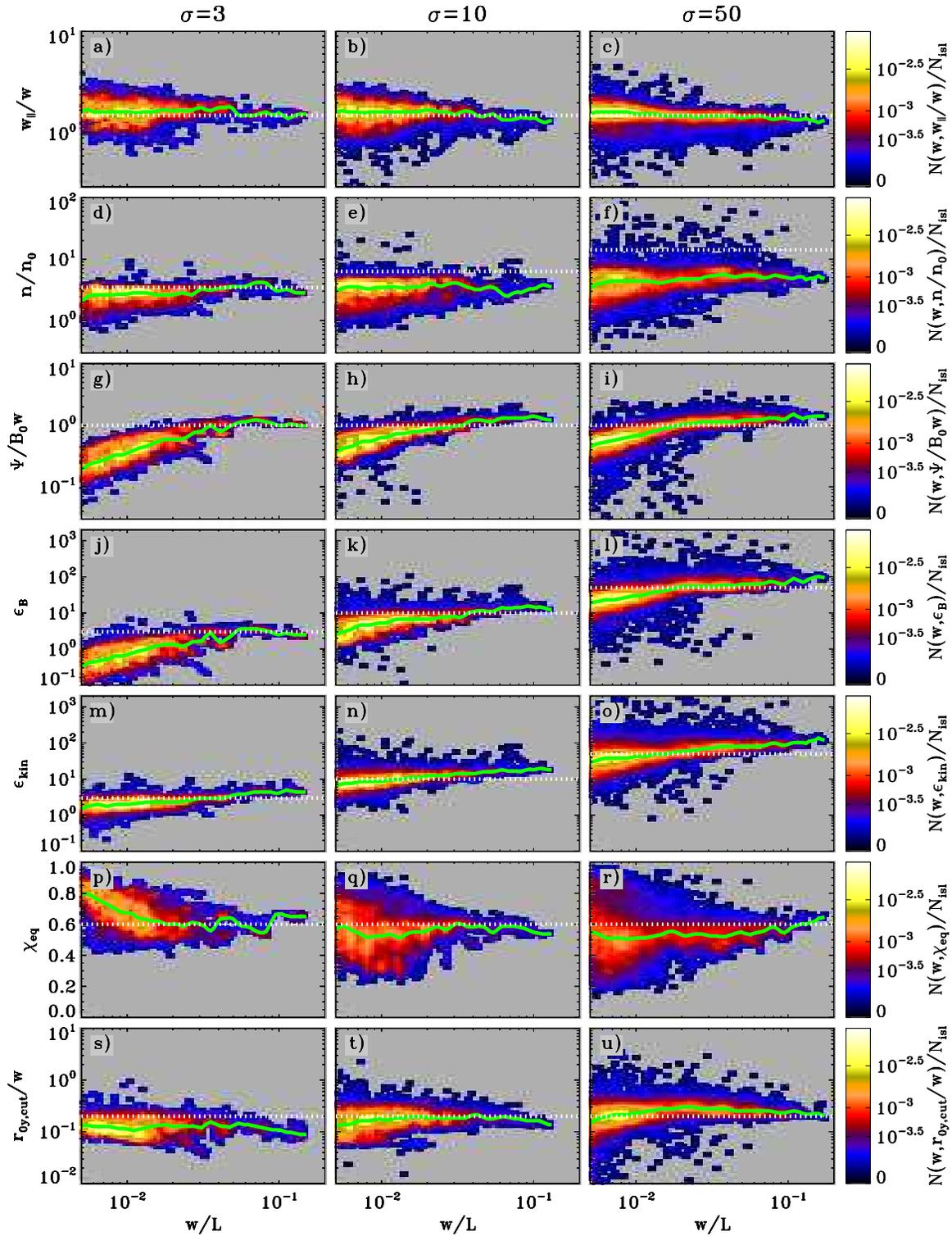}}
 \caption{2D histograms of various plasmoid properties as a function of the width $w$ (in units of $L$), for three  values of the magnetization, as indicated at the top ($\sigma=3$ on the left, $\sigma=10$ in the middle, and $\sigma=50$ on the right). In each bin, the number of plasmoids is normalized to the total number of magnetic islands $N_{\rm isl}$ (see the colorbars on the right). The green solid lines indicate the median values. From top to bottom row, we show: (a)-(c) the ratio of the length $w_\parallel$ (measured in the comoving frame) to the width $w$, which is around $\sim 1.5$ regardless of the magnetization (as indicated by the dotted white lines); (d)-(f) the comoving density $n$ averaged over the plasmoid area, in units of the density $n_0$ far from the current sheet, with the  scaling $n/n_0\sim 2\sqrt{\sigma}$ predicted by \citet{lyubarsky_05} indicated with the dotted white lines; (g)-(i) the plasmoid flux $\Psi$, which approaches the nominal value $B_0 w$ at large sizes (see the dotted white lines); (j)-(l) the magnetic energy fraction $\epsilon_{\rm B}=B^2/8\pi n_0 m c^2$ averaged over the plasmoid area, where $B$ is measured in the plasmoid comoving frame; (m)-(o) the internal energy fraction $\epsilon_{\rm kin}=(\langle\gamma\rangle-1)\, n/n_0$ averaged over the plasmoid area, where $\langle\gamma\rangle$ is the mean particle Lorentz factor in the plasmoid frame; both $\epsilon_{\rm B}$ and $\epsilon_{\rm kin}$ scale as $\propto \sigma$, as indicated with the dotted white lines; (p)-(r) the equipartition parameter $\chi_{\rm eq}$ as defined in \eq{chi}, which lies around $\sim 0.6$ regardless of the magnetization, as indicated by the dotted white lines; (s-u) the Larmor radius $r_{0y,\rm cut}=p_{y,\rm cut}c/eB_0$ of the positrons at the cutoff momentum $p_{y,\rm cut}$ as defined in \eq{pcut}, in units of the plasmoid width; this lies around $\sim 0.2$ regardless of the magnetization, as indicated by the dotted white lines.}
 \label{fig:islfluids}
\end{figure*}

\subsection{Plasmoid Fluid Properties}\label{sec:fluid}
In \fig{islfluids}, we analyze a number of fluid properties of secondary plasmoids, averaged over the plasmoid volume. We investigate how the plasmoid properties depend on the width $w$ (on the horizontal axis, in units of $L$) and on the flow magnetization $\sigma$ (with $\sigma=3$ in the left column, $\sigma=10$ in the middle column and $\sigma=50$ in the right column). Each panel in \fig{islfluids} is a 2D histogram indicating the number of plasmoids with a given fluid property (for example, a given value of the rest-frame density) and a given width, normalized to the overall number of magnetic islands $N_{\rm isl}$. All of the plasmoid properties investigated in \fig{islfluids} refer to comoving quantities. From quantities measured in the simulation frame, the corresponding comoving values can be easily obtained via Lorentz transformations, since we can measure the velocity of each plasmoid (and consequently, its bulk Lorentz factor $\Gamma$). 
We remark again that, while earlier works focused on the structure of primary plasmoids \citep{nalewajko_15}, that are necessarily affected by the prescribed setup of the initial Harris sheet, here we study secondary plasmoids, whose properties only depend on the flow magnetization.

The first row of \fig{islfluids} shows that the plasmoids are quasi-spherical in their rest frame, with a ratio of comoving length $w_\parallel$ to width $w$ that lies around $\sim 1.5$ (as indicated by the dotted white lines in the plot), irrespective of the magnetization $\sigma$. The tendency for sphericity is even more pronouced when the plasmoids approach the terminal four-velocity $\sim \sqrt{\sigma}\,c$.
We have also verified that the plasmoid area is well approximated by $\pi w_\parallel w/4$, as expected for an ellipse with major axis $w_\parallel$ and minor axis $w$.  

The second row of panels shows dependence on $w$ and $\sigma$ of the average comoving density $n=n_{\rm lab}/\Gamma$, in units of the particle number density $n_0$ far from the current sheet. For each value of the magnetization, the comoving number density appears to be nearly independent of the plasmoid size.
The density has a weak dependence on magnetization, varying from $n/n_0\sim 3$ for $\sigma=3$ up to $n/n_0\sim 5$ for $\sigma=50$. This is in apparent disagreement with \citet{lyubarsky_05}'s theory, predicting that the rest-frame density in the fast outflows from relativistic reconnection should scale as $n/n_0\sim 2\sqrt{\sigma}$ (as indicated by the dotted white lines). However, \lyub's scalings only apply to a fully accelerated smooth outflow (i.e., moving with the \alf\ speed). As we will see in \sect{accel}, while most of the plasmoids are successfully accelerated to the \alf\ speed for $\sigma=3$ (where, in fact, the expected scaling is satisfied), only few plasmoids can reach the terminal four-velocity of $\sim\sqrt{\sigma}\,c$ for $\sigma=50$.

We now present a simple argument describing why the scaling $n/n_0\propto \sqrt{\sigma}$ is only to be expected for the fastest moving plasmoids. Pressure balance across the current sheet, between the pressure of hot particles  in the islands and the magnetic pressure of the cold inflow, dictates that the mean plasmoid internal energy fraction should scale as $\epsilon_{\rm kin}\sim \sigma$ (see also the fifth row of panels in \fig{islfluids}). Here, $\epsilon_{\rm kin}=(\langle\gamma\rangle-1)n/n_0$, where $\langle\gamma\rangle$ is the mean comoving particle Lorentz factor. In addition, conservation of the energy per particle requires that $\sigma\sim \Gamma \epsilon_{\rm kin}/(n/n_0)$, where $\Gamma$ is the plasmoid bulk Lorentz factor. In reality, the mean energy per particle is likely to be smaller than $\sim \sigma$ during the acceleration phase, since some ``potential energy'' is still available in the field line tension, which will only be released when the plasmoid reaches the expected terminal velocity. It follows that we would generally expect $\Gamma \epsilon_{\rm kin}/(n/n_0)\lesssim \sigma$. Combined with the pressure balance $\epsilon_{\rm kin}\sim \sigma$, this implies that $n/n_0\gtrsim \Gamma$, with the equaliy being realized only for plasmoids moving at the \alf\ speed, or equivalently with $\Gamma \sim \sqrt{\sigma}$.\footnote{In this whole argument, we have implicitly assumed that $\sigma\gg1$ for the sake of simplicity.} The condition $n/n_0\gtrsim \Gamma$ is  realized for all the plasmoids of our simulations, regardless of the flow magnetization. In particular, we find that the densest plasmoids are typically the fastest ones, fully accelerated up to the \alf\ speed.

The third row of panels in \fig{islfluids} illustrates the dependence on plasmoid size and magnetization of the magnetic flux $\Psi$ (more specifically, of $\Psi/B_0 w$, since the flux is expected to depend linearly on the plasmoid size). The magnetic flux in a given plasmoid is defined as the difference between the vector potential $A_z$ at the plasmoid O-point and at its contour (which is an equipotential surface). Since $A_z$ is Lorentz invariant, the plasmoid flux is also Lorentz invariant. As expected, we find that at large sizes $\Psi/B_0w$ approaches a constant of order unity (as indicated by the dotted white lines). In contrast, at small sizes (typically, $w/L\lesssim 0.02$) the plasmoid flux is systematically lower than the expected value $\sim B_0 w$. As we show below, the same trend is observed for the mean magnetic energy density of small plasmoids.

A similar deficit of magnetic flux at small sizes has been observed in MHD simulations of non-relativistic plasmoid-dominated reconnection \citep[Fig.~4 in][]{loureiro_12}. There, it was attributed to the effect of plasmoid mergers. They envisioned the coalescence as a gradual stripping of the outer layers of the smaller plasmoid, so the magnetic flux in a semi-digested plasmoid would be $\Psi \sim B_0 w^2/w_0$, where $w_0$ is the plasmoid width at the beginning of the merger. Here, we find that the lack of magnetic flux in small plasmoids holds  for both merging and non-merging plasmoids. As we have anticipated in \sect{struct}, the inflow region around a relatively large plasmoid generally displays a lower magnetic content (so, $B\lesssim B_0$), as a result of the field lines piling up on the outskirts of the large plasmoid. A small plasmoid lying in the current sheet in the vicinity of a large plasmoid will have a surface magnetic field $B\lesssim B_0$, as we further demonstrate in \app{profile}. More precisely, the surface magnetic field is weaker for smaller plasmoids. This justifies the trend in magnetic flux in the third row of \fig{islfluids}. 

We have also verified that the transition from $\Psi/B_0 w\lesssim 1$ to $\Psi/B_0 w\sim 1$ always occurs at the same value of $w/L$, regardless of the system length (i.e., for different choices of $L/\rhot$). In fact, as we demonstrate in \sect{size}, the largest islands always reach a width $w/L\sim 0.1-0.2$, regardless of $L/\rhot$. At their surface the magnetic field is comparable to the field $B_0$ far from the current sheet. Since the largest islands are controlling the structure of magnetic field lines near the current sheet (magnetic field lines have to wrap around the biggest islands), one would expect that small islands whose width is a fixed fraction of the size of the largest islands will have a similar suppression of the magnetic flux, independent of the system length $L/\rhot$. This is indeed what we find.

\begin{figure*}
\centering
 \resizebox{0.9\hsize}{!}{\includegraphics{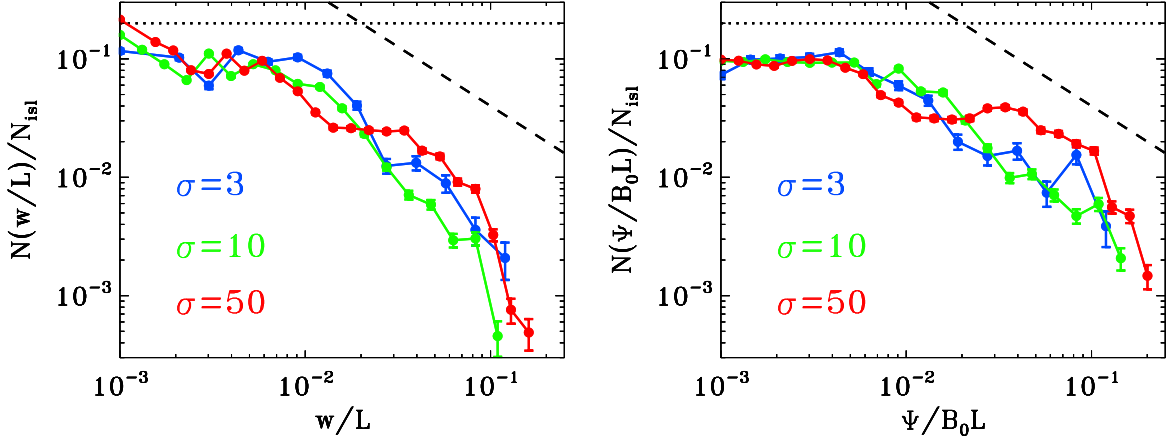}}
 \caption{Plasmoid width (left) and flux (right) cumulatives distribution functions, for three values of the magnetization, as indicated in the legend ($\sigma=3$ in blue, $\sigma=10$ in green, and $\sigma=50$ in red). The width is normalized to the system length $L$ and the flux to $B_0L$, while the histogram (with Poissonian error bars) is normalized to the overall number of plasmoids $N_{\rm isl}$. The corresponding differential distributions can be obtained as $f(w)=dN(w)/dw$ and $f(\Psi)=dN(\Psi)/d\Psi$. The predictions $N(w)\propto w^{-1}$ by \citet{uzdensky_10,loureiro_12} and $N(w)\propto {\rm const}$ by \citet{huang_12} (and similarly for $\Psi$) are plotted as dashed and dotted black lines, respectively.}
 \label{fig:islstat}
\end{figure*}

From the 2D histogram of size and magnetic flux in the third row of \fig{islfluids}, one can find the distribution of plasmoid sizes and magnetic fluxes, by projecting onto the two axes. Our results are shown in \fig{islstat}, where we present the \textit{cumulative} distribution of plasmoid sizes (left panel, with the plasmoid width in units of the system length $L$) and fluxes (right panel, with the flux in units of $B_0 L$). 
In the plots, the dashed lines indicate the prediction $N(w)\propto w^{-1}$ by \citet{uzdensky_10} (and similarly for $\Psi$), whereas the dotted lines indicate the scaling $N\propto$ const suggested by \citet{huang_12}. The main difference among the two models lies in the assumption on the relative velocity between merging plasmoids, which \citet{uzdensky_10}  took to be always $\sim v_A$, whereas \citet{huang_12} allowed for a more detailed dependence on the plasmoid size.

\fig{islstat} suggests that the plasmoid distributions (for both width and magnetic flux) are nearly the same for the three values of magnetization we investigate in this work ($\sigma=3$ in blue, 10 in green and 50 in red). In addition, our findings are fully consistent with the results of MHD simulations of non-relativistic and relativistic reconnection \citep{fermo_10,loureiro_12,huang_12,takamoto_13}. The distributions show a hard slope at small sizes and fluxes (at $w/L\lesssim 0.01$ and $\Psi/B_0L\lesssim 0.01$), resembling the $N\propto$ const prediction of \citet{huang_12}. This is clearer in the magnetic flux distribution than in the size distribution. Ideally, one would need to extend our study to even larger system lengths in order to attain a broader dynamic range in $w$ and $\Psi$, and reliably probe the distributions of small islands (for comparison, the MHD study by \citet{huang_12} extended over six orders of magnitude in $\Psi$). For the magnetic flux, the break at $\Psi/B_0L\sim 0.01$ likely results from the fact that small islands (with $w/L\lesssim 0.02$) display a deficit of magnetic flux as compared to the expected value of $\Psi\sim B_0 w$, as shown in the third row of \fig{islfluids}.

At larger sizes and fluxes ($w/L\gtrsim 0.01$ and $\Psi/B_0L\gtrsim 0.01$), the distributions show a steeper decay, approximately as $N(w)\propto w^{-1}$ and $N(\Psi)\propto \Psi^{-1}$, which is consistent with the model of \citet{uzdensky_10}. The size distribution cuts off at $w/L\sim 0.2$, whereas the flux distribution terminates at $\Psi/B_0 L\sim 0.2$. We find that the reconnection layer might occasionally result in the formation of extraordinarily large plasmoids with $w/L\sim 0.3-0.4$, but their occurrence is extremely rare (not more than once every few tens of $L/c$). Overall, we conclude that the plasmoids routinely produced in the reconnection layer reach at most a width of $w^{\rm max}/L\sim 0.2$. This is in good agreement with the findings of non-relativistic MHD simulations, where the largest plasmoids (defined as ``monster plasmoids'' by \citet{uzdensky_10}) reached a width of $\sim 0.2 \,L$.\footnote{Note that in \citet{loureiro_12} the overall system length was $L$, whereas it is $2L$ in our case. Also, they defined $w$ to be the plasmoid half-width, whereas it is the full width in our work.}


We now proceed to describe the fourth, fifth and sixth rows in \fig{islfluids}. They all illustrate the magnetic and kinetic energy content of plasmoids, as a function of size and magnetization. The comoving magnetic energy fraction in the fourth row of panels is computed as 
 $\epsilon_{\rm B}=\epsilon_{\rm B,lab}-\epsilon_{\rm E,lab}$, where the magnetic energy in the comoving frame is $B^2/8\pi= (B_{\rm lab}^2-E_{\rm lab}^2)/8 \pi$, assuming that the electric field vanishes in the plasmoid comoving frame. Alternatively, by defining $\epsilon_{\parallel,\rm lab}=B_{x,\rm lab}^2/8 \pi n_0 mc^2$ and $\epsilon_{\perp,\rm lab}=(B_{y,\rm lab}^2+B_{z,\rm lab}^2)/8 \pi n_0 mc^2$, the comoving magnetic energy fraction equals $\epsilon_{\rm B}=\epsilon_{\parallel,\rm lab}+\epsilon_{\perp,\rm lab}/\Gamma^2$. We have verified that the two expressions yield nearly identical results. The comoving magnetic energy fraction scales in the large plasmoids as $\epsilon_{\rm B}\sim \sigma$, as  indicated by the dotted white lines and expected from pressure equilibrium. If the magnetic field in the plasmoids were to be the same as the field in the inflow, we would expect $\epsilon_{\rm B}\sim \sigma/2$. It follows that the comoving magnetic field in the large plasmoids is on average $\sim \sqrt{2}\, B_0$. The apparent deficit in magnetic energy in the small plasmoids is directly related to the lack of magnetic flux discussed above, and it ultimately results from the decrease of magnetic field strength in the vicinity of the current sheet, as we describe in \app{profile}. So, small plasmoids are still in pressure equilibrium with their surroundings, but the mean magnetic energy (both inside and outside the small plasmoids) is smaller than for larger plasmoids.
 
We have also measured the ratio of the ``parallel'' and ``perpendicular'' comoving magnetic energies, $\epsilon_{\parallel,\rm lab}/(\epsilon_{\perp,\rm lab}/\Gamma^2)$ (not shown in \fig{islfluids}). We have verified that it is remarkably constant with respect to island size (in particular, it does not show the deficit at small islands of the overall magnetic energy fraction), and it is slightly larger than unity. This is in agreement with the fact that the plasmoids are nearly spherical, just slightly elongated along the direction parallel to the outflow (see the first row in \fig{islfluids}). 
 
The fifth row in \fig{islfluids} illustrates the internal energy content of secondary magnetic islands. The kinetic energy fraction in the plasmoid comoving frame $\epsilon_{\rm kin}$ is computed from quantities measured in the simulation frame as
\be
\epsilon_{\rm kin}=\frac{\epsilon_{\rm kin,lab}-(\Gamma-1)n_{\rm lab}/n_0}{\hat{\gamma}\Gamma^2-(\hat{\gamma}-1)}
\ee
where we have assumed that the particle distribution in the plasmoid comoving frame is isotropic with adiabatic index $\hat{\gamma}$. The adiabatic index is computed iteratively using the \citet{synge_57}'s equation of state, and in all the cases we find that $\hat{\gamma}\sim 4/3$. The assumption of an isotropic particle population is well realized in large plasmoids, as we demonstrate in \sect{spec}.

In analogy to the magnetic energy fraction, the kinetic energy fraction scales as $\sim \sigma$, as  indicated by the dotted white lines and expected from pressure equilibrium. For small islands, we notice the usual deficit, although it is much less dramatic for the kinetic energy fraction than for the magnetic energy fraction, for reasons that will be clarified in \app{profile}.
For large islands, the kinetic energy fraction is slightly larger than the magnetic energy fraction (compare the fourth and fifth rows in \fig{islfluids}). This suggests that secondary plasmoids display a slight dominance of particle energy over magnetic energy.

This is further illustrated in the sixth row of \fig{islfluids}, which shows the equipartition parameter $\chi_{\rm eq}$. Following SPG15, this is defined as
\be\label{eq:chi}
\chi_{\rm eq}=\frac{\int \frac{\varepsilon_{\rm kin}}{\varepsilon_{\rm kin}+\varepsilon_{\rm B}} \varepsilon_{\rm kin} \,dV}{\int \varepsilon_{\rm kin} \,dV}
\ee
where the integral is extended over the volume of the plasmoid, $\varepsilon_{\rm kin}$ is the particle kinetic energy density (as measured in the plasmoid rest frame) and  $\varepsilon_{\rm B}$ is the comoving magnetic energy density. In the case of a proton-electron plasma, $\varepsilon_{\rm kin}$  should only include the kinetic energy of the radiating species, i.e., of the electrons. As appropriate for the fast cooling regime (see SPG15), the ratio $\varepsilon_{\rm kin}/(\varepsilon_{\rm kin} +\varepsilon_{\rm B} )$ in \eq{chi} is weighted with the kinetic energy.

In the case of equipartition between kinetic and magnetic energy densities, we would expect $\chi_{\rm eq}\sim 0.5$. The sixth row in \fig{islfluids} shows that, regardless of the magnetization or the plasmoid size, secondary islands are indeed close to equipartition, with only a slight dominance of the particle kinetic component, leading to $\chi_{\rm eq}\sim 0.6$ (as indicated by the dotted white lines). This is consistent with the results in SPG15, where the equipartition parameter was measured by integrating over the entire plasmoid chain (rather than isolating individual plasmoids, as we are doing here). Finally, we remark that the tendency for $\chi_{\rm eq}\rightarrow 1$ at small island sizes (most significantly for $\sigma=3$, left panel) is just related to the corresponding deficit of magnetic energy that we have discussed above, and that we explain in \app{profile}.

We comment on the bottom-most panel in \fig{islfluids} in the next subsection, where we describe the particle spectrum and anisotropy in secondary plasmoids.

\begin{figure*}
\centering
 \resizebox{\hsize}{!}{\includegraphics{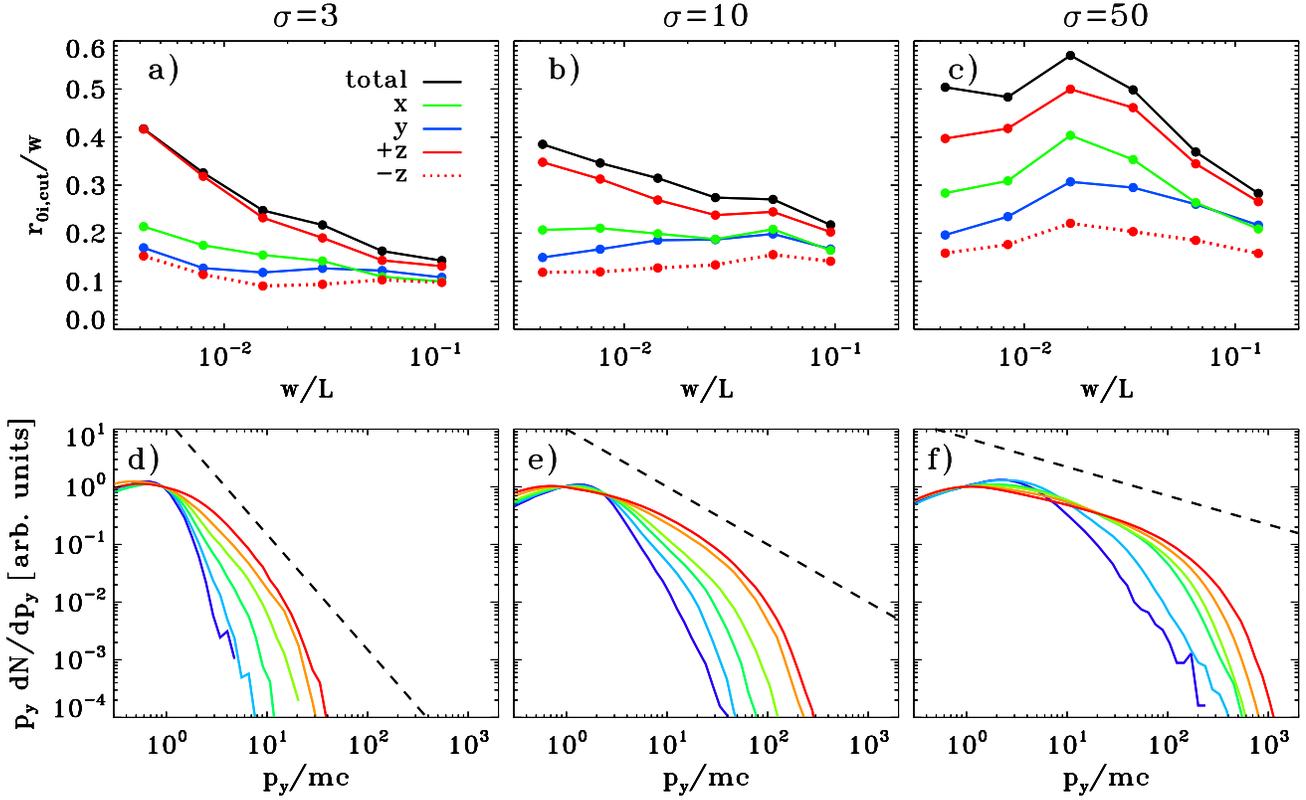}}
 \caption{Spectrum and anisotropy as a function of the plasmoid width, for three values of the magnetization, as indicated at the top ($\sigma=3$ in the left column, $\sigma=10$ in the middle column, and $\sigma=50$ in the right column). Top row: in bins of the plasmoid size $w$ normalized to $L$, we plot  the Larmor radius $r_{0i,\rm cut}=p_{i,\rm cut}c/eB_0$ of the positrons at the comoving cutoff momentum $p_{i,\rm cut}$ ($i=x$ in green, $i=y$ in blue, $i=+z$ in solid red, $i=-z$ in dotted red) or the Larmor radius $r_{0,\rm cut}=p_{\rm cut}c/eB_0$ of the positrons at the total comoving cutoff momentum, regardless of the direction (black solid lines). Each data point represents the mean value among the plasmoids whose width falls in that range. While small plasmoids preserve a significant anisotropy in the direction of the accelerating electric field, large plasmoids have nearly isotropic particle distributions. Bottom row: momentum spectra of $p_{y}$, with the different colors corresponding to the six bins in width indicated in the top panels (from blue for small $w$, to red for large $w$). At large plasmoid widths, the spectrum approaches a power law, with slope $s\sim3$ for $\sigma=3$, $s\sim2$ for $\sigma=10$ and $s\sim1.5$ for $\sigma=50$, as indicated by the dashed lines.}
 \label{fig:spec}
\end{figure*}

\subsection{Particle Spectrum and Anisotropy in Plasmoids}\label{sec:spec}
\fig{spec} describes the properties of the population of particles belonging to individual plasmoids, as a function of the plasmoid size and of the flow magnetization (with $\sigma=3$ in the left column, $\sigma=10$ in the middle column and $\sigma=50$ in the right column).

In the top row, we plot the Larmor radius $r_{0i,\rm cut}=p_{i,\rm cut}c/eB_0$ of the positrons at the comoving cutoff momentum $p_{i,\rm cut}$ along different directions ($i=x$ in green, $i=y$ in blue, $i=+z$ in solid red, $i=-z$ in dotted red) or the Larmor radius $r_{0,\rm cut}=p_{\rm cut}c/eB_0$ of the positrons at the total comoving cutoff momentum, regardless of the direction (black solid lines). The definition of the cutoff momentum is in \eq{pcut}, and the Larmor radius is normalized to the plasmoid width $w$. Each filled circle in the top row of \fig{spec} represents the mean value among the plasmoids whose width (as indicated on the horizontal axis, in units of $L$) falls in that range. 

The Larmor radius $r_{0y,\rm cut}=p_{y,\rm cut}c/eB_0$ (blue lines) measured with the $y$ component of the momentum transverse to the current sheet (which is Lorentz invariant) scales almost linearly with the plasmoid width $w$ (equivalently, $r_{0y,\rm cut}/w$ is a constant). This is also illustrated in the bottom-most row in \fig{islfluids}, where we present the same ratio  $r_{0y,\rm cut}/w$ for all the plasmoids in our simulations, as a 2D histogram. Remarkably, the ratio $r_{0y,\rm cut}/w$ is also nearly insensitive to the magnetization, and always around $\sim 0.2$ (as indicated by the dotted white lines in \fig{islfluids}), with only a slight tendency for a lower value at $\sigma=3$ (see the top left panel in \fig{spec}).

We argue that the ratio $r_{0y,\rm cut}/w$ is an excellent indicator of the confinement capabilities of secondary plasmoids, i.e., of  the highest energy particles that can stay trapped in a plasmoid of width $w$. In fact, the particles inflowing into the reconnection layer are initially accelerated by the reconnection electric field at the X-points (either the primary X-point or the series of secondary X-points in between secondary plasmoids), as described by \citet{zenitani_01} and SS14. At X-points, positrons are accelerated along the $+z$ direction and electrons in the opposite direction. Then, the tension force of the reconnected magnetic field advects the particles away from the X-point along the $x$ direction of the outflow, where they will be trapped in magnetic islands. So, before entering a magnetic island the particle motion preferentially lies in the $xz$ plane. It follows that the outflowing particles can attain an appreciable component of momentum along the $y$ direction only by interaction with the toroidal magnetic loops of the islands. Larger islands will be able to scatter along $y$ higher energy particles, and then to keep them confined within the island contour. 

The value $r_{0y,\rm cut}/w\sim 0.2$ can indeed be understood with a simple argument, based on particle confinement. A particle will stay trapped in a given plasmoid if its full Larmor circle is smaller than the island half-width (the $B_x$ magnetic field switches sign on the two sides of the island, so the comparison is with the plasmoid half-width, rather than its full width). The most constraining condition will apply to the particles at the cutoff momentum $p_{y,\rm cut}$, that need to have $2 \,r_{0y,\rm cut}\lesssim w/2$, or equivalently $r_{0y,\rm cut}/w\lesssim 0.25$. This is remarkably close to the results shown by the blue lines in the top row of \fig{spec}. The fact that the the ratio $r_{0y,\rm cut}/w$ is slightly smaller for $\sigma=3$ than for higher magnetizations is probably due to our definition of $p_{i,\rm cut}$ in \eq{pcut}, which overestimates the exponential cutoff of the momentum distribution by a factor of $n_{\rm cut}-s$ (see the discussion after \eq{pcut}). We always choose $n_{\rm cut}=6$, but the spectral slope $s$ of the momentum distribution is a function of the magnetization (as we explain below), with $s\sim 3$ for $\sigma=3$, $s\sim 2$ for $\sigma=10$ and $s\sim 1.5$ for $\sigma=50$. So, our definition of $p_{i,\rm cut}$ overestimates the true cutoff momentum by a factor of $\sim 3$ for $\sigma=3$ and $\sim 4-5$ for higher magnetizations. So, if we were to measure the Larmor radius with the true cutoff momentum, rather than our proxy in \eq{pcut}, we would obtain that it is a constant fraction of the plasmoid width, remarkably independent of the magnetization.

As shown by the top row in \fig{spec}, the ratio $r_{0x,\rm cut}/w$, with the cutoff momentum $p_{x,\rm cut}$ measured in the plasmoid rest-frame, follows closely the ratio $r_{0y,\rm cut}/w$ for all the values of magnetization we explore. Thus, the ratio $r_{0x,\rm cut}/w$ is also a good indicator of the confinement capabilities of secondary plasmoids. The fact that $r_{0x,\rm cut}/w$ is slighlty larger than $r_{0y,\rm cut}/w$ for small islands (compare the green and blue lines in the top row of \fig{spec}) is probably related to the fact that the plasmoid length is typically larger than its width by a factor of $\sim 1.5$ (see the top row in \fig{islfluids}), so that particles with $x$ momentum moderately higher than the $y$ momentum can still stay confined in a plasmoid of given $w$.

While the positron cutoff momentum along the $-z$ direction follows the same trend as $p_{y,\rm cut}$ and $p_{x,\rm cut}$ (compare the dotted red line with the blue and green lines), the positron momentum along the $+z$ direction of the reconnection electric field shows a distinct behavior (solid red line in the top panels of \fig{spec}). Small islands are highly anisotropic, with $p_{+z,\rm cut}$ appreciably larger than $p_{y,\rm cut}$ and $p_{x,\rm cut}$ \citep[see also][]{cerutti_12b,cerutti_13a,kagan_16}. As a small (and so, fast) island moves along the current sheet, it might stay in phase for a significant time with high-energy particles accelerated at a neighboring X-point, that are now propagating along the current sheet. 
This results in a value of  the positron cutoff momentum $p_{+z,\rm cut}$ (along the accelerating electric field) much larger than $p_{-z,\rm cut}$. For electrons, the opposite anisotropy is observed. 
The sign of the $z$ anisotropy seen in the top row of \fig{spec} is consistent with the sign of the electric field at X-points, suggesting that direct acceleration by the reconnection electric field plays an important role in the early stages of particle acceleration \citep[SS14]{zenitani_01,nalewajko_15}.\footnote{The electric field at the ends of two coalescing islands (i.e., outside of the region in between the two islands) is also oriented along $+z$, but we do not expect small islands to have suffered many mergers.} In contrast, the anti-reconnection electric field at the interface between merging islands is oriented along the $-z$ direction, and it would result in an anisotropy opposite to what is observed in \fig{spec}. This suggests that the anti-reconnection electric field does not play a major role for the acceleration of the particles trapped in small islands. The fact that the $z$ anisotropy in small islands is preferentially induced by the accelerating electric field at X-points is also revealed by the 2D pattern of the $z$ anisotropy (not shown), whose strongest signal appears in the vicinity of the current sheet plane (i.e., for $|y|\ll w$).

Large islands are nearly isotropic, for all the magnetizations we have explored. As we demonstrate in \sect{size}, the transition between small anisotropic islands and large isotropic islands occurs at the same plasma scale, i.e., at a size that is a fixed multiple of $\!\rhot$, for different values of the overall system length $L/\rhot$. It follows that, for a realistic astrophysical system with $L/\rhot\gg1$, all but the smallest islands will be fairly isotropic. Kinetic simulations in small computational domains, due to the lack of a sufficient separation of scales between the plasma scales and the system size, might have artificially over-emphasized the degree of particle anisotropy \citep{cerutti_12b,cerutti_13a,kagan_16}, which is found to be quite low in the big islands of our large-scale simulations.

This also explains why, for a fixed $w/L$, the $z$ anisotropy is more pronouced at higher magnetizations. As we have stated above, the degree of anisotropy depends primarily on $w/\rhot$. Our domain size for $\sigma=10$ is twice as big (in units of $\rhot$) than for $\sigma=50$. This explains why, at the same $w/L$, the case $\sigma=50$ displays a higher degree of $z$ anisotropy than the case $\sigma=10$.

The residual weak anisotropy in large islands is not related to the reconnection electric field, but it is a consequence of the combined effect of the $\nabla B$-drift and the curvature drift, both pointing along $+z$ for positrons (and in the opposite direction for electrons). As a result, the anisotropy is not localized in the vicinity of the $y=0$ plane (as it is the case for small islands), but it is rather uniform over the plasmoid surface (not shown). For a relativistic particle, the magnitude of the drift speed scales proportionally to the ratio between the particle Larmor radius and the island half-width (which we take as a proxy for the scale length $\sim B/|\nabla B|$ or the curvature radius of field lines). This has two consequences. First, for a given island, the anisotropy in $z$ will be maximal for the highest energy particles, since they will have a larger ratio of their Larmor radius to the island size. This trend is indeed observed (but not shown). Second, the black lines in the top panel of \fig{spec} show that the ratio of the Larmor radius at the total cutoff momentum (regardless of direction) and the island size is a decreasing function of $w$. So, the drift speed at the total cutoff momentum will be larger for smaller islands, which explains the trend in anisotropy seen in \fig{spec} (but only at the high-$w$ end, since we have argued that the anisotropy in small islands has a different origin).

The highest energy particles in the largest islands cannot be produced via direct acceleration by the reconnection electric field at X-points \citep[SS14]{guo_14,nalewajko_15}. \citet{werner_16} found that spectrum of particles produced at X-points should cut off at $\sim 4\,\sigma$ (see also Fig.~5 in SS14). Given that our proxy for the cutoff momentum is a factor of a few larger than the true cutoff, it would correspond to a Larmor radius $r_{0,\rm cut}\sim 10 \rhot$, independently from the island size. In large islands, we argue that  the high-energy cutoff of the particle spectrum is populated by particles accelerated during island mergers \citep[SS14]{guo_14,nalewajko_15}.  Let us assume that the accelerating electric field generated during island mergers is $\sim 0.2 B_0$, where we have taken the reconnection rate in between the two merging islands to be of order $\sim 0.15\,c$ and we have considered that the magnetic field in the plasmoids is a factor of $\sim 1.5$ larger than the field $B_0$ in the inflow (see the fourth row in \fig{islfluids}). The potential energy available over a characteristic acceleration length of order $\sim w$ will be $\sim 0.2 e\,B_0 w$, which results in a Larmor radius $r_{0,\rm cut}/w\sim 0.2$. This  simple argument shows that as a result of the merger of two large islands, particles can be accelerated so that to maintain the ratio $r_{0,\rm cut}/w$ close to 0.2, as required by our findings.

By equating $r_{0,\rm cut}\sim 10 \rhot$ (from X-point acceleration) with $r_{0,\rm cut}/w\sim 0.2$ (from acceleration in island mergers), we find a critical island size of $w/\rhot\sim 50$. The highest energy particles in islands with $w/\rhot\lesssim 50$ are primarily accelerated by the reconnection electric field at X-points (and we would expect a high degree of $z$ anisotropy), whereas acceleration in island mergers dominates at $w/\rhot\gtrsim 50$. In units of the system length $L$, we expect the transition to occur at $w/L\sim0.07$ for $\sigma=3$, at $w/L\sim0.04$ for $\sigma=10$ (for our fiducial case $L/\rhot\simeq 1130$) and at $w/L\sim0.1$ for $\sigma=50$.



The bottom row in \fig{spec} shows how the momentum spectrum $p_y dN/dp_y$ of positrons trapped in plasmoids depends on the plasmoid size (different colors, from blue to red, correspond to the bins in width indicated by the filled circles in the top panels). As we demonstrate in \app{speccomp}, most of the high-energy particles reside in plasmoids (more precisely, in the largest plasmoids), so the particle spectrum from plasmoids is an excellent proxy for the spectrum integrated over the whole current sheet (i.e., including the plasmoids as well as the regions in between islands).

The bottom row of \fig{spec} shows that the upper cutoff of the momentum spectrum scales linearly with the island size (the bins in width are logarithmically spaced), in agreement with the blue line in the top row. The spectral shape is a strong function of the island size. Small islands have a momentum distribution that is nearly thermal, since they cannot confine the highest energy particles accelerated at X-points. An extended power-law distribution appears in larger islands, since they can successfully trap all of the particles accelerated at X-points. The power-law slope in the largest islands is asymptoting to $s\sim 3$ for $\sigma=3$, $s\sim 2$ for $\sigma=10$ and $s\sim 1.5$ for $\sigma=50$, as indicated by the dashed black lines in the three bottom panels. Such slopes are consistent with the values quoted in SS14, where the particle spectrum was integrated over the whole current sheet. As we demonstrate in \app{speccomp}, most of the high-energy particles are contained within large islands. It follows that the spectrum measured in SS14, despite being integrated over the whole layer, was actually mostly contributed by the few largest islands. It is then quite natural to expect that the spectral slope of the largest islands in \fig{spec} is comparable to the power-law index found in SS14, for all the magnetizations we explore.

In the regime $\sigma\gg1$, one expects that the mean energy per particle in the plasmoids should scale as $\propto \sqrt{\sigma}$ \citep{lyubarsky_05}. Since small islands have nearly-thermal particle spectra, one would expect the same scaling for the peak momentum, as long as it exceeds unity. In fact, the peak of the blue curve in panel (e) lies at $p_y/mc \sim 1.5$, whereas it is close to $p_y/mc \sim 3$ for the blue line in panel (f). This is consistent with the predicted scaling $\propto \sqrt{\sigma}$.\footnote{For $\sigma=3$ the peak is at even lower values, but since it lies at non-relativistic speeds we would not necessarily expect that it scales as $\propto \sqrt{\sigma}$.} For a given magnetization, the mean energy per particle should be independent of the plasmoid size, which explains why for larger sizes, as the particle spectrum extends to higher momenta, the low-energy cutoff also recedes to lower values. This is more pronounced for $\sigma=10$ and 50 than for $\sigma=3$, since the power-law slope for $\sigma=3$ is so steep that most of the energy content is controlled by the particles at the low-energy cutoff, which is then expected to stay nearly the same as we vary the plasmoid width.

In the bottom row of \fig{spec} we have chosen to display the momentum spectrum along the direction transverse to the current sheet to emphasize the confinement capabilities of plasmoids of different sizes. We find that the comoving momentum spectrum $p_x dN/dp_x$ is nearly identical (not shown; yet, compare the blue and green lines in the top row), and also similar to $[p_z dN/dp_z]_{p_z<0}$, i.e., the $z$ momentum spectrum of positrons having $p_z<0$ (or similarly, of electrons having $p_z>0$). In contrast, the  spectrum $[p_z dN/dp_z]_{p_z>0}$ of positrons having $p_z>0$ shows a harder spectral slope (even for relatively small islands), resulting from particles accelerated by the reconnection electric field at X-points (see \app{speccomp}).


\begin{figure*}
\centering
 \resizebox{\hsize}{!}{\includegraphics{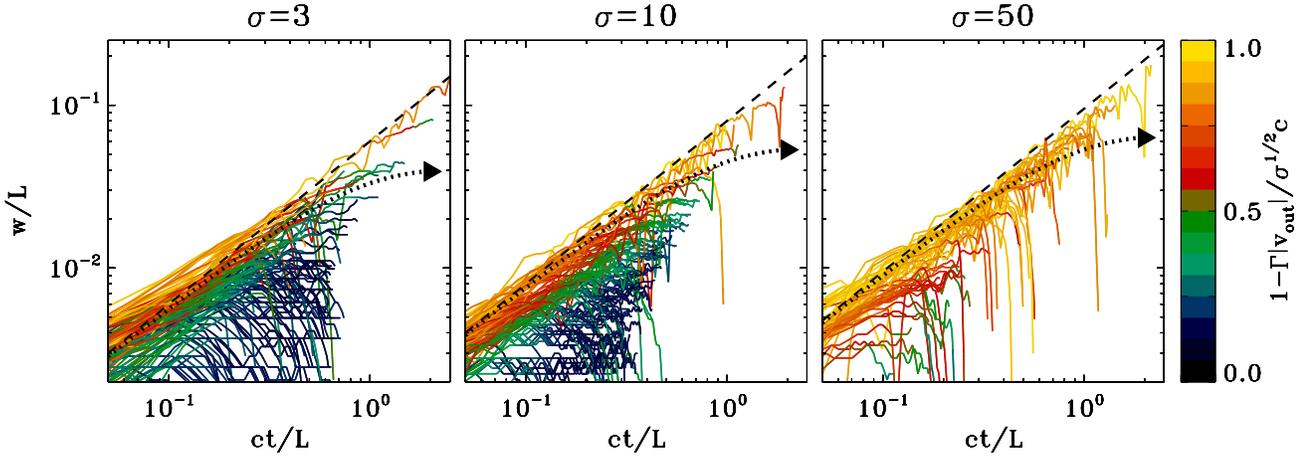}}
 \caption{Growth of the plasmoid width as a function of the proper time, for three values of the magnetization, as indicated at the top ($\sigma=3$ in the left panel, $\sigma=10$ in the middle panel, and $\sigma=50$ in the right panel). For the sake of clarity, we only plot the evolutionary tracks of plasmoids whose lifetime (in the lab frame) is longer than $\simeq0.4 L/c$. Colors indicate the plasmoid four-velocity, relative to the  terminal four-velocity $\sim \sqrt{\sigma}\,c$ predicted by \citet{lyubarsky_05}, as indicated by the colorbar (blue for fast plasmoids, yellow for slow plasmoids). It is apparent that most of the growth happens at slow velocities (yellow or red tracks), and that in this phase the plasmoid width increases linearly with time (with a coefficient of $w/t\simeq0.06\,c$ for $\sigma=3$, $\simeq 0.08\,c$ for $\sigma=10$, and $\simeq 0.1\,c$ for $\sigma=50$, as indicated by the dashed black lines). When the plasmoid speed approaches the \alf\ speed (or equivalently, its four-velocity approaches $\sqrt{\sigma}\,c$), the plasmoid growth terminates, as schematically shown by the curved dotted lines with arrows.}
 \label{fig:islgrowth}
\end{figure*}
\subsection{Plasmoid Growth}\label{sec:growth}
By following individual plasmoids over time, we can quantify the rate at which they grow. This is illustrated in \fig{islgrowth}, where we present the temporal evolution of the plasmoid width (time is measured in the plasmoid rest-frame), for three values of the magnetization, as indicated at the top ($\sigma=3$ in the left panel, $\sigma=10$ in the middle panel, and $\sigma=50$ in the right panel). For the sake of clarity, we only plot the evolutionary tracks of plasmoids whose lifetime (in the lab frame) is longer than $\simeq0.4 L/c$. 

We find that in the plasmoid rest frame, the growth in plasmoid width proceeds at a constant rate
\be\label{eq:growth}
\frac{dw}{c\,dt}=\beta_{\rm g}
\ee	
where the coefficient $\beta_{\rm g}$ has a weak dependence on magnetization. As indicated by the three dashed black lines in \fig{islgrowth}, $\beta_{\rm g}\simeq 0.06$ for $\sigma=3$, $\beta_{\rm g}\simeq 0.08$ for $\sigma=10$ and $\beta_{\rm g}\simeq 0.1$ for $\sigma=50$. For all the values of $\sigma$, we find that $\beta_{\rm g}$ is about half of the reconnection inflow rate $|v_{\rm in}|/c$ (see the top panel in \fig{outflows}). We have also verified that our results are not consistent with the assumption of constant growth in the laboratory frame. Rather, since it is in the plasmoid rest-frame that the growth rate is constant, the increase in plasmoid width in the simulation frame follows $dw/c\,dt_{\rm lab}=\beta_{\rm g}/\Gamma$, where $\Gamma$ is the plasmoid bulk Lorentz factor. In \sect{toy}, we propose a toy model to corroborate our results.
	
The colored tracks in \fig{islgrowth} confirm that the growth of the largest plasmoids proceeds at a constant rate in the comoving frame, following the black dashed lines. The sudden dips that occasionally appear in the tracks (e.g., at $ct/L\sim 2$ in the right panel) are a consequence of the apparent shrinking in the plasmoid size that accompanies island mergers (see the beginning of \sect{chain} in relation to \fig{isltracks}). Soon after the merger, the surviving plasmoid (defined as the larger of the two that merged) recovers its proper width. The smaller of the two merging plasmoids  terminates its life trajectory. This explains why in \fig{islgrowth} smaller plasmoids tend to have shorter lives, since they are more likely to encounter a larger plasmoid that swallows them. 

The color coding in \fig{islgrowth} indicates the plasmoid four velocity, relative to the expected terminal four-velocity $\sim \sqrt{\sigma}\, c$ (more precisely, colors indicate $1-\Gamma|v_{\rm out}|/\sqrt{\sigma}\,c$).  Most of the plasmoid growth happens while the islands are still non-relativistic (yellow in \fig{islgrowth}). When the plasmoids approach the \alf\ speed (so, their four-velocity becomes closer to $\sim \sqrt{\sigma}\, c$), the accretion rate diminishes and the plasmoid width saturates (green and blue in \fig{islgrowth}). As a result, their tracks deviate from the locus of constant growth indicated by the dashed black lines in \fig{islgrowth}  (as sketched by the curved dotted lines with arrows in \fig{islgrowth}). For example, the two plasmoids that live longest in the middle panel (for $\sigma=10$) deviate from the $dw/c\,dt=0.08$ curve when their four velocity reaches $\sim 0.5 \sqrt{\sigma}\,c$, and after this point their width stays nearly unchanged.

For each plasmoid, we have tracked the trajectories of the particles that get trapped within the plasmoid, which allows to clarify how the plasmoid growth proceeds. We find that most of the long-lasting plasmoids are born close to the center of the current sheet. The longer they spend in the vicinity of $x\sim 0$  in their early life (see the position-time tracks in \fig{isltracks}), the larger they will eventually become. While they stay in the vicinity of the center (roughly speaking, within a distance of a few times their size), the islands accrete from both sides of the current sheet. Eventually, the tension force of the field lines pulls them away from the center. 
At a given distance from the current sheet, the speed of a plasmoid is inversely proportional to its size, as we show in \sect{accel}. It follows that while its speed is still non-relativistic, a given plasmoid mostly accretes smaller (so, faster) plasmoids that are trailing in its wake. As it accelerates to relativistic speeds, the accretion rate from the trailing side decreases, since now the plasmoid is moving at the same speed ($\sim c$) as all of the smaller plasmoids in its wake, and the trailing islands cannot catch up with it. At this point, most of the accreted particles come from the region ahead of the plasmoid.


\begin{figure}
\centering
 \resizebox{\hsize}{!}{\includegraphics{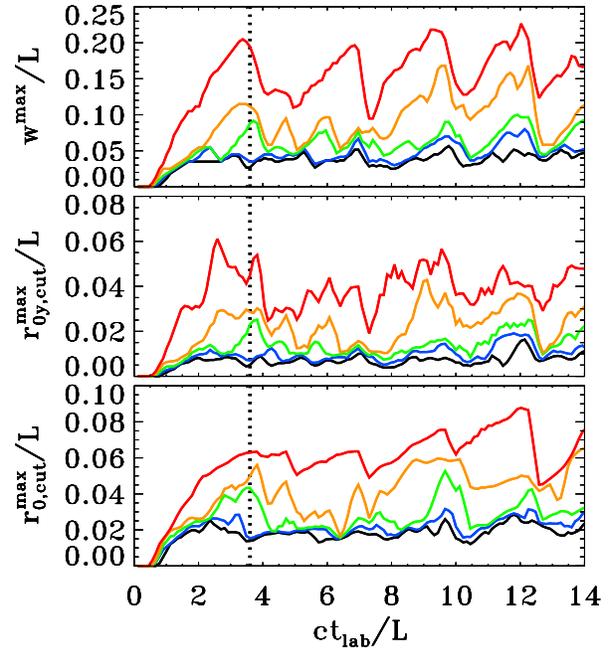}}
 \caption{Temporal evolution of the properties of the five largest plasmoids (from the largest in red to the fifth largest in black), as a function of time in the laboratory frame, for a simulation with $\sigma=10$ and $L\simeq 127\rhot$. Top panel: width of the five largest plasmoids, in units of the system size $L$. Middle panel: the five plasmoids with the largest value of the  positron Larmor radius $r_{0y,\rm cut}=p_{y,\rm cut}c/eB_0$, where $p_{y,\rm cut}$ is the cutoff momentum in the $y$ direction. Bottom panel: same as in the middle panel, but for the total comoving positron momentum, rather than its $y$ component. The dotted black line shows the characteristic timespan ($\sim 3.6\,L/c$) of our fiducial runs. The panels show that the typical recurrence time for the largest plasmoids (having a width $w^{\rm max}\sim 0.2\,L$) is $\sim 2.5 L/c$, and that the temporal evolutions of $r^{\rm max}_{0y,\rm cut}$ and $r^{\rm max}_{0,\rm cut}$ closely follow the time track of $w^{\rm max}$.}
 \label{fig:timesize}
\end{figure}

\fig{timesize} shows that accretion of particles in the current sheet can lead to the formation of ``monster plasmoids'' whose width reaches $w/L\sim 0.2$, in agreement with the findings of MHD simulations of non-relativistic reconnection \citep{loureiro_12}. \fig{timesize} refers to a system with $\sigma=10$ and $L/\rhot\simeq127$, but we have verified that this conclusion holds regardless of $\sigma$ or $L/\rhot$.
Extraordinarily large plasmoids with $w/L\sim 0.3-0.4$ might be occasionally produced, but their occurrence is extremely rare (not more than once every few tens of $L/c$). In fact, in the timespan of $\sim14\lc$ covered by \fig{timesize}, no plasmoid larger than $\sim0.2\,L$ is produced.

The red curve in the top panel of \fig{timesize} shows the width of the largest plasmoid existing in our domain as a function of time. The other curves show the width of the second largest plasmoid (in yellow) down to the fifth largest plasmoid (in black), showing that plasmoids with $w/L\sim 0.05$ occur much more frequently than the monster plasmoid with $w/L\sim 0.2$ (as we have already described in \fig{islstat}). 
The largest plasmoid (red curve in the top panel of \fig{timesize}) grows at a fraction $\sim 0.1$ of the speed of light, in agreement with the conclusions of \fig{islgrowth}. This can be measured from the temporal slope of the red curve in the top panel: in the time interval $1\lesssim ct\lab/L\lesssim 3$, the plasmoid width increases up to $0.2\,L$, as indeed expected for a growth rate of $\sim 0.1\,c$.\footnote{As we demonstrate in \sect{accel}, large plasmoids move at non-relativistic speeds, so no relativistic corrections are required to transform from the comoving to the laboratory time.} The sudden drop in the red curve at $ct\lab/L\sim 3.5$ occurs when the large plasmoid is ejected from the current sheet. At this point, the plasmoid that used to be the second largest (see the yellow curve in the top panel at $ct\lab/L\sim 3.5$) becomes the largest one in the domain (and the third largest becomes the second largest, and so on).

The red curve in the top panel of \fig{timesize} demonstrates that the typical recurrence time of monster plasmoids is $\sim 2.5 \,L/c$ (see the quasi-periodic peaks in the red curve).\footnote{It follows that the simulation timespan of $\sim 3.6\lc$ of our fiducial runs (indicated with a vertical dotted black line in \fig{timesize}) is sufficient to capture the steady state physics of the system, and in particular the occurrence of monster plasmoids.} This is indeed the time needed to grow a plasmoid up to the monster width of $w/L\sim 0.2$, since the growth proceeds at a rate $\sim 0.08\,c$ (see the middle panel in \fig{islgrowth}). In turn, this implies that most of the plasma flowing into the current sheet within this time interval is accreted onto the monster plasmoid that is currently present in the reconnection layer. 
 It is only when the monster plasmoid is ejected from the current sheet that another plasmoid can grow to monster-like sizes.

The middle panel in \fig{timesize} presents the five plasmoids with the five largest values  (from red to black) of the  positron Larmor radius $r_{0y,\rm cut}=p_{y,\rm cut}c/eB_0$, where $p_{y,\rm cut}$ is the cutoff momentum in the $y$ direction. The temporal evolution of the curves in the middle panel displays a remarkable correlation with the lines in the top panel (compare curves of the same color). This is in agreement with the results shown in \fig{islfluids} (bottom-most row) and in \fig{spec} (top row), i.e., that the Larmor radius computed with the cutoff momentum $p_{y,\rm cut}$ scales linearly with the island size. Those two plots showed that this linear relation holds for all sizes, whereas \fig{timesize} only focuses on the largest plasmoids, with size $w/L\gtrsim 0.05$. From the top and middle panels in \fig{timesize}, one can compute that the coefficient of the linear scaling is around $r_{0y,\rm cut}/w\sim 0.2$, in excellent agreement with \fig{islfluids} and \fig{spec}. 

The top row in \fig{spec} showed that the largest plasmoids are nearly isotropic, and the upper cutoff in their momentum spectrum (black lines; regardless of the direction) is comparable to the cutoff momentum in the $y$ direction (blue lines). This is confirmed by the bottom panel in \fig{timesize}, where the curves display the  plasmoids with the five largest values  (from red to black) of the total positron Larmor radius $r_{0,\rm cut}=p_{\rm cut}c/eB_0$. The  evolution of the curves in the bottom panel is well correlated in time with the lines in the middle panel (compare curves of the same color), and their ratio is of order of unity, i.e.,  $r_{0,\rm cut}\gtrsim r_{0y,\rm cut}$, in agreement with the top row in \fig{spec} at the largest sizes ($w/L\gtrsim 0.05$).

\subsubsection{Toy Model for the Plasmoid Growth}\label{sec:toy}
We now present a toy model for the plasmoid growth, in support of the numerical findings presented above.
Let us assume that each plasmoid of size $w$ is accreting material from a ``distance of influence'' $d_w$, as measured along the current sheet in the plasmoid rest-frame. 
An empirical estimate of the distance of influence will be given below. This implies that the number of particles $N_{\rm prt}$ in the plasmoid will grow, in the comoving frame, as 
\be
   \frac{dN_{\rm prt}}{cdt}=2 n_0 \frac{|v_{\rm in}|}{c} d_w
\ee
where the factor of two accounts for accretion from the two sides of the current sheet (the  flux of particles perpendicular to the current sheet is Lorentz invariant). The plasmoid area is $\sim \pi w w_{\parallel}/4\simeq w^2$, where we have used that $w_\parallel\simeq 1.5\,w$ (see the top row in \fig{islfluids}). If $n$ is the plasmoid mean density, the number of particles in the plasmoid will be $N_{\rm prt}\simeq  n w^2$. This implies that the plasmoid width increases at a rate
\be
\frac{dw}{cdt}\simeq \frac{n_0}{n} \frac{|v_{\rm in}|}{c} \frac{d_w}{w}
\ee
where we have assumed that the comoving density $n$ does not significantly change with time. To estimate the distance of influence $d_w$, one can assume that all the plasma that is attracted to a given island will eventually accrete onto that island. The electric current integrated over the surface of an island of size $w$ is proportional to $w$, so its distance of influence will also scale linearly with $w$. By following the trajectories of the particles that will eventually accrete onto a given plasmoid, we can measure the distance of influence in our simulations, and we find that it is of order $d_w\sim 2\,w$ for all the magnetizations we explore. Since $n/n_0\sim 3-5$ regardless of $\sigma$ (second row in \fig{islfluids}), we find that the plasmoid size grows as
\be
\frac{dw}{cdt}\simeq 0.5 \frac{|v_{\rm in}|}{c} 
\ee
which is consistent with our numerical findings.

\begin{figure*}
\centering
 \resizebox{\hsize}{!}{\includegraphics{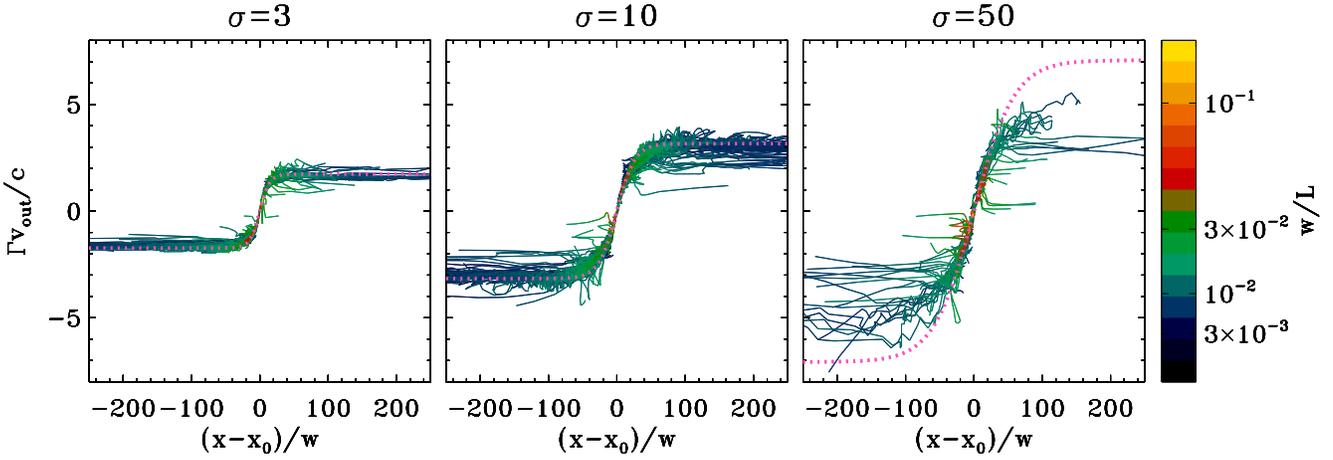}}
 \caption{Plasmoid four-velocity along the outflow direction as a function of distance  $(x-x_0)$ from their birth location $x_0$, normalized to the instantaneous plasmoid width $w$, for three values of the magnetization, as indicated at the top ($\sigma=3$ in the left panel, $\sigma=10$ in the middle panel, and $\sigma=50$ in the right panel). For the sake of clarity, we only plot the tracks of plasmoids whose lifetime (in the lab frame) is longer than $\simeq0.35 L/c$. Once the distance along the current sheet is normalized with the plasmoid width $w$, we show that all the tracks overlap, regardless of the plasmoid size (which is indicated by the colors, see the colorbar on the right). The plasmoid four-velocity follows a universal profile $\Gamma v_{\rm out}/c\simeq \sqrt{\sigma}\,\tanh[0.12(x-x_0)/\sqrt{\sigma}w]$, where the coefficient of $\simeq0.12$ is nearly insensitive to the magnetization (see the pink dotted lines in the three panels, for the three values of $\sigma$).}
 \label{fig:islvel}
\end{figure*}
\subsection{Plasmoid Acceleration}\label{sec:accel}
After growing by accretion, secondary plasmoids are accelerated by the tension force of the reconnected magnetic field up to ultra-relativistic speeds, while they propagate toward the boundaries of the domain (``first they grow, then they go''). We find that, at a given distance from the center of the current sheet, larger plasmoids tend to be slower. We find that the shape of the bulk acceleration profile becomes universal (i.e., the same for all the plasmoids in a given system) if the plasmoid four-velocity $\Gamma v_{\rm out}$ along the outflow direction is measured as a function of the distance $x-x_0$ from their birth location $x_0$, normalized to the instantaneous plasmoid width $w$. This is demonstrated in \fig{islvel}, which shows that, once the distance along the current sheet is in units of the plasmoid width $w$, all the plasmoid tracks overlap, regardless of the plasmoid size (which is indicated by the colors, see the colorbar on the right). 

We find a universal profile of the form (see the dashed pink lines in \fig{islvel})
\be\label{eq:acc}
\Gamma \frac{v_{\rm out}}{c}\simeq \sqrt{\sigma} \tanh\left(\frac{\beta_{\rm acc}}{\sqrt{\sigma}}\frac{x-x_0}{w}\right)
\ee
where $x_0$ is the plasmoid location at its birth, and the dimensionless acceleration rate $\beta_{\rm acc}\simeq 0.12$ is nearly independent of the flow magnetization (with only a minimal tendency to increase for higher reconnection rates, i.e., higher $\sigma$). For $\sigma=10$, we have also verified that the acceleration rate $\beta_{\rm acc}$ does not depend on the overall system length $L/\rhot$, from $L/\rhot\simeq127$ up to $L/\rhot\simeq1130$.

\eq{acc} implies that  high velocities (or equivalently, for $\beta_{\rm acc}|x-x_0|\gtrsim \sqrt{\sigma} w$, the plasmoids approach the terminal four-velocity $\sim\sqrt{\sigma}\,c $. On the other hand, at low velocities (or equivalently, for $\beta_{\rm acc}|x-x_0|\lesssim \sqrt{\sigma} w$) the scaling is 
\be\label{eq:accsimple}
\Gamma \frac{v_{\rm out}}{c}\sim \beta_{\rm acc} \frac{x-x_0}{w}
\ee
This implies that: (\textit{i}) monster plasmoids with $w/L\sim 0.2$ leave the system at trans-relativistic speeds, with $|v_{\rm out}|/c\sim 0.5$; (\textit{ii}) the largest plasmoids capable of reaching the end of the current sheet (i.e., $|x|\sim L$) with the terminal four-velocity $\sim\sqrt{\sigma}\,c $ have a final width of $w/L\sim \beta_{\rm acc}/\sqrt{\sigma}\sim 0.1/\sqrt{\sigma}$, i.e., they are systematically smaller at higher magnetizations.

The scaling in \eq{accsimple} has a simple empirical justification. Quite generally, we can equate the growth time up to a width $w$, which equals $t_{\rm g,lab}\sim\Gamma w/\beta_{\rm g}$ in the lab frame, with the acceleration/propagation time up to a distance $x-x_0$ from the plasmoid birth place, which is $t_{\rm acc,lab}\sim c(x-x_0)/v_{\rm out}$, finding that 
\be\label{eq:accmodel}
\Gamma \frac{v_{\rm out}}{c}\sim \beta_{\rm g} \frac{x-x_0}{w}
\ee
This has the same form as \eq{accsimple}, and by comparison we argue that $\beta_{\rm acc}\sim \beta_{\rm g}\sim 0.1$, as indeed we find in our simulations. 

The right panel in \fig{islvel} shows that for $\sigma=50$ only few islands can approach the terminal dimensionless four-velocity $\sim \sqrt{\sigma}\sim 7$, as compared to the copious number of fast plasmoids observed for $\sigma=3$ and 10. Yet, since regions with four-velocity as fast as $\sim \sqrt{\sigma}\,c\sim 7 \,c$ are present in the reconnection layer, in the fast smooth outflows in between magnetic islands (see the red line in the bottom panel of \fig{outflows}), we anticipate that, for a sufficiently large system, a number of plasmoids will be capable of reaching the expected terminal four-velocity $\sim\sqrt{\sigma}\,c$, even for $\sigma=50$.

The apperent lack of fast plasmoids for $\sigma=50$ is due to three main reasons. First, for a given final plasmoid size $w$ (in units of $\rhot$), acceleration to the terminal speed requires a length $L/\rhot\gtrsim \sqrt{\sigma} \beta_{\rm acc}^{-1} (w/\rhot)$, i.e., larger domains (i.e., larger $L/\rhot$) are needed at higher $\sigma$. For comparison, our domain length for $\sigma=10$ is $L/\rhot\simeq 1130$, which implies that we should use $L/\rhot\sim 2500$ for $\sigma=50$, in order to capture a sufficient number of plasmoids moving at the terminal speed (rather, our domain for $\sigma=50$ is $L/\rhot=505$). Second, as we have described above, the critical size of the largest plasmoid that can reach the terminal \alf\ velocity is $\sim 0.1\,L/\sqrt{\sigma}$, so it is smaller at higher $\sigma$. This implies that the probability for such a plasmoid to accrete onto a larger and slower plasmoid is higher at stronger magnetizations (since the ``target'' plasmoid can range in size from  the monster width of $\sim 0.2\,L$ down to the width of the ``projectile'' plasmoid $\sim 0.1\,L/\sqrt{\sigma}$). 
The small and fast plasmoid disappears into the large and slow one that lies ahead, before reaching the terminal four-velocity. The third reason is the fact that interactions among the plasmoids are stronger at higher magnetizations, as we have discussed at the beginning of \sect{chain}. As a result, a small and fast plasmoid that is formed ahead of a large and slow one gets pulled back, inhibiting its acceleration. For the second and third reasons, we still expect that the number of plasmoids capable of reaching the expected terminal four-velocity $\sim\sqrt{\sigma}\,c$ will be smaller for $\sigma=50$ than for lower magnetizations,  even at the same value of $L/\sqrt{\sigma}\rhot$.

We have explicitly verified that it is the tension of the field lines that is responsible for the plasmoid bulk acceleration. Since the total magnetic and kinetic energy content of a plasmoid with width $w$ scales as $\propto \sigma w^2$ in the plasmoid comoving frame, its momentum in the lab frame will be $\propto \sigma w^2 \Gamma v_{\rm out}$. From \eq{growth} and \eq{accsimple}, it follows that the force exerted on the plasmoid in the simulation frame (i.e., the time derivative of its momentum in the lab frame) scales as $\propto \sigma w v_{\rm out}/c$, until the plasmoid reaches the terminal four-velocity. By measuring the tension force exerted on each plasmoid by the bundle of field lines lying between the plasmoid contour and its trailing X-point, we have successfully assessed that it is indeed the magnetic field tension that drives the plasmoid bulk acceleration. 

We now present a toy model to support why the magnetic tension force should scale as $\propto w v_{\rm out}/c$, or equivalently (using \eq{accsimple}), as $\propto w \min[1,\beta_{\rm acc}x/w]$, assuming for the sake of simplicity that the plasmoid starts at $x_0\sim 0$. For a plasmoid sufficiently far from the center, the accelerating force will be provided by a cone of field lines trailing behind the plasmoid. The opening angle of the cone is $\sim |v_{\rm in}|/c$ (as predicted by \citet{lyubarsky_05}, this is the inclination angle of the field lines) and its apex lies at a distance $d_{\rm apex}\sim (c/|v_{\rm in}|) \,w/2$ behind the plasmoid. For a plasmoid sufficiently far from the center, all of the field lines between the apex and the plasmoid will contribute to provide the accelerating force, which would then scale as $\propto w$. In contrast, if the apex lies beyond the center of the current sheet (i.e., to the other side, as compared to the plasmoid location), only a fraction of the field lines would be available for acceleration (in the special case of a plasmoid that lies close to the center, the tension force is expected to vanish). This happens when the plasmoid distance from the center is $x\lesssim d_{\rm apex}$, in which case only a fraction $\sim x/d_{\rm apex}$ of the field lines will be employed for acceleration. Putting everything together, the tension force should scale as $\propto \min[x,d_{\rm apex}]\propto w\min[1,2\,(|v_{\rm in}|/c) \,x/w ]$, in good agreement with our numerical results.

\begin{figure}
\centering
\includegraphics[width=0.45\textwidth]{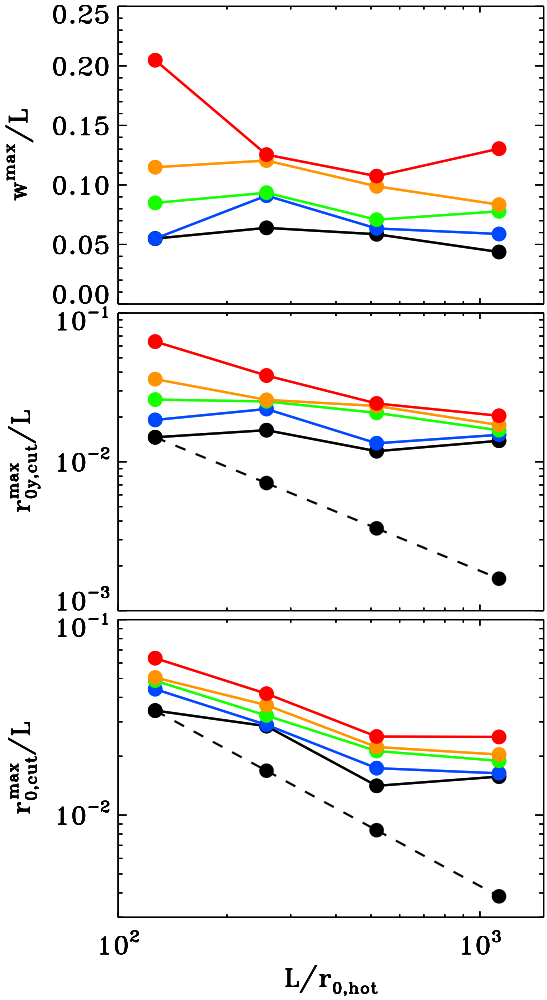}
 \caption{Scalings of plasmoid properties with respect to the overall system length $L$, as measured in units of $\rhot$, from $L/\rhot\simeq127$ up to $L/\rhot\simeq 1130$. In all the cases, the simulation timespan is $\sim 3.6 \lc$. Top panel: the red line shows the plasmoid that, in the course of its history, reaches the largest width; the yellow line shows the second largest, and so on until the fifth largest plasmoid (black line). The width of the few largest plasmoids approaches a fixed fraction of the system size $L$, which is independent of the length $L$ itself. Middle panel:  the red line shows the plasmoid that, in the course of its history, reaches the largest value of the positron Larmor radius $r_{0y,\rm cut}=p_{y,\rm cut}c/eB_0$, where $p_{y,\rm cut}$ is the cutoff momentum in the $y$ direction. Similarly, the yellow line shows the plasmoid that reaches the second largest value, and so on until the fifth largest value (black line). At large $L$, the maximum Larmor radius approaches a fixed fraction of the overall system length $L$, independently of $L$. For comparison, the dashed black line shows the scaling $r_{0y,\rm cut}^{\rm max}/L\propto L^{-1}$ expected if the maximum cutoff $p_{y,\rm cut}^{\rm max}$ were to saturate. Bottom panel: same as in the middle panel, but for the total comoving positron momentum, rather than its $y$ component. This can be regarded as the Hillas criterion for relativistic reconnection.}
 \label{fig:scalesize}
\end{figure}

\section{Dependence on the System Length}\label{sec:size}
In this section, we describe how our results depend on the overall length of the system $L$, measured in units of the Larmor radius $\rhot=\sqrt{\sigma}\comp$ of the particles heated/accelerated by reconnection.\footnote{We remind that the overall extent of our simulation domain in the $x$ direction of the outflow  is actually $2\,L$.} We focus on our fiducial magnetization $\sigma=10$ and we vary the system length $L/\rhot$ from 127 up to 1130. In all the cases, the simulation timespan is $\sim 3.6 \lc$, for fair comparison among the different values of $L/\rhot$.

We find that both the plasmoid growth rate $\beta_{\rm g}$ and the acceleration rate $\beta_{\rm acc}$ are nearly identical over the range of $L/\rhot$ that we explore. The plasmoid fluid properties presented in \fig{islfluids} do not depend on the overall system length. As we have anticipated in \sect{fluid}, the value of $w/L$ where the lack of magnetic flux starts to appear (i.e., smaller islands have a deficit of magnetic flux, as compared to the scaling $\Psi/B_0w\sim 1$ realized at larger widths) is always $w/L\sim 0.02$ regardless of the system length.

\fig{scalesize} quantifies how the size of the largest plasmoids and the maximum energy of accelerated particles depend on the system length $L$. In the top panel, the red line shows the plasmoid that, in the course of its history, reaches the largest width; the yellow line shows the second largest, and so on until the fifth largest plasmoid (black line). The width of the largest plasmoid (red curve) is affected by the limited timespan of our simulations. In fact, \fig{timesize} shows, for the case $L/\rhot\simeq127$, that a timespan of $\sim 3.6\lc$ (as indicated by the dotted vertical black line in \fig{timesize}) is barely sufficient to capture the full growth of the largest plasmoid. From the second to the fifth largest plasmoids (yellow to black lines in \fig{scalesize}), we are not affected by the limited  timespan of our simulations, since the number of plasmoids increases at smaller widths. The corresponding curves in the top panel of \fig{scalesize} show that the width of the few largest plasmoids is a fixed fraction of the system length $L$, irrespective of $L/\rhot$. 

We find that the linear scaling between plasmoid width $w$ and Larmor radius $r_{0y,\rm cut}$ of the particles at the cutoff $y$ momentum holds regardless of the system length. Since the largest plasmoids have $w/L\sim 0.05-0.2$ independently of the system length (top panel), it is then not surprising that the largest values of the  positron Larmor radius $r_{0y,\rm cut}$ scale linearly with the system size, as shown in the middle panel of \fig{scalesize}. The normalization of the curves in the middle panel is consistent with the relation $r_{0y,\rm cut}/w\sim 0.2$ discussed in \sect{spec} and with the ratios  $w/L$ presented in the top panel (in fact, the slight decay of the red curve in the middle panel is to be correlated with the red line  in the top panel). This suggests that the maximum $y$ momentum of particles accelerated in reconnection scales linearly with the system size $L$. In contrast, if reconnection were to give a maximal value of the positron $y$ momentum that stays constant with $L$, we would expect that $r_{0y,\rm cut}^{\rm max}/L\propto L^{-1}$ in the middle panel of \fig{scalesize}. This is shown as a dashed black line, and it is clearly inconsistent with our data.

These conclusions are further supported by the bottom panel in \fig{scalesize}. There, we perform a similar analysis as in the middle panel, but for the total comoving positron momentum, rather than its $y$ component. All the curves (from red for the largest value of the Larmor radius $r_{0,\rm cut}$, down to black for the fifth largest) display a similar trend, decreasing as $r_{0,\rm cut}^{\rm max}/L\propto L^{-1}$ for $L/\rhot\lesssim 300$ (compare with the dashed black line), and flattening out for $L/\rhot\gtrsim 300$. We attribute the transition to a change in the dominant mechanism for particle acceleration. At relatively small system sizes, particle acceleration by the reconnection electric field at X-points dominates \citep[e.g.,][]{zenitani_01}. The accelerated positrons move preferentially along the $+z$ direction of the electric field (and electrons in the opposite direction). The particle distribution is highly anisotropic, and the total momentum is primarily controlled by its $z$ component (see the top row in \fig{spec}). This explains why the decreasing trend in the bottom panel of \fig{scalesize}, which is driven by the $z$ momentum, does not appear in the middle panel, where the Larmor radius $r_{0y,\rm cut}$ is computed using the $y$ momentum. At this stage (i.e., for $L/\rhot\lesssim 300$), the maximum energy of accelerated particles is controlled by the acceleration capabilities of X-points. As anticipated in \sect{spec}, the upper cutoff in the momentum spectrum of particles accelerated at X-points corresponds to a Larmor radius $r_{0,\rm cut}\sim 10 \rhot$, regardless of the system length $L$. This explains why for $L/\rhot\lesssim 300$ the Larmor radius of the highest energy particles scales as $r_{0,\rm cut}^{\rm max}/L\propto L^{-1}$. This is in agreement with the results by \citet{werner_16}.

For larger systems (i.e., at $L/\rhot\gtrsim 300$), particle acceleration during island mergers plays a more and more dominant role.  The potential energy available during the merger of two islands of width $\sim w$ is $\sim 0.2 e\,B_0 w$,\footnote{We have taken the reconnection rate in between the two merging islands to be of order $\sim 0.15\,c$ and we have considered that the magnetic field in the plasmoids is a factor of $\sim 1.5$ larger than the field $B_0$ in the inflow (see the fourth row in \fig{islfluids}).} which results in a Larmor radius $r_{0,\rm cut}\sim 0.2\,w$. By equating $r_{0,\rm cut}\sim 10 \rhot$ (from X-point acceleration) with $r_{0,\rm cut}\sim 0.2\,w$ (from acceleration in mergers), we find that the transition between X-point acceleration and acceleration governed by island mergers should occur at a critical island width of $w/\rhot\sim 50$. Then, remembering that $w/L\sim 0.2$ for the largest islands (which will give the highest energy particles), we find that the transition width $w/\rhot\sim 50$ should correspond to a critical domain length of $L/\rhot\sim 250$, in agreement with the break in the curves of \fig{scalesize} (bottom panel).

For $L/\rhot\gtrsim 300$, the ratio $r_{0,\rm cut}/L$ is expected to remain constant with $L$, since the energy of the highest energy particles accelerated in island mergers scales linearly with the width $w$ of the largest plasmoids, which in turn is proportional to the system size $L$ (see the top panel). Simulations in a small domain \citep{werner_16} are not able to reach this asymptotic limit, beyond the scaling $r_{0,\rm cut}^{\rm max}/L\propto L^{-1}$ that holds at small $L$.\footnote{In the case of untriggered reconnection studied with periodic boundary conditions the same requirement of $L/\rhot\gtrsim 300$ should be imposed over the distance in between two neighboring primary islands, resulting in a much more constraining condition on the overall box length (which typically includes many primary plasmoids).} Rather, we find that in the limit $L/\rhot\gg1$ of astrophysical interest, the highest energy particles accelerated by reconnection have a Larmor radius $r_{0,\rm cut}\sim 0.03 L$, regardless of $L/\rhot$ (here, we have implicitly assumed that the highest energy particles in the system are contained in magnetic islands, as we demonstrate in \app{speccomp}). This can be regarded as the \textit{Hillas criterion for relativistic reconnection}.

\begin{figure}
\centering
\includegraphics[width=0.5\textwidth]{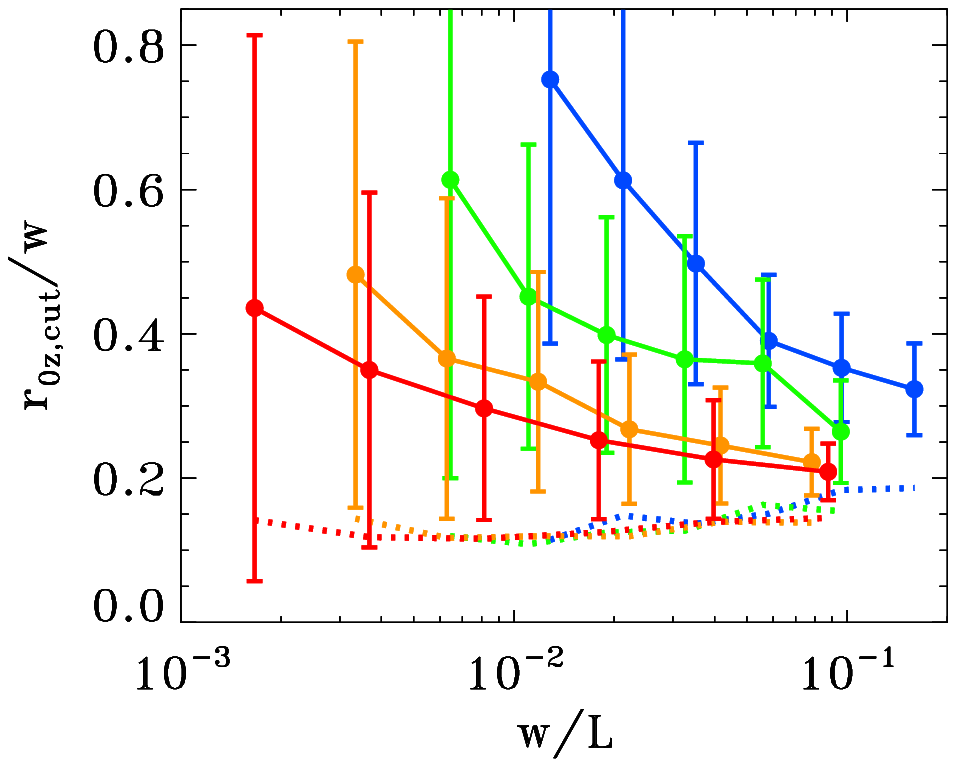}
 \caption{Positron anisotropy in the $z$ direction, as a function of the plasmoid size $w$ and for different values of the system length, $L/\rhot\simeq127$ (blue), $L/\rhot\simeq257$ (green), $L/\rhot\simeq 518$ (yellow) and $L/\rhot\simeq 1130$ (red). In all the cases, the simulation timespan is $\sim 3.6 \lc$. Each of the filled circles indicates the  value of $r_{0i,\rm cut}/w$ with $i=+z$ (i.e., along the reconnection electric field), averaged among the plasmoids whose width falls in that range. Error bars indicate the standard deviation. In contrast, the dotted lines show the mean values of $r_{0i,\rm cut}/w$ with $i=-z$. For each choice of $L$, larger plasmoids are closer to isotropy. Also, small system lengths $L$, where the largest plasmoids are not much bigger than $\rhot$, tend to over-emphasize the degree of anisotropy.}
 \label{fig:scaleaniso}
\end{figure}

We conclude this section by investigating the dependence of the $z$ anisotropy of the particle distribution on the system length $L$. In \fig{scaleaniso}, we quantify the $z$ anisotropy by means of the Lamor radius of positrons at the upper cutoff of their $z$ momentum spectrum, considering only positrons with $p_z>0$ (solid lines with error bars) or $p_z<0$ (dotted lines). The positron Larmor radius is plotted as a function of the island size $w/L$ for different system lengths ($L/\rhot\simeq127$ in blue, $L/\rhot\simeq257$ in green, $L/\rhot\simeq 518$ in yellow and $L/\rhot\simeq 1130$ in red). For each value of $L/\rhot$, the positron anisotropy decreases with increasing plasmoid width, as already discussed in \sect{spec}. More interestingly, the degree of anisotropy at fixed $w/L$ is significantly lower for larger system lengths $L/\rhot$. Rather than being dependent on $w/L$, the level of $z$ anisotropy seems to be a function of $w/\rhot$, i.e., of the plasmoid width normalized to plasma scales (rather than to the system length $L$). This is suggested in \fig{scaleaniso} by the fact that different solid curves appear to overlap, if we were to shift them along the horizontal axis by the corresponding value of $L/\rhot$ (this, in fact, is equivalent to measuring $w$ in units of $\rhot$). This results from the fact that in small systems the process of particle acceleration is dominated by the X-point stage, that occurs on plasma scales $\sim \rhot$.

The pronounced anisotropy observed at small $L/\rhot$ still bears memory of the anisotropy of particles accelerated at X-points \citep{cerutti_12b,cerutti_13a,kagan_16}. In contrast, in large systems most of the particles accelerated at X-points are efficiently isotropized in magnetic islands (SS14), resulting in a lower degree of anisotropy. It follows that the kinetic beaming effect described by \citet{cerutti_12b,cerutti_13a,kagan_16}, due to the strong anisotropy of particles accelerated by reconnection, tends to be important only in small systems. Since realistic astrophysical systems typically have lengths $L/\rhot\gg1$, we conclude that the comoving particle distribution in the largest plasmoids, which are likely to give the brightest emission signatures, is nearly isotropic.

\section{Summary and Discussion}\label{sec:summary}
In this work, we have performed a suite of 2.5D PIC simulations of anti-parallel relativistic reconnection in pair plasmas --- 2D in space, but all three components of velocities and electromagnetic fields are tracked --- to study the long-term evolution of the system, independently of the initial setup of the current sheet. We explore a range of flow magnetizations (from $\sigma=3$ to $\sigma=50$), focusing on unprecedentedly large-scale systems with length $L\gg\rhot$, where $\rhot=\sqrt{\sigma}\comp$ is the Larmor radius of particles heated/accelerated by reconnection. We find that a self-consistent by-product of the asymptotic physics is the continuous formation of a chain of plasmoids/magnetic islands, generated by the secondary tearing instability \citep{uzdensky_10}. We argue that such plasmoids, quasi-spherical structures filled with high-energy particles and magnetic fields, can play a dominant role in the high-energy emission from relativistic astrophysical sources, such as PWNe and jets in AGNs and GRBs. We first summarize our findings on the reconnection physics and then briefly discuss their astrophysical implications.

\subsection{Reconnection Plasma Physics Summary}
We have fully characterized the plasmoid properties as a function of their width $w$ (transverse to the reconnection layer) and the flow magnetization $\sigma$, and our main conclusions can be summarized as follows:
\bi
\item the plasmoids are nearly spherical, with length along the current sheet that is a factor of $\sim 1.5$ larger than the width. They are moderately denser than the inflowing plasma (a factor of a few, with only a moderate dependence on $\sigma$), with magnetic field strength averaged over the plasmoid volume that is $\sim 50\%$ higher than the value $B_0$ the inflow region. Both the magnetic  and the kinetic energy density in the plasmoids scale linearly with the magnetization $\sigma$. The plasmoids are nearly in equipartition between particles and magnetic fields, with only a moderate dominance of the particle kinetic content, most pronounced in small plasmoids ($w/L\lesssim 0.02$).
\item Our choice of absorbing/outflow boundary conditions in the $x$ direction of the outflow --- as opposed to the common choice of periodic boundaries --- allows to follow the system for many light crossing times, and to assess the statistical distributions of plasmoid width $w$ and magnetic flux $\Psi$. For large plasmoids ($\Psi/B_0 L\gtrsim 0.01$) the differential distributions of $w$ and $\Psi$ both follow a $-2$ power-law distribution. Smaller plasmoids  ($\Psi/B_0 L\lesssim 0.01$) have harder distributions, with power-law slope around $-1$. The width distribution cuts off at $w/L\sim 0.2$ (and correspondingly, the flux distribution at $\Psi/B_0 L\sim 0.2$). The results of our fully-kinetic simulations are consistent with MHD simulations of non-relativistic \citep{loureiro_12,huang_12}  and relativistic \citep{takamoto_13} reconnection.
\item We identify the particles belonging to each plasmoid, and we quantify their particle spectrum and anisotropy. We find that the Larmor radius $r_{0y,\rm cut}=p_{y,\rm cut}c/eB_0$ measured with the cutoff momentum $p_{y,\rm cut}$ along the $y$ direction transverse to the current sheet scales almost linearly with the plasmoid width $w$, with the same constant of proportionality $r_{0y,\rm cut}/w\sim 0.2$ at all magnetizations. This corresponds to the particles with the highest $y$ momentum being barely confined in the plasmoids (i.e., a confinement criterion). The particle population is roughly isotropic in the $xy$ plane. In small islands, a strong degree of anisotropy in the positron distribution is observed along the $+z$ direction of the reconnection electric field (electrons have the opposite anisotropy), suggesting that direct acceleration at X-points plays an important role in the early stages of particle acceleration \citep[SS14]{zenitani_01,nalewajko_15}. In contrast, large islands are nearly isotropic, and the highest energy particles they contain are accelerated during island mergers \citep[SS14]{guo_14,guo_15a}. The transition between small anisotropic islands and large isotropic islands occurs at $w/\rhot\sim 50$, regardless of the overall system length $L$. It follows that, for a realistic astrophysical system with $L/\rhot\gg1$, all but the smallest islands will be fairly isotropic. In contrast, in small computational domains, due to the lack of a sufficient separation of scales between the plasma scales and the system size, one might artificially over-emphasize the degree of particle anisotropy (as in \citealt{cerutti_12b,cerutti_13a,kagan_16}).
\item Small islands have a momentum distribution that is nearly thermal, since they cannot successfully confine the highest energy particles accelerated at X-points. In contrast, large islands have power-law momentum distributions, with a power-law slope $s\sim 3$ for $\sigma=3$, $s\sim 2$ for $\sigma=10$ and $s\sim 1.5$ for $\sigma=50$. Such slopes are consistent with the values quoted in SS14, where the particle spectrum accounted for all the particles in the current sheet (and not just the particles trapped in plasmoids). This is expected, since we find that nearly all of the highest energy particles produced by reconnection are contained in the few largest plasmoids.
\item By following the trajectory of individual plasmoids over time, we find that the life of secondary plasmoids from birth to adulthood is characterized by two phases: \textit{first they grow, then they go}. They are born on microscopic plasma scales, and they grow by accretion, with a constant comoving growth rate $\beta_{\rm g}=dw/c\,dt\sim 0.1$ that has only a weak dependence on magnetization. For all the values of $\sigma$ we explore, we find that $\beta_{\rm g}$ is about half of the reconnection inflow rate $|v_{\rm in}|/c$, which slightly increases from $|v_{\rm in}|/c\simeq 0.1$ for $\sigma=3$ up to $|v_{\rm in}|/c\simeq 0.2$ for $\sigma=50$. The weak dependence of $|v_{\rm in}|/c$ on magnetization is consistent with the predictions of \citet{lyubarsky_05}'s model of relativistic reconnection. 
\item Occasionally, a plasmoid in the reconnection layer can reach a ``monster'' width of $w\sim 0.2\,L$, consistent with the results of MHD simulations of non-relativistic reconnection \citep{loureiro_12}. The size of the monster plasmoid is always a fixed fraction of the system length $L$, for different choices of $L/\rhot$, and their typical recurrence time is $\sim 2.5\, L/c$. Monster plasmoids have nearly isotropic particle distributions and they contain the highest energy particles in the system. For sufficiently large domains ($L/\rhot\gtrsim 300$), we show that the Larmor radius of the highest energy particles is $r_{0,\rm cut}^{\rm max}\sim 0.03\,L$, i.e., a constant fraction of the system size. This can be regarded as the \textit{Hillas criterion for relativistic reconnection}. In contrast, simulations in a small domain with $L/\rhot\lesssim 300$ \citep{werner_16} would not be able to reach this asymptotic limit, beyond the scaling $r_{0,\rm cut}^{\rm max}/L\propto L^{-1}$ that we observe at small $L$, which is expected from the fact that early particle acceleration at X-points yields a maximum Larmor radius $r_{0,\rm cut}^{\rm max}\sim 10\rhot$ regardless of $L$.
\item After their growth, the plasmoids are accelerated toward the boundaries of the domain by the tension force of the magnetic field lines. We find that the bulk four-velocity of the accelerating plasmoids follows a universal profile $\Gamma v_{\rm out}/c\simeq \sqrt{\sigma}\,\tanh[\beta_{\rm acc}(x-x_0)/\sqrt{\sigma}w]$, where the acceleration rate $\beta_{\rm acc}\simeq0.12$ is nearly insensitive to the flow magnetization. Here, $x_0$ is the plasmoid location at birth. This implies that  at large distances ($\beta_{\rm acc}|x-x_0|\gtrsim \sqrt{\sigma} w$), the plasmoid four-velocity approaches $\sim\sqrt{\sigma}\,c $, i.e., the plasmoid moves at nearly the \alf\ speed $v_A=\sqrt{\sigma/(\sigma+1)}\,c$. This is indeed the outflow speed from relativistic reconnection predicted by \citet{lyubarsky_05}, as we have verified for all the magnetizations we explore (earlier studies in smaller domains could not capture this asymptotic limit, see \citealt{cerutti_13a,guo_15a,kagan_16}). On the other hand, at small distances ($\beta_{\rm acc}|x-x_0|\lesssim \sqrt{\sigma} w$), the scaling is $\Gamma v_{\rm out}/c\sim \beta_{\rm acc}(x-x_0)/w$. It follows that (\textit{i}) monster plasmoids with $w/L\sim 0.2$ leave the system at trans-relativistic speeds; (\textit{ii}) the largest plasmoids capable of reaching the end of the current sheet with the terminal four-velocity $\sim\sqrt{\sigma}\,c $ will have a final width of $w/L\sim \beta_{\rm acc}/\sqrt{\sigma}$, i.e., they are systematically smaller at higher magnetizations.
\ei
We conclude with a few caveats. In this work we have only focused on electron-positron reconnection, but we claim that all of our results will hold for electron-proton reconnection, since in the relativistic regime $\sigma\gg1$ the field dissipation results in nearly equal amounts of energy transferred to protons as to electrons, as we have demonstrated in SPG15. So, the mean energy per particle of the two species is nearly the same, as it is the case for an electron-positron plasma.
From a numerical point of view, electron-ion reconnection is much more demanding than electron-positron reconnection, since the system needs to have a length (in cells) larger by $\sqrt{m_i/m_e}$ in each direction (here $m_i$ and $m_e$ are the proton and electron masses), and the evolution needs to be followed for a factor $\sqrt{m_i/m_e}$ longer. So, it will be even more challenging to reach the asymptotic state described in this work, where the largest islands are nearly isotropic, the outflow speed reaches the expected terminal velocity, and the maximum energy of accelerated particles increases linearly with the system size. Studies in small systems will tend to artificially over-emphasize the importance of effects that are only appropriate at microscopic plasma scales, and irrelevant for $L/\rhot\gg1$ systems of astrophysical interest.

Also, we have only explored the case of anti-parallel fields, i.e., without a guide field perpendicular to the alternating fields. For stronger guide fields, one expects that the efficiency of reconnection will be reduced and the plasmoids will become more magnetically-dominated, as we have shown in SPG15. A complete investigation of the plasmoid properties in guide-field reconnection will be presented elsewhere. Finally, our simulations are two-dimensional. As shown in SS14, the long-term evolution of 3D anti-parallel reconnection is remarkably similar to the 2D physics, both in terms of the dynamics of the reconnection layer and the efficiency of particle acceleration. Still, the structure of plasmoids/flux ropes in 3D remains to be investigated.

\subsection{Astrophysical Implications}
A large recent volume of research in the field has revealed that
relativistic reconnection is a highly dynamical process that involves
a broad range of physical processes on very different timescales; all of which can have direct
observational signatures.  Consider a reconnection layer of length $L$
where magnetic energy is dissipated in a magnetized fluid of total volume
$\sim L^3$ (for simplicity we take all the scales of the problem to be
comparable). Reconnection proceeds on a global timescale 
$t_{\rm rec}\sim L/|v_{\rm in}|\sim 10\,L/c$ until it exhausts all the available 
energy in the system. The dissipated energy ends up in ultra-relativistic
particles that are mostly contained in plasmoids. The emission from 
the layer, therefore, comes in bursts or flares whose duration is closely related 
to the time it takes for the plasmoids to grow and leave the layer.
The largest plasmoids take a time $\sim L/c$ to form, and they leave the 
reconnection layer at mildly relativistic speeds ($\Gamma |v_{\rm out}|/c\sim
1$). Powerful flares are therefore expected from these plasmoids
on a timescale  $t_{\rm l} \simless L/c$. Smaller plasmoids are accelerated to
relativistic speeds $\Gamma \sim \sqrt{\sigma}$ and they radiate 
anisotropically, as a result of their bulk motion. When the observer lies along the current sheet,
small plasmoids are viewed as very powerful and extremely 
fast evolving $t_{\rm s}\ll L/c$ emitters. 

The detailed shape and variety of flares expected from a current sheet 
and their implications for blazar jets are presented elsewhere
(Petropoulou, Giannios \& Sironi, in prep.) Here, we estimate
the maximum energies that particles can achieve in reconnection
layers in blazar jets and during flares from the Crab Nebula.  
Let us assume a flow with a bulk motion $\Gamma_{\rm b}$
that beams its emission towards the observer with corresponding
Doppler factor $\delta \sim \Gamma_{\rm b}$. Consider a major flare
from this flow powered by a large plasmoid. The observed duration
of the flare is $t_{\rm f}\sim L/c \delta$, which constrains the size of
the reconnection layer. From our ``Hillas criterion for reconnection," that dictates the maximum
particle energy associated with this plasmoid, we find that the 
gyro-radius of the highest energy particles is $r_g\sim 0.03\,L$. The energy of the cosmic ray particles 
is $E_{\rm CR}=\Gamma_{\rm b} e Br_g\sim 0.03 \Gamma_{\rm b} \delta eB c t_{\rm f}$,
were the magnetic field $B$ is measured in the rest frame of the flow.

Blazar jets show major flares on a timescale of hours/days 
(as a reference value, we set $t_{\rm f}=10^5t_5\,$s).  
The ``blazar zone'' is characterized by typical $\Gamma_{\rm b}\sim \delta \simmore 10$
while the magnetic field strength is $B\sim 1G$ \citep[e.g.,][]{celotti_08}.\footnote{This value of magnetic field is typical only for leptonic models. In hadronic models $B\sim 100\,$G.}
Protons present in the reconnecting plasma can be accelerated to 
$E_{\rm CR}\sim 3\times 10^{18}\Gamma_{1}\delta_{1}B_0\,t_5$ eV. Here, $\Gamma_1=\Gamma_{\rm b}/10$, $\delta_1=\delta/10$ and $B_0=B/1\unit{G}$.
Therefore, protons can possibly reach the highest observed energies 
of $E\sim 10^{20}$ eV for sufficiently fast jet flows $\Gamma_{\rm b}\simmore 10$
with field strengths $B\simmore 1$G.

The Fermi and AGILE satellites have detected a number of $\sim$day
long flares at GeV energies from the Crab Nebula, which surprisingly falsify the
widely-believed Òstandard candleÓ nature of the high-energy Crab
emission. During these events the Crab nebula $\gamma$-ray flux 
above 100 MeV exceeded its average value by a factor of several or 
higher \citep{abdo_11,buehler_12}.
Fermi acceleration at the termination shock of the Crab nebula fails
to explain the observed GeV flares. In contrast, rapid
conversion of magnetic field energy into particle energy via 
magnetic reconnection has been recently proposed to explain the Crab flares \citep[e.g.,][]{cerutti_13a,cerutti_13b}.
For magnetic field strengths in the nebula of $B\sim$~several mG,
  $t_{\rm f}=10^5t_5\,$s and allowing for a modest relativistic motion 
of the emitting plasmoid $\Gamma_{\rm b}\simeq \delta\simeq 2$, 
we find that pairs can be accelerated up to $E_{\rm CR}\sim 1$PeV. 
This energy is sufficient to potentially explain the extreme
synchrotron peak during the Crab flares. Radiative cooling has, however,
to be taken into account self-consistently  when one considers the 
maximum attainable particle energy. This is left for future work.

 \section*{Acknowledgments}
MP is supported by NASA 
through Einstein Postdoctoral 
Fellowship grant number PF3~140113 awarded by the Chandra X-ray 
Center, which is operated by the Smithsonian Astrophysical Observatory
for NASA under contract NAS8-03060. We thank Katya Giannios for suggestions on the title. The simulations were performed on XSEDE resources under contract No. TG-AST120010, and on NASA High- End Computing (HEC) resources through the NASA Advanced Supercomputing (NAS) Division at Ames Research Center. We also gratefully acknowledge access to the PICSciE-OIT High Performance Computing Center and Visualization Laboratory at Princeton University.
\bibliography{blob}

\begin{thebibliography}{}

\bibitem[\protect\citeauthoryear{{Abdo}}{{Abdo}}{2011}]{abdo_11}
{Abdo} A.~A. e.~a.,  2011, Science, 331, 739

\bibitem[\protect\citeauthoryear{{Bai}, {Caprioli}, {Sironi} \&
  {Spitkovsky}}{{Bai} et~al.}{2015}]{bai_15}
{Bai} X.-N.,  {Caprioli} D.,  {Sironi} L.,    {Spitkovsky} A.,  2015, \apj,
  809, 55

\bibitem[\protect\citeauthoryear{{Begelman}}{{Begelman}}{1998}]{begelman_98}
{Begelman} M.~C.,  1998, \apj, 493, 291

\bibitem[\protect\citeauthoryear{{Belyaev}}{{Belyaev}}{2015}]{belyaev_15}
{Belyaev} M.~A.,  2015, \na, 36, 37

\bibitem[\protect\citeauthoryear{{Bessho} \& {Bhattacharjee}}{{Bessho} \&
  {Bhattacharjee}}{2005}]{bessho_05}
{Bessho} N.,  {Bhattacharjee} A.,  2005, Physical Review Letters, 95, 245001

\bibitem[\protect\citeauthoryear{{Bessho} \& {Bhattacharjee}}{{Bessho} \&
  {Bhattacharjee}}{2007}]{bessho_07}
{Bessho} N.,  {Bhattacharjee} A.,  2007, Physics of Plasmas, 14, 056503

\bibitem[\protect\citeauthoryear{{Bessho} \& {Bhattacharjee}}{{Bessho} \&
  {Bhattacharjee}}{2010}]{bessho_10}
{Bessho} N.,  {Bhattacharjee} A.,  2010, Physics of Plasmas, 17, 102104

\bibitem[\protect\citeauthoryear{{Bessho} \& {Bhattacharjee}}{{Bessho} \&
  {Bhattacharjee}}{2012}]{bessho_12}
{Bessho} N.,  {Bhattacharjee} A.,  2012, \apj, 750, 129

\bibitem[\protect\citeauthoryear{{Buehler} \& al.}{{Buehler} \&
  al.}{2012}]{buehler_12}
{Buehler} R.,  al. 2012, \apj, 749, 26

\bibitem[\protect\citeauthoryear{{Buneman}}{{Buneman}}{1993}]{buneman_93}
{Buneman} O., , 1993, {in ``Computer Space Plasma Physics'', Terra Scientific,
  Tokyo, 67}

\bibitem[\protect\citeauthoryear{{Celotti} \& {Ghisellini}}{{Celotti} \&
  {Ghisellini}}{2008}]{celotti_08}
{Celotti} A.,  {Ghisellini} G.,  2008, \mnras, 385, 283

\bibitem[\protect\citeauthoryear{{Cerutti}, {Philippov}, {Parfrey} \&
  {Spitkovsky}}{{Cerutti} et~al.}{2015}]{cerutti_15}
{Cerutti} B.,  {Philippov} A.,  {Parfrey} K.,    {Spitkovsky} A.,  2015,
  \mnras, 448, 606

\bibitem[\protect\citeauthoryear{{Cerutti}, {Werner}, {Uzdensky} \&
  {Begelman}}{{Cerutti} et~al.}{2012}]{cerutti_12b}
{Cerutti} B.,  {Werner} G.~R.,  {Uzdensky} D.~A.,    {Begelman} M.~C.,  2012,
  \apjl, 754, L33

\bibitem[\protect\citeauthoryear{{Cerutti}, {Werner}, {Uzdensky} \&
  {Begelman}}{{Cerutti} et~al.}{2013a}]{cerutti_13a}
{Cerutti} B.,  {Werner} G.~R.,  {Uzdensky} D.~A.,    {Begelman} M.~C.,  2013a,
  \apj, 770, 147

\bibitem[\protect\citeauthoryear{{Cerutti}, {Werner}, {Uzdensky} \&
  {Begelman}}{{Cerutti} et~al.}{2013b}]{cerutti_13b}
{Cerutti} B.,  {Werner} G.~R.,  {Uzdensky} D.~A.,    {Begelman} M.~C.,  2013b,
  ArXiv:astro-ph/1311.2605

\bibitem[\protect\citeauthoryear{{Daughton} \& {Karimabadi}}{{Daughton} \&
  {Karimabadi}}{2007}]{daughton_07}
{Daughton} W.,  {Karimabadi} H.,  2007, Physics of Plasmas, 14, 072303

\bibitem[\protect\citeauthoryear{{Daughton}, {Scudder} \&
  {Karimabadi}}{{Daughton} et~al.}{2006}]{daughton_06}
{Daughton} W.,  {Scudder} J.,    {Karimabadi} H.,  2006, Physics of Plasmas,
  13, 072101

\bibitem[\protect\citeauthoryear{{Drake}, {Swisdak}, {Che} \& {Shay}}{{Drake}
  et~al.}{2006}]{drake_06}
{Drake} J.~F.,  {Swisdak} M.,  {Che} H.,    {Shay} M.~A.,  2006, \nat, 443, 553

\bibitem[\protect\citeauthoryear{{Drenkhahn} \& {Spruit}}{{Drenkhahn} \&
  {Spruit}}{2002}]{drenkhahn_02a}
{Drenkhahn} G.,  {Spruit} H.~C.,  2002, \aap, 391, 1141

\bibitem[\protect\citeauthoryear{{Fermo}, {Drake} \& {Swisdak}}{{Fermo}
  et~al.}{2010}]{fermo_10}
{Fermo} R.~L.,  {Drake} J.~F.,    {Swisdak} M.,  2010, Physics of Plasmas, 17,
  010702

\bibitem[\protect\citeauthoryear{{Giannios}}{{Giannios}}{2008}]{giannios_08}
{Giannios} D.,  2008, \aap, 480, 305

\bibitem[\protect\citeauthoryear{{Giannios}}{{Giannios}}{2013}]{giannios_13}
{Giannios} D.,  2013, \mnras, 431, 355

\bibitem[\protect\citeauthoryear{{Giannios}, {Uzdensky} \&
  {Begelman}}{{Giannios} et~al.}{2009}]{giannios_09}
{Giannios} D.,  {Uzdensky} D.~A.,    {Begelman} M.~C.,  2009, \mnras, 395, L29

\bibitem[\protect\citeauthoryear{{Giannios}, {Uzdensky} \&
  {Begelman}}{{Giannios} et~al.}{2010}]{giannios_10b}
{Giannios} D.,  {Uzdensky} D.~A.,    {Begelman} M.~C.,  2010, \mnras, 402, 1649

\bibitem[\protect\citeauthoryear{{Guo}, {Li}, {Daughton} \& {Liu}}{{Guo}
  et~al.}{2014}]{guo_14}
{Guo} F.,  {Li} H.,  {Daughton} W.,    {Liu} Y.-H.,  2014, Physical Review
  Letters, 113, 155005

\bibitem[\protect\citeauthoryear{{Guo}, {Li}, {Li}, {Daughton}, {Zhang},
  {Lloyd-Ronning}, {Liu}, {Zhang} \& {Deng}}{{Guo} et~al.}{2016}]{guo_16}
{Guo} F.,  {Li} X.,  {Li} H.,  {Daughton} W.,  {Zhang} B.,  {Lloyd-Ronning} N.,
   {Liu} Y.-H.,  {Zhang} H.,    {Deng} W.,  2016, \apjl, 818, L9

\bibitem[\protect\citeauthoryear{{Guo}, {Liu}, {Daughton} \& {Li}}{{Guo}
  et~al.}{2015}]{guo_15a}
{Guo} F.,  {Liu} Y.-H.,  {Daughton} W.,    {Li} H.,  2015, \apj, 806, 167

\bibitem[\protect\citeauthoryear{{Hesse} \& {Zenitani}}{{Hesse} \&
  {Zenitani}}{2007}]{hesse_zenitani_07}
{Hesse} M.,  {Zenitani} S.,  2007, Physics of Plasmas, 14, 112102

\bibitem[\protect\citeauthoryear{{Huang} \& {Bhattacharjee}}{{Huang} \&
  {Bhattacharjee}}{2012}]{huang_12}
{Huang} Y.-M.,  {Bhattacharjee} A.,  2012, Physical Review Letters, 109, 265002

\bibitem[\protect\citeauthoryear{{Jaroschek}, {Lesch} \&
  {Treumann}}{{Jaroschek} et~al.}{2004}]{jaroschek_04}
{Jaroschek} C.~H.,  {Lesch} H.,    {Treumann} R.~A.,  2004, \apjl, 605, L9

\bibitem[\protect\citeauthoryear{{Kagan}, {Milosavljevi{\'c}} \&
  {Spitkovsky}}{{Kagan} et~al.}{2013}]{kagan_13}
{Kagan} D.,  {Milosavljevi{\'c}} M.,    {Spitkovsky} A.,  2013, \apj, 774, 41

\bibitem[\protect\citeauthoryear{{Kagan}, {Nakar} \& {Piran}}{{Kagan}
  et~al.}{2016}]{kagan_16}
{Kagan} D.,  {Nakar} E.,    {Piran} T.,  2016, ArXiv e-prints

\bibitem[\protect\citeauthoryear{{Kagan}, {Sironi}, {Cerutti} \&
  {Giannios}}{{Kagan} et~al.}{2015}]{kagan_15}
{Kagan} D.,  {Sironi} L.,  {Cerutti} B.,    {Giannios} D.,  2015, Space Science
  Reviews

\bibitem[\protect\citeauthoryear{{Kirk} \& {Skj{\ae}raasen}}{{Kirk} \&
  {Skj{\ae}raasen}}{2003}]{kirk_sk_03}
{Kirk} J.~G.,  {Skj{\ae}raasen} O.,  2003, \apj, 591, 366

\bibitem[\protect\citeauthoryear{{Liu}, {Li}, {Yin}, {Albright}, {Bowers} \&
  {Liang}}{{Liu} et~al.}{2011}]{liu_11}
{Liu} W.,  {Li} H.,  {Yin} L.,  {Albright} B.~J.,  {Bowers} K.~J.,    {Liang}
  E.~P.,  2011, Physics of Plasmas, 18, 052105

\bibitem[\protect\citeauthoryear{{Liu}, {Guo}, {Daughton}, {Li} \&
  {Hesse}}{{Liu} et~al.}{2015}]{liu_15}
{Liu} Y.-H.,  {Guo} F.,  {Daughton} W.,  {Li} H.,    {Hesse} M.,  2015,
  Physical Review Letters, 114, 095002

\bibitem[\protect\citeauthoryear{{Loureiro}, {Samtaney}, {Schekochihin} \&
  {Uzdensky}}{{Loureiro} et~al.}{2012}]{loureiro_12}
{Loureiro} N.~F.,  {Samtaney} R.,  {Schekochihin} A.~A.,    {Uzdensky} D.~A.,
  2012, Physics of Plasmas, 19, 042303

\bibitem[\protect\citeauthoryear{{Lyubarsky} \& {Kirk}}{{Lyubarsky} \&
  {Kirk}}{2001}]{lyubarsky_kirk_01}
{Lyubarsky} Y.,  {Kirk} J.~G.,  2001, \apj, 547, 437

\bibitem[\protect\citeauthoryear{{Lyubarsky} \& {Liverts}}{{Lyubarsky} \&
  {Liverts}}{2008}]{lyubarsky_liverts_08}
{Lyubarsky} Y.,  {Liverts} M.,  2008, \apj, 682, 1436

\bibitem[\protect\citeauthoryear{{Lyubarsky}}{{Lyubarsky}}{2003}]{lyubarsky_03}
{Lyubarsky} Y.~E.,  2003, \mnras, 345, 153

\bibitem[\protect\citeauthoryear{{Lyubarsky}}{{Lyubarsky}}{2005}]{lyubarsky_05}
{Lyubarsky} Y.~E.,  2005, \mnras, 358, 113

\bibitem[\protect\citeauthoryear{{Lyutikov} \& {Blandford}}{{Lyutikov} \&
  {Blandford}}{2003}]{lyutikov_03}
{Lyutikov} M.,  {Blandford} R.,  2003, ArXiv:astro-ph/0312347

\bibitem[\protect\citeauthoryear{{Lyutikov} \& {Uzdensky}}{{Lyutikov} \&
  {Uzdensky}}{2003}]{lyutikov_uzdensky_03}
{Lyutikov} M.,  {Uzdensky} D.,  2003, \apj, 589, 893

\bibitem[\protect\citeauthoryear{{Melzani}, {Walder}, {Folini}, {Winisdoerffer}
  \& {Favre}}{{Melzani} et~al.}{2014}]{melzani_14}
{Melzani} M.,  {Walder} R.,  {Folini} D.,  {Winisdoerffer} C.,    {Favre}
  J.~M.,  2014, \aap, 570, A112

\bibitem[\protect\citeauthoryear{{Murphy}, {Young}, {Shen}, {Lin} \&
  {Ni}}{{Murphy} et~al.}{2013}]{murphy_13}
{Murphy} N.~A.,  {Young} A.~K.,  {Shen} C.,  {Lin} J.,    {Ni} L.,  2013,
  Physics of Plasmas, 20, 061211

\bibitem[\protect\citeauthoryear{{Nalewajko}, {Uzdensky}, {Cerutti}, {Werner}
  \& {Begelman}}{{Nalewajko} et~al.}{2015}]{nalewajko_15}
{Nalewajko} K.,  {Uzdensky} D.~A.,  {Cerutti} B.,  {Werner} G.~R.,
  {Begelman} M.~C.,  2015, \apj, 815, 101

\bibitem[\protect\citeauthoryear{{P{\'e}tri} \& {Lyubarsky}}{{P{\'e}tri} \&
  {Lyubarsky}}{2007}]{petri_lyubarsky_07}
{P{\'e}tri} J.,  {Lyubarsky} Y.,  2007, \aap, 473, 683

\bibitem[\protect\citeauthoryear{{Romanova} \& {Lovelace}}{{Romanova} \&
  {Lovelace}}{1992}]{romanova_92}
{Romanova} M.~M.,  {Lovelace} R.~V.~E.,  1992, \aap, 262, 26

\bibitem[\protect\citeauthoryear{{Sironi}, {Petropoulou} \&
  {Giannios}}{{Sironi} et~al.}{2015}]{sironi_15}
{Sironi} L.,  {Petropoulou} M.,    {Giannios} D.,  2015, \mnras, 450, 183

\bibitem[\protect\citeauthoryear{{Sironi} \& {Spitkovsky}}{{Sironi} \&
  {Spitkovsky}}{2009}]{sironi_spitkovsky_09}
{Sironi} L.,  {Spitkovsky} A.,  2009, \apj, 698, 1523

\bibitem[\protect\citeauthoryear{{Sironi} \& {Spitkovsky}}{{Sironi} \&
  {Spitkovsky}}{2011a}]{sironi_spitkovsky_11b}
{Sironi} L.,  {Spitkovsky} A.,  2011a, \apj, 741, 39

\bibitem[\protect\citeauthoryear{{Sironi} \& {Spitkovsky}}{{Sironi} \&
  {Spitkovsky}}{2011b}]{sironi_spitkovsky_11a}
{Sironi} L.,  {Spitkovsky} A.,  2011b, \apj, 726, 75

\bibitem[\protect\citeauthoryear{{Sironi} \& {Spitkovsky}}{{Sironi} \&
  {Spitkovsky}}{2012}]{sironi_spitkovsky_12}
{Sironi} L.,  {Spitkovsky} A.,  2012, Computational Science and Discovery, 5,
  014014

\bibitem[\protect\citeauthoryear{{Sironi} \& {Spitkovsky}}{{Sironi} \&
  {Spitkovsky}}{2014}]{ss_14}
{Sironi} L.,  {Spitkovsky} A.,  2014, \apjl, 783, L21

\bibitem[\protect\citeauthoryear{{Sironi}, {Spitkovsky} \& {Arons}}{{Sironi}
  et~al.}{2013}]{sironi_13}
{Sironi} L.,  {Spitkovsky} A.,    {Arons} J.,  2013, \apj, 771, 54

\bibitem[\protect\citeauthoryear{{Spitkovsky}}{{Spitkovsky}}{2005}]{spitkovsky_05}
{Spitkovsky} A.,  2005, in {T.~Bulik, B.~Rudak, \& G.~Madejski} ed.,
  Astrophysical Sources of High Energy Particles and Radiation Vol.~801 of AIP
  Conf. Ser., {Simulations of relativistic collisionless shocks: shock
  structure and particle acceleration}.
p.~345

\bibitem[\protect\citeauthoryear{{Spruit}}{{Spruit}}{2010}]{spruit_10}
{Spruit} H.~C.,  2010, in {Belloni} T.,  ed., Lecture Notes in Physics, Berlin
  Springer Verlag Vol.~794 of Lecture Notes in Physics, Berlin Springer Verlag,
  {Theory of Magnetically Powered Jets}.
p.~233

\bibitem[\protect\citeauthoryear{{Spruit}, {Daigne} \& {Drenkhahn}}{{Spruit}
  et~al.}{2001}]{spruit_01}
{Spruit} H.~C.,  {Daigne} F.,    {Drenkhahn} G.,  2001, \aap, 369, 694

\bibitem[\protect\citeauthoryear{{Synge}}{{Synge}}{1957}]{synge_57}
{Synge} J.~L., , 1957, {The Relativistic Gas (North-Holland, Amsterdam), 33}

\bibitem[\protect\citeauthoryear{{Takamoto}}{{Takamoto}}{2013}]{takamoto_13}
{Takamoto} M.,  2013, \apj, 775, 50

\bibitem[\protect\citeauthoryear{{Thompson}}{{Thompson}}{1994}]{thompson_94}
{Thompson} C.,  1994, \mnras, 270, 480

\bibitem[\protect\citeauthoryear{{Thompson}}{{Thompson}}{2006}]{thompson_06}
{Thompson} C.,  2006, \apj, 651, 333

\bibitem[\protect\citeauthoryear{{Usov}}{{Usov}}{1994}]{usov_94}
{Usov} V.~V.,  1994, \mnras, 267, 1035

\bibitem[\protect\citeauthoryear{{Uzdensky}, {Loureiro} \&
  {Schekochihin}}{{Uzdensky} et~al.}{2010}]{uzdensky_10}
{Uzdensky} D.~A.,  {Loureiro} N.~F.,    {Schekochihin} A.~A.,  2010, Physical
  Review Letters, 105, 235002

\bibitem[\protect\citeauthoryear{{Werner}, {Uzdensky}, {Cerutti}, {Nalewajko}
  \& {Begelman}}{{Werner} et~al.}{2016}]{werner_16}
{Werner} G.~R.,  {Uzdensky} D.~A.,  {Cerutti} B.,  {Nalewajko} K.,
  {Begelman} M.~C.,  2016, \apjl, 816, L8

\bibitem[\protect\citeauthoryear{{Yin}, {Daughton}, {Karimabadi}, {Albright},
  {Bowers} \& {Margulies}}{{Yin} et~al.}{2008}]{yin_08}
{Yin} L.,  {Daughton} W.,  {Karimabadi} H.,  {Albright} B.~J.,  {Bowers} K.~J.,
     {Margulies} J.,  2008, Physical Review Letters, 101, 125001

\bibitem[\protect\citeauthoryear{{Zenitani} \& {Hoshino}}{{Zenitani} \&
  {Hoshino}}{2001}]{zenitani_01}
{Zenitani} S.,  {Hoshino} M.,  2001, \apjl, 562, L63

\bibitem[\protect\citeauthoryear{{Zenitani} \& {Hoshino}}{{Zenitani} \&
  {Hoshino}}{2005}]{zenitani_05b}
{Zenitani} S.,  {Hoshino} M.,  2005, Physical Review Letters, 95, 095001

\bibitem[\protect\citeauthoryear{{Zenitani} \& {Hoshino}}{{Zenitani} \&
  {Hoshino}}{2007}]{zenitani_07}
{Zenitani} S.,  {Hoshino} M.,  2007, \apj, 670, 702

\bibitem[\protect\citeauthoryear{{Zenitani} \& {Hoshino}}{{Zenitani} \&
  {Hoshino}}{2008}]{zenitani_08}
{Zenitani} S.,  {Hoshino} M.,  2008, \apj, 677, 530

\end{thebibliography}

\appendix

\section[]{Plasmoid Spatial Profiles}\label{app:profile}
\begin{figure*}
\centering
 \resizebox{0.85\hsize}{!}{\includegraphics{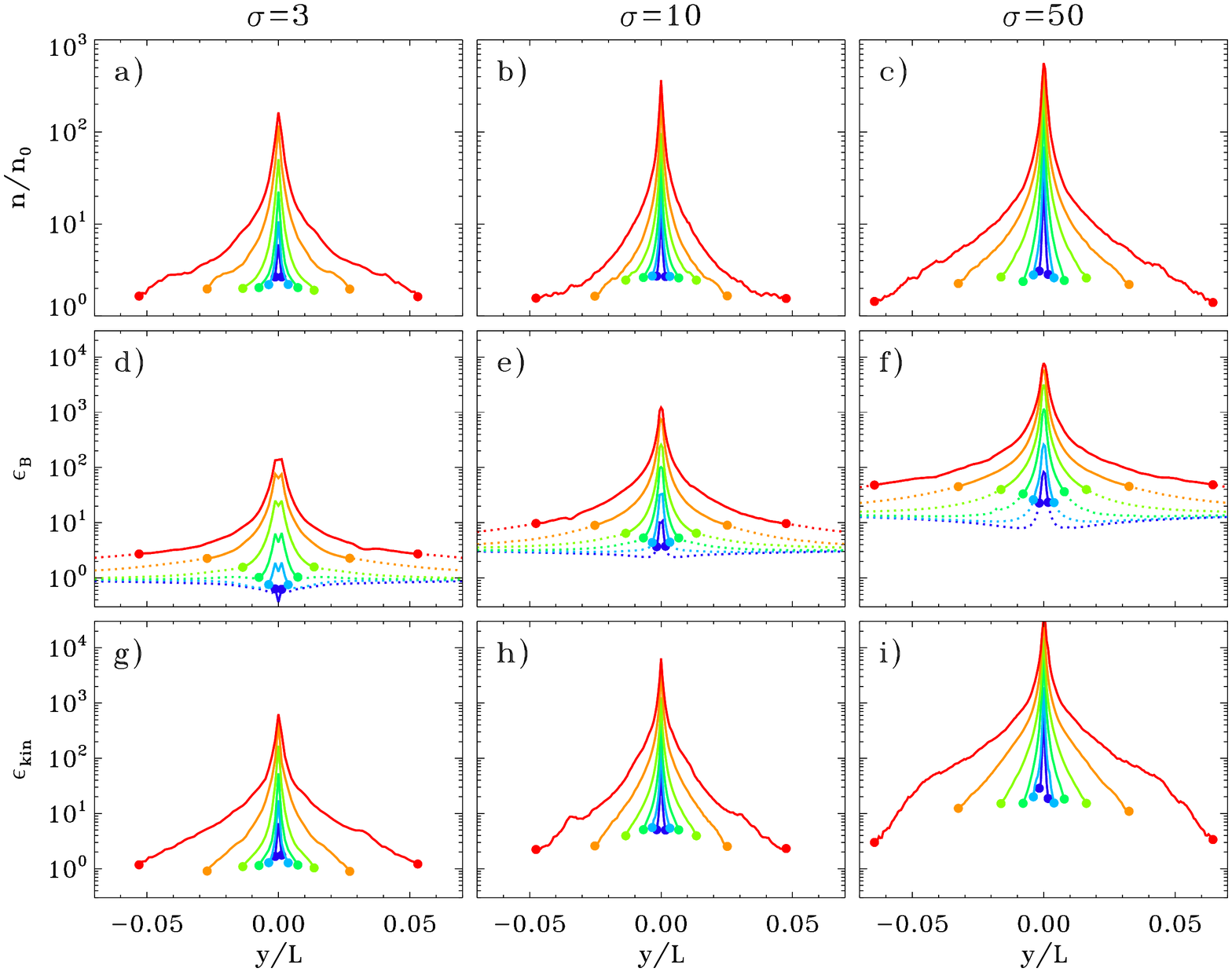}}
 \caption{1D profiles of various plasmoid properties, along the direction $y$ perpendicular to the current layer, for three values of the magnetization, as indicated at the top ($\sigma=3$ in the left column, $\sigma=10$ in the middle column, and $\sigma=50$ in the right column). In each panel, curves of different colors correspond to a different bin of plasmoid width $w$ (from blue to red in order of increasing $w$), and each line is obtained by averaging the $y$-profile at the plasmoid $x$-location over all the plasmoids whose width falls in that range. The corresponding plasmoid width can be read from the separation, along the $y$ direction, between the two filled circles, that denote the plasmoid boundaries along $y$. The bins in width are the same as in \fig{spec}. From the top to the bottom row, we plot: (a)-(c) the comoving density $n$, in units of the density $n_0$ far from the current sheet; (d)-(f) the magnetic energy fraction $\epsilon_{\rm B}=B^2/8\pi n_0 m c^2$, where $B$ is measured in the plasmoid comoving frame; (g)-(i) the internal energy fraction $\epsilon_{\rm kin}=(\langle\gamma\rangle-1)\, n/n_0$, where $\langle\gamma\rangle$ is the mean particle Lorentz factor in the plasmoid frame. In the second row of panels, we plot the profile of $\epsilon_{\rm B}$ outside the plasmoids with dotted lines. Here, $B\sim |B_x|$, which is Lorentz invariant, so the $\epsilon_{\rm B}$ profile outside the plasmoids does not depend on the plasmoid velocity, and the profiles of plasmoids with different speeds can be meaningfully averaged. This argument does not hold for $n$ or $\epsilon_{\rm kin}$.}
 \label{fig:profile}
\end{figure*}
\fig{profile} presents the spatial profiles of various comoving quantities along the $y$ direction transverse to the current sheet, for the three values of  magnetization that we explore in this work ($\sigma=3$ on the left, $\sigma=10$ in the middle and $\sigma=50$ on the right). In each panel, curves of different colors correspond to a different bin of plasmoid width $w$ (from blue to red in order of increasing $w/L$), and each line is obtained by averaging the $y$-profile at the plasmoid $x$-location over all the plasmoids whose width falls in that range. The corresponding plasmoid width can be read from the separation, along the $y$ direction, between the two filled circles, that denote the plasmoid boundaries along $y$.  

In the top row, we show the $y$-profiles of the plasmoid comoving density, in units of the particle number density $n_0$ far from the current sheet. By plotting on a log-log scale (not shown), we can measure that  the density scales with distance from the current sheet roughly as $n\propto (|y|/w)^{-1.0}$, with only a slight tendency for steeper profiles at higher magnetizations. This scaling implies that the plasmoid mass is mostly sensitive to the value of the density at the plasmoid outskirts. The top row in \fig{profile} shows that the density in the outskirts is a weak function of the magnetization, which explains why the scaling $\propto \sqrt{\sigma}$ expected for the mean density in relativistic reconnection \citep{lyubarsky_05} is not realized in the second row of \fig{islfluids}.\footnote{Yet, such a scaling seems to be fulfilled by the peak density (i.e., at $y\sim 0$) of the largest islands (compare the three red curves for different values of $\sigma$ in the top row of \fig{profile}).} As we have argued in \sect{fluid}, the scaling $n\propto \sqrt{\sigma}$ is to be expected only for the fastest plasmoids, that can reach the terminal four-velocity $\sim \sqrt{\sigma}\,c$. As described in \sect{accel}, only few plasmoids approach $\sim \sqrt{\sigma}\,c$ for $\sigma=50$, which justifies why the density in the plasmoid outskirts in the top row of \fig{profile} does not appreciably increase between $\sigma=10$ and $\sigma=50$.

The second row in \fig{profile} illustrates the $y$-profile of the comoving magnetic energy fraction. Since the slice along $y$ where we compute the profiles goes through the plasmoid center, the magnetic field $B$ is dominated by its $x$-component, which is Lorentz invariant. For this reason, we can extend the profiles in the second row of \fig{profile} outside of the plasmoid contour (see the dotted lines), without ambiguity (i.e., the profiles of plasmoids with different speeds can be meaningfully averaged). By plotting on a log-log scale (not shown), we can measure that  the magnetic energy scales as $\epsilon_{\rm B}\propto (|y|/w)^{-1.2}$, regardless of the magnetization. This scaling implies that most of the magnetic energy is contributed by the plasmoid outskirts (filled circles in the plot). Since \fig{profile} shows that the magnetic energy at the plasmoid boundary scales as $\propto \sigma$ (e.g., see the red curves, for the largest islands), the same scaling is expected for the volume-averaged magnetic energy fraction, as indeed found in the fourth row of \fig{islfluids}. 

For a given magnetization, \fig{profile} reveals that the magnetic field at the plasmoid boundary has a strong dependence on the island width, with smaller islands systematically residing in regions with weaker fields. As described in \sect{struct}, this trend is due to field lines wrapping around a large island, so that the smaller islands in its vicinity will preferentially lie in a region where the field lines are less densely spaced, and so the field is weaker. The fact that smaller islands are more likely to reside in ``wells'' of magnetic energy, as revealed by the second row of \fig{profile}, explains the apparent lack of magnetic flux and magnetic energy seen in \fig{islfluids} (third and fourth row) at small plasmoid widths.

The bottom row of panels in \fig{profile} describes the profiles of the comoving particle kinetic energy. On a log-log scale (not shown), we find that  the kinetic energy fraction scales with distance from the sheet as $\epsilon_{\rm kin}\propto (|y|/w)^{-1.4}$, regardless of the magnetization. By comparing with the density profile, we find that the mean kinetic energy per particle scales with distance as $\propto  (|y|/w)^{-0.4}$, with only a slight tendency for a harder slope at higher $\sigma$. This scaling for the mean energy per particle (i.e., for the temperature) is consistent with adiabatic heating, since for an ultra-relativistic gas it should depend on the density as $\propto n^{1/3}\propto (|y|/w)^{-0.3}$.
The scaling $\epsilon_{\rm kin}\propto (|y|/w)^{-1.4}$ of the kinetic energy fraction is slightly steeper than the magnetic energy profile, so that the island core will typically be dominated by the particle kinetic energy, whereas the magnetic energy will govern the plasmoid outskirts. 

As in the case of the magnetic energy fraction, the scaling $\epsilon_{\rm kin}\propto (|y|/w)^{-1.4}$
implies that most of the kinetic energy is contributed by the plasmoid outskirts. There, $\epsilon_{\rm kin}$ scales as $\propto \sigma$ (e.g., see the red curves, for the largest islands), so the same scaling is expected for the volume-averaged kinetic energy fraction, as indeed found in the fifth row of \fig{islfluids}. Unlike the magnetic energy fraction, the value of the kinetic energy at the plasmoid boundary does not appreciably depend on the plasmoid size, at fixed $\sigma$ (see the filled circles in the bottom row). In turn, this explains why the kinetic energy fraction in the fifth row of \fig{islfluids} is nearly constant with respect to the plasmoid width (especially as compared to the magnetic energy fraction in the fourth row of \fig{islfluids}, which shows a significant deficit at small sizes).

\section[]{Comparison of the Spectrum in Plasmoids with the Overall Spectrum}\label{app:speccomp}
\begin{figure*}
\centering
 \resizebox{0.8\hsize}{!}{\includegraphics{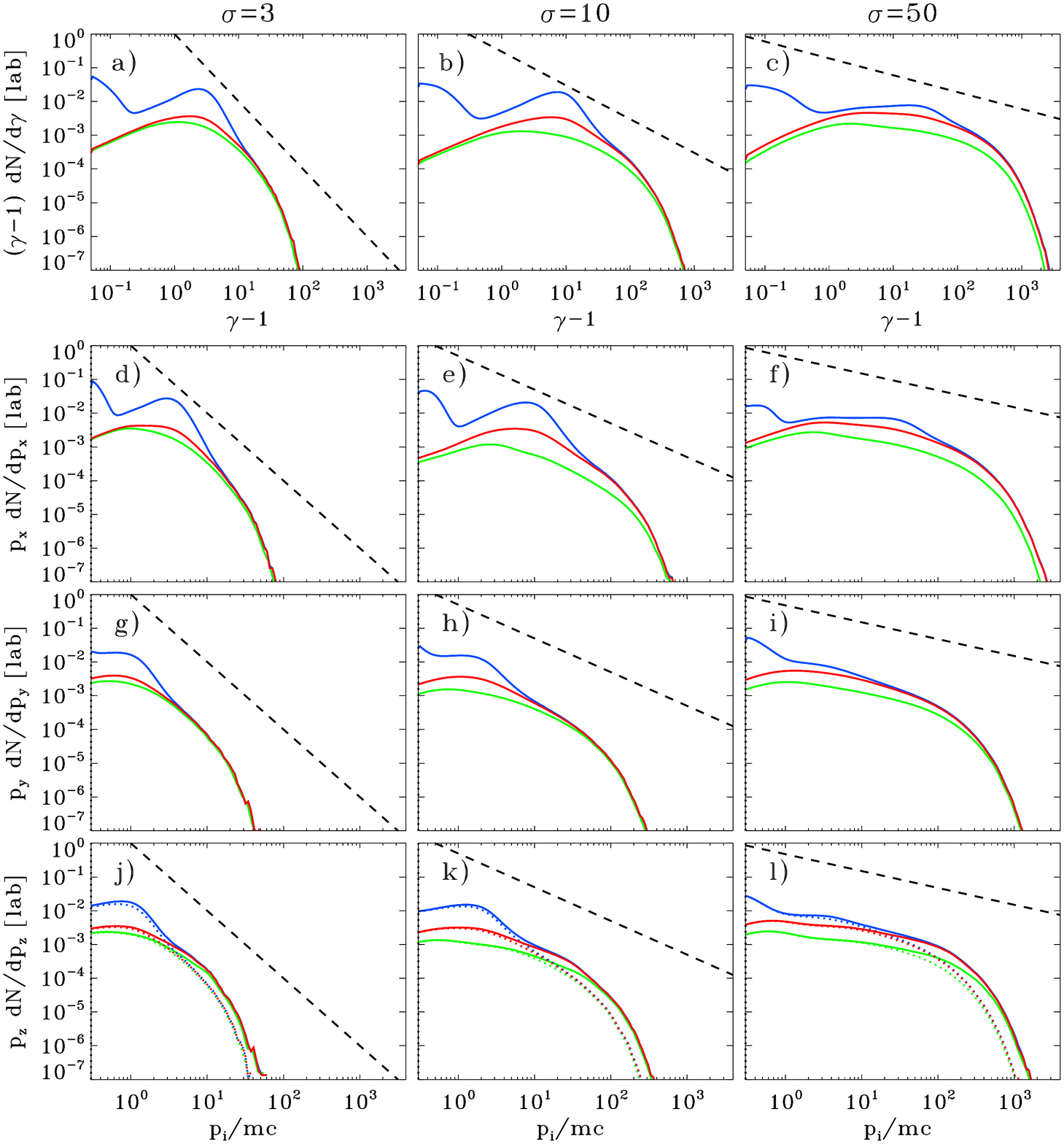}}
 \caption{Positron energy and momentum spectra in the laboratory frame, for three values of the magnetization, as indicated at the top ($\sigma=3$ in the left column, $\sigma=10$ in the middle column, and $\sigma=50$ in the right column). In each panel, we plot the overall spectrum from the current sheet (blue lines; integrated over the whole reconnection layer and over the timespan of our simulations), the spectrum of plasmoids (red lines; integrated over all plasmoids at all times), and the cumulative spectrum from the five largest plasmoids (green lines; integrated over the history of the five largest plasmoids). We plot: (a)-(c) the positron energy  spectrum in the laboratory frame; (d)-(f) the positron $p_x$  spectrum in the laboratory frame;  (g)-(i) the positron $p_y$ spectrum in the laboratory frame;  (j)-(l) the positron $p_z$ spectrum in the laboratory frame, differentiating between positrons with $p_z>0$ (solid curves) and $p_{z}<0$ (dotted curves). For reference, we plot with black dashed lines a power law with slope $s=3$ for $\sigma=3$, $s=2$ for $\sigma=10$ and $s=1.5$ for $\sigma=50$. The plot shows that the highest energy particles are contained within the largest plasmoids.}
 \label{fig:speccomp}
\end{figure*}

\fig{speccomp} illustrates how the positron spectrum in the plasmoids compares with the positron spectrum in the overall current sheet (i.e., including also the region in between plasmoids). The spectra are measured in the laboratory frame for the three values of magnetization that we explore in this work ($\sigma=3$ in the left column, $\sigma=10$ in the middle column, and $\sigma=50$ in the right column). In each panel, we plot the overall spectrum from the current sheet (blue lines; integrated over the whole reconnection layer and over the timespan of our simulations), the spectrum of plasmoids (red lines; integrated over all the plasmoids at all times), and the cumulative spectrum from the five largest plasmoids (green lines; integrated over the history of the five largest plasmoids).

Both the energy spectra in the top row and the momentum spectra in the other panels suggest that the highest energy particles always belong to a plasmoid. In all the panels, the blue and red curves overlap at the high-energy end. Even more dramatically, most of the high-energy particles are contained in the few largest plasmoids (compare with the green curves, which only account for the five largest plasmoids). It follows that the maximum energy of particles accelerated in reconnection is identical to the maximum energy of particles contained in plasmoids, which validates the generality of the Hillas criterion discussed in \sect{size}.

\fig{speccomp} confirms what we had anticipated in \sect{spec}, i.e., that the spectral slope of the largest plasmoids (green curves) asymptotes to $s\sim 3$ for $\sigma=3$, $s\sim 2$ for $\sigma=10$ and $s\sim 1.5$ for $\sigma=50$, as indicated by the dashed black lines. Such slopes are consistent with the values quoted in SS14, where the particle spectrum was integrated over the whole current sheet. As \fig{speccomp} suggests, this is because the spectrum integrated over the whole layer is actually dominated by the few largest islands. It is then quite natural to expect that the spectral slope of the largest islands in \fig{speccomp} is comparable to the power-law index found in SS14.

In the particle energy spectra of the top row, the excess of particles at intermediate energies in the blue line (as compared to the red line) is due to hot particles in the reconnection outflow that do not belong to plasmoids.  They move with a bulk Lorentz factor $\propto \sqrt{\sigma}$ and their mean temperature also scales as $\propto \sqrt{\sigma}$. Overall, this implies that their mean energy in the lab frame should scale as $\propto \sigma$, as indeed observed in the top row of \fig{speccomp} (see the peak at $\gamma-1\sim 3$ for $\sigma=3$, at $\gamma-1\sim 7$ for $\sigma=10$ and at $\gamma-1\sim 30$ for $\sigma=50$). A signature of this bump at intermediate energies is also seen in the spectrum of small plasmoids (compare the red curve in panel (b) with the green line, showing that such bump is absent in the spectrum of the largest plasmoids). In simulations with periodic boundaries, where the spectrum at any given time is dominated by the particle content in the largest primary islands, one would expect that such bump at intermediate energies would be buried underneath the broad non-thermal spectrum of the largest islands, as it was indeed the case in SS14.

Since the particles that populate the bump at intermediate energies are preferentially moving along the $x$ direction of the outflow, a similar signature should appear in the second row of \fig{speccomp}, as it is indeed observed. In the $y$ and $z$ momentum spectra, i.e., in the directions perpendicular to the bulk outflow, one should still see the thermal component of this particle population, whose temperature scales as $\propto \sqrt{\sigma}$. In fact, this explains the bump appearing in the third and bottom rows at trans-relativistic momenta, with a clear tendency for the peak momentum to increase with magnetization (the peak is located at $\sim 0.8\, mc$ for $\sigma=3$, at $\sim 1.5\, mc$ for $\sigma=10$ and at $\sim 3\, mc$ for $\sigma=50$).

In the bottom row, we present the positron $p_z$ spectrum in the laboratory frame, differentiating between positrons with $p_z>0$ (solid curves) and $p_{z}<0$ (dotted curves). This shows that the momentum spectrum in the $+z$ direction of the reconnection electric field is slightly harder than along the $-z$ direction (or along the $y$ direction). The asymmetry between the $p_z>0$ and $p_z<0$ momentum spectra is most pronounced at the highest energies, in agreement with the argument in \sect{spec}, i.e., that the anisotropy in large islands is driven by the curvature and $\nabla B$ drift speed, which for a fixed island width is an increasing function of the particle energy.
 

\end{document}